\documentclass[11pt]{amsart}
\usepackage[margin=1in]{geometry}
\usepackage{amssymb}
\usepackage{mathrsfs}
\usepackage[hidelinks]{hyperref}
\date{}
\usepackage{mathtools}
\usepackage{tikz}
\allowdisplaybreaks[1]
\newcommand{\dd}{\,\mathrm d}

\numberwithin{equation}{section}
\theoremstyle{definition}
\newtheorem{theorem}{Theorem}[section]
\newtheorem{lemma}[theorem]{Lemma}
\newtheorem{proposition}[theorem]{Proposition}
\newtheorem{definition}[theorem]{Definition}
\theoremstyle{remark}
\newtheorem*{rema*}{Remark}
\theoremstyle{definition}
\newtheorem*{theoB}{Theorem B}
\newtheorem*{theoC}{Theorem C}
\newcommand{\diff}{\mathrm{d}}

\title[An extension principle with finite Doppler exponent]{An extension principle for Einstein--Maxwell--charged scalar field system with finite Doppler exponent}
\author{Xinliang An}
\address{Department of Mathematics, National University of Singapore, Singapore}
\email{matax@nus.edu.sg}
\author{Haoyang Chen}
\address{School of Mathematical Sciences, Nankai University, China}
\email{chenhy@nankai.edu.cn}
\author{Taoran He}
\address{Institut des Hautes \'Etudes Scientifiques, Bures-sur-Yvette, France}
\email{taoran@ihes.fr}


\begin{document}

\begin{abstract}

We prove a new extension principle for the spherically symmetric Einstein--Maxwell--charged scalar field system. Rather than imposing conditions for mass ratio $2m/r$ as in Christodoulou's small-mass-ratio criterion or as in An--Chen--Singh's proof \cite{AnChenSingh} with $2m/r<3/8$, here we work through a proposal by An-Tan \cite{AnTan} with finite Doppler exponent (blueshift). By conducting two successive blow-up analyses, followed by a stationary comparison and a rigidity argument, we establish the uniform boundness of $\alpha'$ with $\alpha'=\partial_v\alpha/\partial_vr$ and $\alpha=\partial_v(r\phi)/\partial_vr$. This provides the desired extension principle in An-Tan \cite{AnTan} and completes the picture of weak cosmic censorship developed therein.

\vspace{2mm} 
As we were about to release our preprint, Cicortas communicated to us that together with his collaborators, they concomitantly used a different approach to obtain the corresponding extension
principle with finite blueshift (Doppler exponent) for the uncharged Einstein-scalar field system. Hence, we decided to put the current paper as an appendix to An-Tan \cite{AnTan}, thereby completing the proof of the weak cosmic censorship for the Einstein-Maxwell-charged scalar field system. At the same time, we are working with Cicortas and his collaborators to write a new preprint containing both approaches for the
Einstein-scalar field system.

\par
\end{abstract}

\maketitle

\setcounter{tocdepth}{2}
\tableofcontents
\medskip

\section{Introduction}

Extension principles describe when a solution of the Einstein equations
can be continued beyond a possible singular boundary. In spherical
symmetry, the center, where the area radius $r$ vanishes, is a natural
place to study this question. For scalar-field collapse, the Hawking
mass $m$ and the dimensionless ratio $\mu=2m/r$ provide geometric
quantities with which to formulate continuation criteria.

In this paper we prove a $C^1$ extension principle for the
Einstein--real scalar field and the Einstein--Maxwell--charged scalar
field systems under uniform finite upper bound for the Doppler
exponent (blueshift). The bound is imposed throughout a characteristic
triangle ending at a possible first singular point of the center.
Apart from the absence of trapped spheres, no smallness condition on
the mass ratio is required. Our argument starts from the obtained pointwise estimates of
An--Chen--Singh \cite{AnChenSingh}, and conducts two successive
blow-up analysis, followed by a stationary comparison and a
rigidity argument.

\subsection[The main theorems]{The main theorems}

We use double null coordinates $(u,v)$ in which the center is $u=v$ and the initial outgoing cone is $u=0$. For $a<b$, write
\[
 \mathcal D(a,b)=\{(u,v):a\leq u\leq v<b\}.
\]
Fix $0<v_*<\infty$. Prescribe regular characteristic data of class $C^1$ on $0\leq v\leq v_*+\varepsilon_0$ for some $\varepsilon_0>0$, normalized by $\partial_vr(0,v)=1/2$, with
\[
 \max_{0\leq v\leq v_*}\mu(0,v)<1.
\]
We consider a development that is regular on every closed subtriangle with terminal parameter less than $v_*$. Thus $(v_*,v_*)$ is a first possible singular point at the center. The regularity class and the characteristic datum are specified in Section~\ref{sec:statements}.

For the real scalar field and charged scalar field, respectively, define
\begin{equation}\label{eq:doppler}
\begin{aligned}
 \gamma_0(u,v)&=\int_0^u
       \frac{-\partial_ur}{r}\frac{\mu}{1-\mu}(u',v)\,\mathrm du',\\
 \gamma_Q(u,v)&=\int_0^u
       \frac{-\partial_ur}{r}\frac{\mu-Q^2/r^2}{1-\mu}(u',v)\,\mathrm du',
\end{aligned}
\end{equation}
where $Q$ is the enclosed electric charge. All suprema in the following statements are taken over the entire triangle $\mathcal D(0,v_*)$.

\begin{theoB}[An--Chen--He: real scalar field]\label{thm:real}
Consider the spherically symmetric Einstein--real scalar field system in the characteristic setting above. Assume
\[
 \partial_ur<0,\qquad \partial_vr>0,\qquad
 0\leq\mu<1\quad\hbox{at positive radius},\qquad
 \sup_{\mathcal D(0,v_*)}\gamma_0<\infty.
\]
Then $\alpha'=\partial_v\alpha/\partial_vr$, with $\alpha=\partial_v(r\phi)/\partial_vr$, is uniformly bounded on $\mathcal D(0,v_*)$. For some $\xi>0$, the solution extends as a regular $C^1$ solution to $\mathcal D(0,v_*+\xi)$, agreeing with the prescribed continuation of the original characteristic data.
\end{theoB}

\begin{samepage}
{
\begin{theoC}[An--Chen--He: charged scalar field]\label{thm:charged}
Consider the spherically symmetric Einstein--Maxwell--charged scalar field system with vanishing scalar mass in the same characteristic setting. Assume
\[
 \partial_ur<0,\qquad \partial_vr>0,\qquad
 0\leq\mu<1\quad\hbox{at positive radius},\qquad
 \sup_{\mathcal D(0,v_*)}\gamma_Q<\infty.
\]
Then $\alpha'=\partial_v\alpha/\partial_vr$, with $\alpha=\partial_v(r\phi)/\partial_vr$, is uniformly bounded on $\mathcal D(0,v_*)$. For some $\xi>0$, the solution extends as a regular $C^1$ solution to $\mathcal D(0,v_*+\xi)$, agreeing with the prescribed continuation of the original characteristic data.
\end{theoC}
}
\end{samepage}

\subsection{Previous work}

Christodoulou developed the characteristic existence and uniqueness
theory for the spherically symmetric Einstein--scalar field system
\cite{Chr86a,Chr86b,Chr87}, established criteria for trapped-surface
formation \cite{Chr91}, and constructed naked singularities
\cite{Chr94}. Together with his theory of solutions of bounded variation
\cite{Chr93} and the instability theorem \cite{Chr99}, these results form the foundation and provide the
framework for establishing the weak cosmic censorship. In particular, his extension principle gives continuation when the mass
ratio $\mu=2m/r$ is uniformly small near a possible first singular
center. The Doppler exponent $\gamma_0$ also enters his analysis of the
instability of naked singularities.

\vspace{1mm}

An--Tan \cite{AnTan} generalized all these arguments for the Einstein--Maxwell--charged scalar field system with several new features explored and established. To make the whole picture complete, they proposed an extension criterion with finite Doppler exponent.

\vspace{1mm}

Toward this desired extension principle with finite Doppler exponent, An--Chen--Singh \cite{AnChenSingh} made progress and
proved the extension principle with assuming $\mu<3/8$. Besides this extension principle, An-Chen-Singh's arguments also obtain various pointwise estimates for scale-invariant quantities, and prove higher-order spacetime identities
with self-similar weights to control the outgoing variation of the scalar field. The additional restriction on $\mu$ is used in the bulk positivity of those identities. 

\vspace{1mm}

Instead of the spacetime identities, with obtained scale-invariant estimates in \cite{AnChenSingh}, the current paper 
conducts two successive blow-up analyses, followed by a stationary comparison and a rigidity argument. These blow-up analyses give the final push and we establish the desired extension principle with finite Doppler exponent as proposed and conjectured in \cite{AnTan} by An-Tan. For the blow-up analysis, this paper is inspired by the works of Kenig-Merle \cite{KenigMerle} and Koch-Nadirashvili-Seregin-\v{S}ver\'ak,\cite{KNSS}.

\vspace{2mm}
In the end of this subsection, we also mention the works of \cite{LiuLi} by Li-Liu, \cite{LiInterior} by Li on studying the instability mechanism of the spherically symmetric Einstein-scalar field system, and \cite{Cicortas2026} by Cicortas on proving the weak cosmic censorship for the circularly symmetric Einstein-scalar field system in $2+1$ dimensions.   

\subsection{New Ingredients in the Proof}

Since Christodoulou's results, the quantities
\[
 \alpha=\frac{\partial_v(r\phi)}{\partial_vr},\qquad
 \alpha'=\frac{\partial_v\alpha}{\partial_vr}
\]
are adapted to the characteristic formulation.
The pointwise estimates in the previous work have controlled
$r^{\delta}\alpha'$ with $\delta>0$ rather than $\alpha'$. Our proof in this paper directly aims to prove the uniform bound of $|\alpha|'$. 

\vspace{2mm}

We prove a uniform bound for
$\alpha'$ by excluding the arising limiting solutions if it is unbounded. Our new arguments can be divided into three steps:  

\begin{enumerate}
	\item 
\smallskip
\noindent\textit{First blow-up and a nontrivial ancient limit.}
If $|\alpha'|$ is unbounded, we rescale the spacetime through
successively larger and larger values so that the rescaled $\alpha'$ is bounded
by one. The pointwise estimates and the transport equation for
$\alpha'$ give spacetime convergence with expanding domains, including at the
regular center. The transport equation also prevents the scalar
field from becoming constant in the limit, so that the limiting
solution has positive Hawking mass somewhere. This produces a
nontrivial ancient solution, defined for all negative times.

\item
\smallskip
\noindent\textit{Second blow-up at the mass scale.}
The first limiting solution need not have bounded mass in its remote
past. We select earlier outgoing cones and rescale by a Hawking mass
scale, retaining a fixed positive amount of mass within a bounded
radius. The mass monotonicities give a uniform upper bound for the
rescaled mass, while the available time intervals expand in both
directions. After proving strong convergence of the incoming scalar
flux, we obtain an eternal solution with bounded, nonzero mass.

\item
\smallskip
\noindent\textit{Stationary comparison and rigidity.}
The eternal solution has asymptotic time-independent scalar field and
mass, which determine a stationary solution that can be compared near infinity.
Estimates for their differences give rigidity near infinity, and
uniqueness along the characteristics further extends this equality to every
positive radius. The stationary scalar equation and the global scalar
bound then force the mass to vanish, contradicting the positive mass
retained in the second blow-up analysis. 

\end{enumerate}

\noindent These together contradict the unboundness of $\alpha'$ in the beginning.

\vspace{1mm}

For the charged system, the additional estimates in An-Tan \cite{AnTan} are used before
and during the first normalization. Control of $Q$ recovers the
nonnegative incoming integral involving $\mu$ from the signed
Doppler integral. After the first blow-up, the rescaled coupling tends to zero and the electromagnetic fields vanish locally in the limit. The ancient limit is therefore neutral, although its
scalar field may be complex. Both systems subsequently follow the
same mass-rescale and rigidity arguments.

\subsection{Organization of the paper}

Section~\ref{sec:statements} fixes the geometric conventions,
regularity class and basic estimates from the Doppler bound.
Section~\ref{sec:estimates} proves the pointwise estimates used
in the limiting arguments. Section~\ref{sec:first} constructs
the ancient solution and establishes its past asymptotics.
Section~\ref{sec:mass-concentration} proves concentration at the
mass scale. Section~\ref{sec:second} carries out the second
normalization, constructs the eternal limit, and establishes its large-radius
asymptotics. Section~\ref{sec:rigidity} constructs the stationary
comparison, proves exterior and interior equality, and obtains
a bound for the original outgoing derivative $\alpha'$ by rigidity.
Section~\ref{sec:continuation} deduces endpoint regularity and
continuation through the original terminal center from this bound.
Appendix~\ref{characteristic-appendix} supplies the local
characteristic constructions used in the continuation step.

\subsection*{AI Disclosure}
In this paper, the successive two-step blow-up analyses was finalized through exchanging ideas with AI. We had the idea of the first blow-up analysis, the AI input for the 2nd blow-up analysis is crucial for this approach.

\subsection*{Acknowledgments}
\begingroup
\emergencystretch=1em
Xinliang An is supported by MOE Tier 1 grant A-8002933-00-00. Haoyang Chen is supported by the
Fundamental Research Funds for the Central Universities
(050-63263077). Taoran He is supported by the ERC grant
ERC-2023 AdG 101141855 BlaHSt.
\par\endgroup

\section[Geometric setup and preliminary estimates]{Geometric setup and preliminary estimates}
\label{sec:statements}\label{equations-section}

In this section, we introduce the geometric and gauge conventions used throughout this article, and list the reduced Einstein-scalar field system under spherical symmetry. We then use the Doppler bound to derive preliminary estimates in Section~\ref{Subsec:Dopplerbound}.
\subsection[The equations and the regularity class]{The equations and the regularity class}

We use the scalar normalization of An--Tan \cite{AnTan}. In particular, the scalar terms in the two Hawking mass equations carry the coefficient $2\pi$. In double null coordinates,
\[
 g=-\Omega^2\,\mathrm du\,\mathrm dv+r^2\,\mathrm d\sigma_{\mathbb S^2}^2,
\]
where $\mathrm d\sigma_{\mathbb S^2}^2$ is the round metric on the unit sphere. For a massless complex scalar field $\phi$ and coupling constant $\mathfrak e\in\mathbb R$, choose $A_v=0$ and set
\begin{equation}\label{basic-definitions}
\begin{gathered}
 D_u=\partial_u+\mathrm i\mathfrak e A_u,\qquad
 Q=-2r^2\Omega^{-2}\partial_vA_u,\\
 \mu=\frac{2m}{r},\qquad
 1-\mu=-4\Omega^{-2}\partial_ur\,\partial_vr.
\end{gathered}
\end{equation}
The mass $m$ is the Hawking mass, without electromagnetic renormalization. We fix the remaining gauge freedom by requiring $A_u=0$ on the center. The real scalar field system is obtained by setting
$\mathfrak e=Q=A_u=0$ and taking $\phi$ to be real. We write
$\gamma=\gamma_Q$ in the charged case and $\gamma=\gamma_0$ in the real
case, with both exponents defined in \eqref{eq:doppler}.

Following in An--Tan \cite{AnTan}, we introduce the scalar datum $\alpha$ and its
normalized derivative $\alpha'$ as below:
\begin{equation}\label{alpha-definitions}
 \alpha=\frac{\partial_v(r\phi)}{\partial_vr},\qquad
 \alpha'=\frac{\partial_v\alpha}{\partial_vr},\qquad
 \alpha-\phi=\frac{r\partial_v\phi}{\partial_vr}.
\end{equation}
%The prime has this meaning throughout the paper. Ordinary radial derivatives are written explicitly.

We adopt the radius-weighted regularity convention of An--Tan \cite{AnTan}. This allows us to formulate regularity at the center without requiring
an unweighted second derivative of $\phi$ there. 
\begin{definition}\label{regularity-definition}
A regular $C^1$ solution on a closed characteristic triangle satisfies $\Omega>0$, $\partial_ur<0$, $\partial_vr>0$, and $0\leq\mu<1$ at positive radius. The functions
\[
 \Omega,\quad \partial_ur,\quad \partial_vr,\quad \phi,\quad
 r\partial_u\phi,\quad r\partial_v\phi
\]
are $C^1$ up to the center; in the charged case, so are $A_u$ and $Q/r$. The reduced equations below hold at positive radius, and at $u=v$,
\[
 r=m=Q=A_u=0,\qquad \partial_ur=-\partial_vr,\qquad
 \Omega=2\partial_vr.
\]
Quotients at the center take their continuous values. Characteristic data are of class $C^1$ if $\alpha(0,\cdot)$ is $C^1$, whereas the remaining initial fields satisfy the outgoing constraints and the center conditions.
\end{definition}

The prescribed data extend through $v_*$, and we fix
\begin{equation}\label{initial-margin}
 \mu_*:=\max_{0\leq v\leq v_*}\mu(0,v)<1,
 \qquad \sup_{\mathcal D(0,v_*)}\gamma\leq G<\infty.
\end{equation}

With the conventions \eqref{basic-definitions}, the reduced field
equations used
throughout the paper are
\begin{align}
 \partial_u\partial_vr
 &=\frac{\partial_ur\,\partial_vr}{r}
       \frac{\mu-Q^2/r^2}{1-\mu},\label{mixed-radius}\\
 \partial_um
 &=\frac{2\pi r^2(1-\mu)|D_u\phi|^2}{\partial_ur}
       +\frac{Q^2\partial_ur}{2r^2},\label{incoming-mass}\\
 \partial_vm
 &=\frac{2\pi r^2(1-\mu)|\partial_v\phi|^2}{\partial_vr}
       +\frac{Q^2\partial_vr}{2r^2},\label{outgoing-mass}\\
 D_u\partial_v\phi
 &=-\frac{\partial_vr}{r}D_u\phi
   -\frac{\partial_ur}{r}\partial_v\phi
   -\mathrm i\mathfrak e\frac{Q\phi\partial_ur\,\partial_vr}
     {r^2(1-\mu)},\label{wave-equation}\\
 \partial_uQ
 &=-4\pi\mathfrak e r^2\operatorname{Im}
    (\phi\overline{D_u\phi}),\qquad
 \partial_vQ=4\pi\mathfrak e r^2\operatorname{Im}
    (\phi\overline{\partial_v\phi}),\label{maxwell-equations}\\
 \partial_vA_u
 &=\frac{2Q\partial_ur\,\partial_vr}{r^2(1-\mu)}.
        \label{potential-equation}
\end{align}

At positive radius, $\partial_vr>0$ allows us to use \emph{Bondi
coordinates} $(u,r)$ in place of $(u,v)$. In this chart, $\partial_ur$ and $\partial_vr$ continue to denote the original null derivatives. The center value $\partial_vr(u,u)$ is always written in double null coordinates. We write $\left.\partial_u\right|_r$ for the Bondi time
derivative and use $\left.\partial_r\right|_u=(\partial_vr)^{-1}\partial_v$ for radial differentiation. 

In Bondi coordinates, the scalar averages and outgoing constraints take the form
\begin{align}
 \phi(u,r)&=\frac1r\int_0^r\alpha(u,s)\,\mathrm ds,
 &\alpha-\phi&=\frac1r\int_0^r s\alpha'(u,s)\,\mathrm ds,
                                                   \label{scalar-averages}\\
 \left.\frac{\partial m}{\partial r}\right|_u
 &=2\pi(1-\mu)|\alpha-\phi|^2+\frac{Q^2}{2r^2},
                                                   \label{radial-mass}\\
 \left.\frac{\partial Q}{\partial r}\right|_u
 &=4\pi\mathfrak e r\operatorname{Im}
                         (\phi\overline{\alpha-\phi}).
                                                   \label{radial-charge}
\end{align}
The scalar equation \eqref{wave-equation} also gives the following transport equations
\begin{align}
 D_u(r\partial_v\phi)
 &=-\partial_vr D_u\phi
   -\mathrm i\mathfrak e\frac{Q\phi\partial_ur\,\partial_vr}
                               {r(1-\mu)},\label{outgoing-scalar-transport}\\
 \partial_v(rD_u\phi)
 &=-\partial_ur\partial_v\phi
   +\mathrm i\mathfrak e\frac{Q\phi\partial_ur\,\partial_vr}
                               {r(1-\mu)}.
                                                   \label{incoming-scalar-transport}
\end{align}
Note that The electromagnetic terms have opposite signs because
$[D_u, \partial_v]=-\mathrm i\mathfrak e\partial_vA_u$.

We shall also use
\begin{align}
 \partial_u\alpha+\mathrm i\mathfrak e
 \left(A_u+\frac{Q\partial_ur}{r(1-\mu)}\right)\alpha
 &=\frac{-\partial_ur}{r(1-\mu)}
       \left(\mu-\frac{Q^2}{r^2}-\mathrm i\mathfrak e Q\right)
       (\alpha-\phi),\label{alpha-transport}\\
 \partial_v\left(\frac{-\partial_ur}{1-\mu}\right)
 &=\frac{4\pi r(-\partial_ur)|\partial_v\phi|^2}
           {\partial_vr(1-\mu)},\label{raychaudhuri-ratio}\\
 (-\partial_ur)(u,r)
 &=\partial_vr(u,u)
   \exp\left[\int_0^r
       \frac{\mu-Q^2/s^2}{1-\mu}(u,s)\frac{\mathrm ds}{s}\right].
         \label{center-speed}
\end{align}
All these identities follow from \eqref{mixed-radius}--
\eqref{potential-equation} and the center conditions.

\subsection[Consequences of the Doppler bound]{Consequences of the Doppler bound}\label{Subsec:Dopplerbound}
We now use the Doppler bound in \eqref{initial-margin} to control the null derivatives of $r$ and obtain preliminary estimates for $\phi$ and $Q$. In the sequel,
unless otherwise stated, constants $C$ are independent of $v_*$ and depend only on the fixed
initial scalar datum and its first derivative, the positive lower bound for $1-\mu$ on the initial cone, $G$, and $|\mathfrak e|$. We also note that $C$ might vary in different estimates.

\begin{proposition}[Consequences of the Doppler bound]\label{optical-proposition}
Under the hypotheses of \hyperref[thm:charged]{Theorem~B} or \hyperref[thm:real]{Theorem~C},
\begin{align}
 1-\mu&\geq(1-\mu_*)\exp(-G),\label{uniform-margin}\\
 \tfrac12\exp(-G)&\leq\partial_vr
             \leq\frac1{2(1-\mu_*)},\label{outgoing-speed-bounds}\\
 \int_0^u\frac{r|D_u\phi|^2}{-\partial_ur}(u',v)\,\mathrm du'
 &\leq\frac{G-\log(1-\mu_*)}{4\pi},\label{incoming-energy}\\
 |\phi|^2&\leq C\left(1+\log\frac{r(0,v_*)}{r}\right),\label{logarithmic-scalar}\\
 Q^2&\leq C\mu r^4
       \left(1+\log\frac{r(0,v_*)}{r}\right),\label{preliminary-charge}\\
 \int_0^u\frac{-\partial_ur}{r}\frac{\mu}{1-\mu}(u',v)
 \,\mathrm du'&\leq C.\label{nonnegative-optical-bound}
\end{align}
Moreover, $-\partial_ur\geq C^{-1}$ and
$C^{-1}(v-u)\leq r(u,v)\leq C(v-u)$.
\end{proposition}

\begin{proof}
The incoming mass equation gives
\begin{equation}\label{logarithmic-margin}
 \partial_u\log\frac1{1-\mu}
 =-\frac{4\pi r|D_u\phi|^2}{-\partial_ur}
  +\frac{-\partial_ur}{r}\frac{\mu-Q^2/r^2}{1-\mu}.
\end{equation}
Integration of this equation yields \eqref{uniform-margin} and
\eqref{incoming-energy}. Since $1-\mu\leq1$, it also gives
$\gamma\geq\log(1-\mu_*)$. On the other hand,
\eqref{mixed-radius} and the initial normalization give the exact
identity
\begin{equation}\label{doppler-speed}
 \partial_vr(u,v)=\tfrac12\exp[-\gamma(u,v)].
\end{equation}
Combining \eqref{doppler-speed} with the upper and lower bounds for $\gamma$ proves \eqref{outgoing-speed-bounds}.

To estimate $\phi$, we integrate $D_u\phi$ along an incoming
characteristic using the integrating factor $e^{i\int\mathfrak e A_u \dd u}$ of modulus one.
The Cauchy--Schwarz inequality with weights
$r/(-\partial_ur)$ and $(-\partial_ur)/r$ gives
\eqref{logarithmic-scalar}. The initial scalar is bounded by the
radial average of the prescribed $\alpha$.
Using \eqref{radial-charge}, \eqref{radial-mass}, and $m(u,r=0)=0$, we obtain
\begin{equation}\label{charge-energy-bound}
 Q^2(u,r)\leq8\pi\mathfrak e^2 m(u,r)
        \int_0^r\frac{s^2|\phi(u,s)|^2}{1-\mu(u,s)}\,\mathrm ds.
\end{equation}
This proves \eqref{preliminary-charge}. Consequently,
\[
 \int_0^u\frac{(-\partial_ur)Q^2}{r^3(1-\mu)}(u',v)
 \,\mathrm du'
 \leq C\int_0^{r(0,v_*)}
       s\left(1+\log\frac{r(0,v_*)}{s}\right)\mathrm ds\leq C.
\]
Adding this nonnegative integral to $\gamma$ gives
\eqref{nonnegative-optical-bound}.

Finally, \eqref{raychaudhuri-ratio} and its center value render
$(-\partial_ur)/(1-\mu)\geq\partial_vr(u,u)$. Together with
\eqref{uniform-margin} and \eqref{outgoing-speed-bounds}, this shows
$-\partial_ur\geq C^{-1}$. Integrating \eqref{outgoing-speed-bounds}
from the center along an outgoing cone gives the stated bounds for
$r(u,v)$.
\end{proof}

\section[Pointwise estimates]{Pointwise estimates}
\label{sec:estimates}\label{radius-estimates-section}

We collect the pointwise estimates needed for the first normalization.
The fixed-rectangle argument, the cubic mass estimate, and the
integrating factor for $\alpha'$ are due to An--Chen--Singh
\cite[Section~3]{AnChenSingh}. We recall their conclusions and refer
to that work for the repeated real scalar field estimates, giving
the additional estimates required by the electromagnetic terms
below. We also prove a differentiated metric estimate needed for
compactness.

\subsection[Stability on a fixed characteristic rectangle]{Stability on a fixed characteristic rectangle}

The first estimate compares nearby outgoing scalar transports on a
rectangle whose incoming side is fixed.

\begin{lemma}[Stability on a fixed rectangle]\label{rectangle-lemma}
There is a constant $\delta_0>0$ with the following property.
Fix $0\leq u_0<v_0<v_*$ and $0<\delta'\leq\delta_0$, and set
\begin{equation}\label{rectangle-width}
 \delta=\frac{\delta'r(u_0,v_0)}
 {1+\sup_{0\leq u\leq u_0}|\alpha(u,v_0)-\phi(u,v_0)|}.
\end{equation}
Then $v_0-\delta>u_0$, and on
$[0,u_0]\times[v_0-\delta,v_0]$,
\begin{align}
 &|r\partial_v\phi(u,v)-r\partial_v\phi(u,v_0)|\notag\\
 &\qquad\leq C\delta'
 \left(1+\sup_{0\leq u'\leq u_0}
                  |\alpha(u',v_0)-\phi(u',v_0)|\right).
                                                    \label{rectangle-conclusion}
\end{align}
The constants are independent of the chosen rectangle.
\end{lemma}

\begin{proof}
For a real scalar field this is the fixed-rectangle lemma of
\cite{AnChenSingh}. We retain its comparison argument.
The additional terms to be estimated here are the $Q^2$ terms in
the geometric equations, the $Q\phi$ terms in the two scalar
transports, and the potential $A_u$ in $D_u$.

For this proof, write
\[
 F=1+\sup_{0\leq u\leq u_0}|\alpha(u,v_0)-\phi(u,v_0)|,
 \qquad \delta=\delta'r(u_0,v_0)/F.
\]
As in the cited proof, assume $|r\partial_v\phi|\leq C_0F$ on the
rectangle, with $C_0$ fixed sufficiently large. Constants below may
depend on this fixed bootstrap constant. Proposition~\ref{optical-proposition}
gives $v_0-\delta>u_0$ for small $\delta_0$, and
\[
 |r(u,v)-r(u,v_0)|\leq C\delta,\qquad
 C^{-1}r(u,v_0)\leq r(u,v)\leq Cr(u,v_0).
\]

\smallskip\noindent
\emph{Geometric comparisons from \cite{AnChenSingh}.}
The estimates used in the real scalar field proof are
\begin{align}
 \left|\log\frac{-\partial_ur(u,v)}{-\partial_ur(u,v_0)}\right|
 &\leq\frac{C\delta}{r(u,v_0)},\notag\\
 |m(u,v)-m(u,v_0)|&\leq C\delta F^2,\notag\\
 \left|\frac{-\partial_ur}{1-\mu}(u,v)
       -\frac{-\partial_ur}{1-\mu}(u,v_0)\right|
 &\leq C\delta F^2
       \frac{-\partial_ur(u,v_0)}{r(u,v_0)}.
 \label{rectangle-geometric}
\end{align}
We check how the additional $Q^2$ terms enter these three
comparisons. By \eqref{preliminary-charge},
\[
 \frac{Q^2}{r^2}
 \leq Cr^2\left(1+\log\frac{r(0,v_*)}{r}\right)
 \leq Cr(0,v_*)^2\leq C.
\]
For the first comparison, \eqref{mixed-radius} gives
\begin{align*}
 \left|\log\frac{-\partial_ur(u,v)}{-\partial_ur(u,v_0)}\right|
 &\leq\int_v^{v_0}\frac{\partial_vr}{r(1-\mu)}
                  \left(\mu+\frac{Q^2}{r^2}\right)(u,v')
                  \,\mathrm dv'\\
 &\leq C\int_v^{v_0}\frac{\partial_vr}{r}(u,v')\,\mathrm dv'
 \leq\frac{C\delta}{r(u,v_0)}.
\end{align*}
Since $\delta/r(u,v_0)\leq\delta_0$, the incoming speeds
$-\partial_ur(u,v')$ and $-\partial_ur(u,v_0)$ are comparable
throughout the segment.

For the second comparison, the additional term
$Q^2\partial_vr/(2r^2)$ in the mass equation contributes at most
\[
 \int_v^{v_0}\frac{Q^2\partial_vr}{2r^2}(u,v')\,\mathrm dv'
 \leq C\int_v^{v_0}\partial_vr(u,v')\,\mathrm dv'
 \leq C\delta.
\]
The scalar term is bounded using the bootstrap assumption, as in
\cite{AnChenSingh}. Thus \eqref{outgoing-mass} yields
\begin{align*}
 |m(u,v_0)-m(u,v)|
 &\leq 2\pi\int_v^{v_0}\frac{1-\mu}{\partial_vr}
                   |r\partial_v\phi|^2(u,v')\,\mathrm dv'
       +C\delta\\
 &\leq C\delta F^2+C\delta\leq C\delta F^2.
\end{align*}

The third comparison follows from \eqref{raychaudhuri-ratio}
and the incoming-speed comparison above exactly as in
\cite{AnChenSingh}, since the ratio equation has no explicit $Q$ term.

\smallskip\noindent
\emph{Charge, potential, and the incoming scalar transport.}
The Maxwell equation, the bootstrap bound, and
\eqref{logarithmic-scalar} give
\[
 |\partial_vQ|\leq Cr|\phi|\,|r\partial_v\phi|
 \leq CFr\left(1+\log\frac{r(0,v_*)}{r}\right)^{1/2}.
\]
Integration across the rectangle, together with
$|\partial_v\phi|\leq C F/r$, yields
\begin{align}
 |Q(u,v)-Q(u,v_0)|
 &\leq C\delta F r(u,v_0)
       \left(1+\log\frac{r(0,v_*)}{r(u,v_0)}\right)^{1/2},
       \notag\\
 |\phi(u,v)-\phi(u,v_0)|
 &\leq\frac{C\delta F}{r(u,v_0)}.
 \label{rectangle-scalar}
\end{align}
The second estimate is the scalar difference estimate already used
in \cite{AnChenSingh}; the first is the additional Maxwell estimate.
Next, \eqref{potential-equation} and the identity
\[
 \partial_v\left(\frac{Q\phi}{r}\right)
 =\frac{\partial_vQ}{r}\phi+\frac Qr\partial_v\phi
       -\frac{Q\phi\,\partial_vr}{r^2}
\]
give
\begin{align}
 |A_u(u,v)-A_u(u,v_0)|
 &\leq C\delta(-\partial_ur)(u,v_0)
       \left(1+\log\frac{r(0,v_*)}{r(u,v_0)}\right)^{1/2},
       \notag\\
 \left|\frac{Q\phi}{r}(u,v)-\frac{Q\phi}{r}(u,v_0)\right|
 &\leq C\delta F
       \left(1+\log\frac{r(0,v_*)}{r(u,v_0)}\right).
 \label{rectangle-potential}
\end{align}

In \eqref{incoming-scalar-transport}, the additional source satisfies
\[
 \left|\frac{\mathfrak e Q\phi\,\partial_ur\,\partial_vr}
                   {r(1-\mu)}\right|
 \leq Cr(-\partial_ur)\left(1+\log\frac{r(0,v_*)}{r}\right)
 \leq C\frac{-\partial_ur}{r}.
\]
Consequently,
\[
 \int_v^{v_0}
 \left|\frac{\mathfrak e Q\phi\,\partial_ur\,\partial_vr}
                   {r(1-\mu)}\right|(u,v')\,\mathrm dv'
 \leq C\delta\frac{-\partial_ur(u,v_0)}{r(u,v_0)}.
\]
This is absorbed in the bound of \cite{AnChenSingh}, since
$F\geq1$. The incoming scalar
comparison retains the same form, with $D_u\phi$ in place of
$\partial_u\phi$:
\begin{align}
 |D_u\phi(u,v)-D_u\phi(u,v_0)|
 &\leq C\delta F\frac{-\partial_ur(u,v_0)}{r(u,v_0)^2}
       +\frac{C\delta|D_u\phi(u,v_0)|}{r(u,v_0)}.
 \label{rectangle-incoming}
\end{align}

\smallskip\noindent
\emph{The additional charge term in the outgoing speed.}
We compare
\[
 \partial_u\log\partial_vr
 =\frac{2m\partial_ur}{r^2(1-\mu)}
       -\frac{Q^2\partial_ur}{r^3(1-\mu)}
\]
on the two incoming segments. For the first term,
\eqref{rectangle-geometric} gives the difference bound of
\cite{AnChenSingh}:
\[
 \left|\frac{2m\partial_ur}{r^2(1-\mu)}(u,v)
       -\frac{2m\partial_ur}{r^2(1-\mu)}(u,v_0)\right|
 \leq C\delta F^2\frac{-\partial_ur(u,v_0)}{r(u,v_0)^2}.
\]
For the second term, expand the differences of the three factors
$Q^2$, $\partial_ur/(1-\mu)$, and $r^{-3}$. Using
\eqref{preliminary-charge}, \eqref{rectangle-geometric}, and
\eqref{rectangle-scalar}, we obtain
\begin{align*}
 &\left|\frac{Q^2\partial_ur}{r^3(1-\mu)}(u,v)
       -\frac{Q^2\partial_ur}{r^3(1-\mu)}(u,v_0)\right|\\
 &\qquad\leq C\delta F^2(-\partial_ur)(u,v_0)
       \left(1+\log\frac{r(0,v_*)}{r(u,v_0)}\right).
\end{align*}
The two terms are integrated using, respectively,
\begin{align}
 \int_0^u\frac{-\partial_ur}{r^2}(u',v_0)\,\mathrm du'
 &\leq\frac1{r(u,v_0)},\notag\\
 \int_0^u(-\partial_ur)
       \left(1+\log\frac{r(0,v_*)}{r}\right)(u',v_0)\,\mathrm du'
 &\leq\int_0^{r(0,v_*)}
       \left(1+\log\frac{r(0,v_*)}{s}\right)\mathrm ds
 =2r(0,v_*).
 \label{rectangle-primitives}
\end{align}
Since $\partial_vr(0,v)=1/2$ and $\partial_vr$ has uniform
two-sided bounds, this proves
\begin{equation}\label{rectangle-outgoing}
 |\partial_vr(u,v)-\partial_vr(u,v_0)|
 \leq\frac{C\delta F^2}{r(u,v_0)}.
\end{equation}

\smallskip\noindent
\emph{Outgoing scalar comparison and closure of the bootstrap.}
Subtracting \eqref{outgoing-scalar-transport} gives
\begin{align*}
 &\partial_u\bigl[r\partial_v\phi(u,v_0)-r\partial_v\phi(u,v)\bigr]
 +\mathrm i\mathfrak e A_u(u,v_0)
       \bigl[r\partial_v\phi(u,v_0)-r\partial_v\phi(u,v)\bigr]\\
 &=-\bigl[\partial_vrD_u\phi(u,v_0)-\partial_vrD_u\phi(u,v)\bigr]\\
 &\quad-\mathrm i\mathfrak e
       \left[\frac{Q\phi\partial_ur\,\partial_vr}{r(1-\mu)}(u,v_0)
             -\frac{Q\phi\partial_ur\,\partial_vr}{r(1-\mu)}(u,v)\right]\\
 &\quad-\mathrm i\mathfrak e[A_u(u,v_0)-A_u(u,v)]r\partial_v\phi(u,v).
\end{align*}
The first term on the right is the product difference appearing
in \cite{AnChenSingh}. The last two terms are the additional charge and potential
differences.

For the first term, \eqref{rectangle-incoming} and
\eqref{rectangle-outgoing} give
\begin{align*}
 &\int_0^u
  |\partial_vrD_u\phi(u',v_0)-\partial_vrD_u\phi(u',v)|\,\mathrm du'\\
 &\quad\leq C\delta F^2\int_0^u
                  \frac{|D_u\phi|}{r}(u',v_0)\,\mathrm du'
       +C\delta F\int_0^u
                  \frac{-\partial_ur}{r^2}(u',v_0)\,\mathrm du'
 \leq\frac{C\delta F^2}{r(u,v_0)}.
\end{align*}
The last inequality is the weighted Cauchy--Schwarz estimate of
\cite{AnChenSingh}, using \eqref{incoming-energy}.

For the charge difference, expand the three factors $Q\phi/r$,
$\partial_ur/(1-\mu)$, and $\partial_vr$. Equations
\eqref{rectangle-geometric}, \eqref{rectangle-potential}, and
\eqref{rectangle-outgoing} give
\begin{align*}
 &\left|\frac{Q\phi\partial_ur\,\partial_vr}{r(1-\mu)}(u,v)
       -\frac{Q\phi\partial_ur\,\partial_vr}{r(1-\mu)}(u,v_0)\right|\\
 &\qquad\leq C\delta F^2(-\partial_ur)(u,v_0)
       \left(1+\log\frac{r(0,v_*)}{r(u,v_0)}\right).
\end{align*}
The potential difference satisfies, separately,
\[
 |A_u(u,v_0)-A_u(u,v)|\,|r\partial_v\phi(u,v)|
 \leq C\delta F(-\partial_ur)(u,v_0)
       \left(1+\log\frac{r(0,v_*)}{r(u,v_0)}\right)^{1/2}.
\]
Thus \eqref{rectangle-primitives} bounds the incoming integral
of these two additional terms by $C\delta F^2$.
The initial scalar difference is at most $C\delta$ by the $C^1$
datum. Integrating the difference equation now yields
\[
 |r\partial_v\phi(u,v_0)-r\partial_v\phi(u,v)|
 \leq C\delta F^2\left(1+\frac1{r(u,v_0)}\right)
 \leq C\delta'F,
\]
because $\delta=\delta'r(u_0,v_0)/F$ and
$r(u_0,v_0)\leq r(u,v_0)\leq r(0,v_*)$.
This is \eqref{rectangle-conclusion}.
At $v=v_0$, $|r\partial_v\phi|\leq C F$ by the outgoing speed bound.
Choose $C_0$ larger than twice this fixed constant, and then
$\delta_0$ small enough that $C\delta_0\leq C_0/4$.
The last estimate improves the bootstrap to
$|r\partial_v\phi|\leq 3C_0F/4$.
\end{proof}

\subsection[The cubic mass estimate and bounded scalar fields]{The cubic mass estimate and bounded scalar fields}

\begin{proposition}[The cubic mass bound]\label{scalar-bounds-proposition}
Under the hypotheses of \hyperref[thm:charged]{Theorem~B} or \hyperref[thm:real]{Theorem~C},
\begin{align}
 |\alpha-\phi|^3&\leq C\mu,\label{cubic-bound}\\
 |\alpha|+|\phi|+
       \left|\frac{rD_u\phi}{-\partial_ur}\right|&\leq C,
    \label{scalar-bounds}\\
 Q^2&\leq C\mu r^4,\qquad
       \frac{|A_u|}{-\partial_ur}\leq Cr.\label{improved-charge}
\end{align}
\end{proposition}

\begin{proof}
Lemma~\ref{rectangle-lemma} and the mass argument in
\cite{AnChenSingh} give first $|\alpha-\phi|\leq C$
and then \eqref{cubic-bound}; the argument applies unchanged.

We next bound $\alpha$ and $\phi$, and use the resulting bound for
$\phi$ to improve the preliminary charge estimate.
The coefficient of $\alpha$ on the left of
\eqref{alpha-transport} is purely imaginary, so its integrating
factor has modulus one. Since $|\alpha-\phi|\leq C$, integration
along an incoming cone gives
\begin{align*}
 |\alpha(u,v)|
 &\leq |\alpha(0,v)|
  +C\int_0^u\frac{-\partial_ur}{r(1-\mu)}
       \left(\mu+\frac{Q^2}{r^2}+|\mathfrak e Q|\right)(u',v)
          \,\mathrm du'.
\end{align*}
The $\mu$ term is bounded by \eqref{nonnegative-optical-bound}.
For the two charge terms, \eqref{preliminary-charge} gives
\begin{align*}
 &\int_0^u\frac{-\partial_ur}{r(1-\mu)}
       \left(\frac{Q^2}{r^2}+|\mathfrak e Q|\right)(u',v)
          \,\mathrm du'\\
 &\quad\leq C\int_0^u(-\partial_ur)r
       \left[1+\log\frac{r(0,v_*)}{r}
              +\left(1+\log\frac{r(0,v_*)}{r}\right)^{1/2}\right]
                     (u',v)\,\mathrm du'\\
 &\quad\leq C\int_0^{r(0,v_*)}s
       \left[1+\log\frac{r(0,v_*)}{s}
              +\left(1+\log\frac{r(0,v_*)}{s}\right)^{1/2}\right]
                     \,\mathrm ds\leq C.
\end{align*}
Thus $|\alpha|\leq C$. The first identity in
\eqref{scalar-averages} then gives
\[
 |\phi(u,r)|\leq\frac1r\int_0^r|\alpha(u,s)|\,\mathrm ds\leq C.
\]
Substituting this uniform bound in \eqref{charge-energy-bound}
removes the logarithmic factor in \eqref{preliminary-charge}:
\[
 Q^2(u,r)
 \leq8\pi\mathfrak e^2m(u,r)
       \int_0^r\frac{s^2|\phi(u,s)|^2}{1-\mu(u,s)}\,\mathrm ds
 \leq C m(u,r)r^3\leq C\mu(u,r)r^4.
\]

The incoming scalar derivative bound is also proved for the real
scalar field in \cite{AnChenSingh}, by integrating the
wave equation from the center and using the radial monotonicity
of $-\partial_ur$. In the charged case this monotonicity need not
hold; we give the modification below.
By \eqref{raychaudhuri-ratio}, for $0\leq s\leq r$ on an outgoing cone,
\begin{equation}\label{speed-ratio}
 \frac{(-\partial_ur)(u,s)}{(-\partial_ur)(u,r)}\leq C.
\end{equation}
The improved charge bound gives
\[
 \left.\frac{\partial}{\partial r}\right|_u
          \log(-\partial_ur)
       =\frac{\mu-Q^2/r^2}{r(1-\mu)}\geq-Cr.
\]
Thus the negative part of this logarithmic derivative is at most
$Cr$. Using its signed radial variation and \eqref{speed-ratio},
we obtain
\begin{equation}\label{speed-variation}
 \frac1{(-\partial_ur)(u,r)}
       \int_0^r\left|
          \left.\frac{\partial(-\partial_ur)}{\partial s}\right|_u
               (u,s)\right|\,\mathrm ds\leq1+Cr^2.
\end{equation}
The ordered wave equation implies
\begin{equation}\label{incoming-ratio-equation}
 \partial_v\left(\frac{D_u(r\phi)}{\partial_ur}\right)
 =\frac{\partial_vr}{r}\frac{\mu-Q^2/r^2}{1-\mu}
       \left(\phi-\frac{D_u(r\phi)}{\partial_ur}\right)
   +\frac{\mathrm i\mathfrak e Q\partial_vr}{r(1-\mu)}\phi.
\end{equation}
Equivalently,
\[
 \left.\frac{\partial}{\partial r}\right|_u
       \left((-\partial_ur)\frac{D_u(r\phi)}{\partial_ur}\right)
 =
 \left.\frac{\partial(-\partial_ur)}{\partial r}\right|_u\phi
       +\frac{\mathrm i\mathfrak e Q(-\partial_ur)}{r(1-\mu)}\phi.
\]
The quantity inside the radial derivative has center value
$(-\partial_ur)(u,u)\phi(u,u)$.
Integrating this radial identity from $s=0$ to $s=r$ along the
outgoing cone with $u$ fixed, and dividing by $(-\partial_ur)(u,r)$,
we bound $D_u(r\phi)/\partial_ur$ using
\eqref{speed-ratio}--\eqref{speed-variation}, $|\phi|\leq C$,
and $|Q|\leq Cr^2$. Since
$D_u(r\phi)/\partial_ur=\phi+rD_u\phi/\partial_ur$,
the bound for $rD_u\phi/(-\partial_ur)$ follows.
Finally, since $A_u$ vanishes at the center,
\eqref{potential-equation}, \eqref{speed-ratio}, and $|Q|\leq Cr^2$
give
\[
 \frac{|A_u(u,r)|}{(-\partial_ur)(u,r)}
 \leq 2\int_0^r\frac{|Q(u,s)|}{s^2(1-\mu(u,s))}
          \frac{(-\partial_ur)(u,s)}{(-\partial_ur)(u,r)}
                 \,\mathrm ds
 \leq C\int_0^r\mathrm ds=Cr.
\]
\end{proof}

\subsection[Differentiated equations and exact multipliers]{Differentiated equations and exact multipliers}

\begin{lemma}[Differentiated multipliers]\label{multiplier-lemma}
At positive radius,
\begin{align}
 D_u\bigl((\partial_vr)^2\alpha'\bigr)
 ={}&\frac{(-\partial_ur)(\partial_vr)^2}{r^2(1-\mu)}
 \Bigl\{
 4\pi(1-Q^2/r^2)|\alpha-\phi|^2(\alpha-\phi)
 -(3\mu-5Q^2/r^2)(\alpha-\phi)\notag\\
 &\quad-8\pi\mathfrak e Q
       \operatorname{Im}(\phi\overline{\alpha-\phi})(\alpha-\phi)
 +\mathrm i\mathfrak e Q
       [3(\alpha-\phi)+\phi+4\pi\phi|\alpha-\phi|^2]\notag\\
 &\quad+4\pi\mathrm i\mathfrak e^2r^2\phi
                \operatorname{Im}(\phi\overline{\alpha-\phi})
 \Bigr\},\label{scalar-multiplier}\\
 \partial_u\left(\frac{\partial_v^2r}{\partial_vr}\right)
 ={}&\frac{(-\partial_ur)\partial_vr}{r^2(1-\mu)}
 \left[
 2\mu-\frac{4Q^2}{r^2}
 -4\pi(1-Q^2/r^2)|\alpha-\phi|^2
 +8\pi\mathfrak e Q
       \operatorname{Im}(\phi\overline{\alpha-\phi})
 \right].\label{metric-multiplier}
\end{align}
Consequently,
\begin{equation}\label{critical-derivative-bounds}
 |r\alpha'|+
 \left|\frac{r\partial_v^2r}{(\partial_vr)^2}\right|\leq C.
\end{equation}
\end{lemma}

\begin{proof}
Differentiating \eqref{alpha-transport} in $v$ and using the
integrating factor $(\partial_vr)^2$ of \cite{AnChenSingh} gives
\eqref{scalar-multiplier}. Similarly, differentiating
$\partial_u\log\partial_vr$ in $v$ and substituting the field
equations gives \eqref{metric-multiplier}.

The neutral terms in \eqref{scalar-multiplier} are estimated exactly
as in \cite{AnChenSingh}: their numerator is bounded by
$C\mu$, using \eqref{cubic-bound}, and
\[
 r(u,v)\int_0^u\frac{\mu(-\partial_ur)}{r^2(1-\mu)}(u',v)
       \,\mathrm du'
 \leq\int_0^u\frac{\mu(-\partial_ur)}{r(1-\mu)}(u',v)
       \,\mathrm du'\leq C.
\]
Every charged term in the braces, divided by $r^2(1-\mu)$, is
bounded by $C$ by \eqref{scalar-bounds}--\eqref{improved-charge}.
Integrating with the covariant factor of modulus one bounds their
contribution by $C\int_0^u(-\partial_ur)(u',v)\,\mathrm du'\leq C$.
Together with the initial $C^1$ bound, this proves $|r\alpha'|\leq C$.

The bracket in \eqref{metric-multiplier} is uniformly bounded.
Its initial value is zero because $\partial_vr(0,v)=1/2$.
The bound
\[
 r(u,v)\int_0^u\frac{-\partial_ur}{r^2}(u',v)\,\mathrm du'\leq1
\]
and the two-sided bounds for $\partial_vr$ prove the second
inequality in \eqref{critical-derivative-bounds}.
\end{proof}

\section{The first limiting solution and its past asymptotics}
\label{sec:first}\label{first-limit-section}

We use the pointwise estimates of Sections~\ref{sec:statements}
and~\ref{sec:estimates} to extract an ancient solution. We then
establish its asymptotics as $u\to-\infty$ in
Section~\ref{subsec:ancient-past}.

\subsection{First normalization and compactness}
\label{subsec:first-compactness}

In this subsection, $r,m,\phi,\alpha$ and the other unindexed fields
denote the original solution from the preceding sections.
We first prove the following compactness statement.

\begin{proposition}[An ancient neutral limit]\label{ancient-proposition}
If $|\alpha'|$ is unbounded for the original solution
satisfying \hyperref[thm:charged]{Theorem~B} or \hyperref[thm:real]{Theorem~C}, there is a regular neutral
scalar field solution $(r_\infty,m_\infty,\phi_\infty)$ on
\[
 \{(u,v):-\infty<u\leq v\leq0\}.
\]
Writing $\alpha_\infty=\partial_v(r_\infty\phi_\infty)/\partial_vr_\infty$
and $\alpha'_\infty=(\partial_vr_\infty)^{-1}\partial_v\alpha_\infty$,
the limit satisfies $|\alpha'_\infty|\leq1$, the bounds
\eqref{cubic-bound}, \eqref{scalar-bounds},
\eqref{critical-derivative-bounds}, and
\begin{equation}\label{ancient-bounds}
 \begin{gathered}
 1-\mu_\infty\geq C^{-1},\qquad C^{-1}\leq\partial_vr_\infty\leq C,\\
 \int_{-\infty}^{v}\frac{-\partial_ur_\infty}{r_\infty}
             \frac{\mu_\infty}{1-\mu_\infty}(u,v)\,\mathrm du\leq C.
 \end{gathered}
\end{equation}
Here $\mu_\infty=2m_\infty/r_\infty$. The center is regular,
and $m_\infty(u,0)>0$ for some $u<0$.
\end{proposition}

\begin{rema*}
The scalar field is real in the setting of \hyperref[thm:charged]{Theorem~B}.
In the setting of \hyperref[thm:real]{Theorem~C} it may be complex, but its
coupling, charge and electromagnetic potential are zero.
\end{rema*}

\begin{proof}
\textit{Step 1. Normalization.}
For every sufficiently large integer $j$, choose the first $b_j$ for
which
\[
 \max_{0\leq u\leq v\leq b_j}|\alpha'(u,v)|=j.
\]
The maximum is continuous and nondecreasing in $b$ on compact
subintervals of $[0,v_*)$. Hence $b_j\to v_*$, and a maximizing
point lies on $v=b_j$ in the original coordinates.
Define the rescaled fields, using $(u,v)$ for their null coordinates, by
\begin{equation}\label{first-normalization}
\begin{aligned}
 r_j(u,v)&=j\,r(b_j+u/j,b_j+v/j),\\
 m_j(u,v)&=j\,m(b_j+u/j,b_j+v/j),\\
 \phi_j(u,v)&=\phi(b_j+u/j,b_j+v/j),\\
 \alpha_j(u,v)&=\alpha(b_j+u/j,b_j+v/j),\\
 \Omega_j^2(u,v)&=\Omega^2(b_j+u/j,b_j+v/j).
\end{aligned}
\end{equation}
The fields on the right belong to the original solution and are
evaluated at the displayed corresponding points.
The neutral scaling symmetry preserves the field equations.
For the charged system set, at corresponding points,
\[
 Q_j=jQ,\qquad
 A_{u,j}=A_u,\qquad
 \mathfrak e_j=\mathfrak e/j.
\]
Substitution in \eqref{mixed-radius}--\eqref{potential-equation}
gives the charged system with coupling $\mathfrak e_j$.
At the same pair of corresponding points,
\[
 \alpha'_j=j^{-1}\alpha',\qquad
 D_{u,j}\phi_j=j^{-1}D_u\phi,\qquad
 |r_j\alpha'_j|=|r\alpha'|.
\]
Here the primes and null derivatives on the two sides are taken
in the respective coordinates.
The rescaled solutions are controlled on
$\{-jb_j\leq u\leq v\leq0\}$, where $|\alpha'_j|\leq1$.
The bounds for $\mu_j$, $\alpha_j$, $\phi_j$, the normalized
scalar derivatives and the incoming integrals have the same
constants as for the original solution, by the scaling invariance
of \eqref{uniform-margin}--\eqref{incoming-energy},
\eqref{nonnegative-optical-bound}, and \eqref{scalar-bounds}.
At a normalized maximum
$(u,0)$, we have $|\alpha'_j(u,0)|=1$. The scaling-invariant
bound \eqref{critical-derivative-bounds} therefore gives
\[
 r_j(u,0)=|r_j\alpha'_j|(u,0)\leq C.
\]
Since $r_j(u,v)\geq C^{-1}(v-u)$, at the same point
\[
 |u|=-u\leq C r_j(u,0)\leq C.
\]
Thus the normalized maxima remain in a fixed bounded interval
of $v=0$, independently of $j$.

For $T<\infty$, write
\[
 K_T=\{(u,v):-T\leq u\leq v\leq0\}.
\]
For all sufficiently large $j$, $K_T$ lies in the $j$th domain.
Integration from the center along a fixed outgoing cone gives
\[
 r_j(u,v)=\int_u^v\partial_vr_j(u,v')\,\mathrm dv'
          \leq C(v-u)\leq CT,\qquad (u,v)\in K_T.
\]
All compactness statements below are
uniform on each fixed $K_T$; the index after which $K_T$ is
contained in the domain may depend on $T$.

\textit{Step 2. Bounds up to the center.}
We first establish the vanishing orders at the center and bounds
for the scalar derivatives and null derivatives of $r_j$.
Along each fixed outgoing cone of the $j$th solution, we use its
own radius as parameter, denoted by $r$ or $s$.
Since $\partial_r(r\phi_j)=\alpha_j$ and $r\phi_j$ vanishes at
the regular center,
\[
 \phi_j(u,r)=\frac1r\int_0^r\alpha_j(u,s)\,\mathrm ds,
 \qquad r>0.
\]
Moreover, $\partial_r\alpha_j=\alpha'_j$ and $|\alpha'_j|\leq1$,
so $|\alpha_j(u,r)-\alpha_j(u,s)|\leq r-s$ for $0\leq s\leq r$.
Consequently,
\[
 \begin{aligned}
 |\alpha_j(u,r)-\phi_j(u,r)|
 &\leq\frac1r\int_0^r
       |\alpha_j(u,r)-\alpha_j(u,s)|\,\mathrm ds\\
 &\leq\frac1r\int_0^r(r-s)\,\mathrm ds=\frac r2.
 \end{aligned}
\]
Thus $|\alpha_j-\phi_j|\leq r_j/2$, also at the center by continuity.
Fix a finite radius bound $R>0$. Constants $C_R$ below may depend
on $R$, but not on $j$; on a fixed $K_T$, we may take $R=CT$ by
Step~1. The regular center conditions give $Q_j(u,0)=m_j(u,0)=0$
in the radial parametrization. Using $|\phi_j|\leq C$ and
$\mathfrak e_j=\mathfrak e/j$, integration of
\eqref{radial-charge} gives
\[
 \begin{aligned}
 |Q_j(u,r)|
 &\leq\frac{4\pi|\mathfrak e|}{j}
       \int_0^r s|\phi_j(u,s)|\,|\alpha_j-\phi_j|(u,s)\,\mathrm ds\\
 &\leq\frac Cj\int_0^r s^2\,\mathrm ds
       \leq\frac{Cr^3}{j}.
 \end{aligned}
\]
Substituting this bound into \eqref{radial-mass}, we obtain,
for $0<r\leq R$,
\[
 \begin{aligned}
 0\leq m_j(u,r)
 &=\int_0^r\left[2\pi(1-\mu_j)|\alpha_j-\phi_j|^2
                    +\frac{Q_j^2}{2s^2}\right](u,s)\,\mathrm ds\\
 &\leq C\int_0^r\left(s^2+\frac{s^4}{j^2}\right)\,\mathrm ds
       \leq C\left(r^3+\frac{r^5}{j^2}\right)
       \leq C_Rr^3.
 \end{aligned}
\]
Since $\mu_j=2m_j/r_j$, these estimates give, on $r_j\leq R$,
\begin{equation}\label{first-center-orders}
 |Q_j|\leq C_Rr_j^3/j,\qquad
 0\leq m_j\leq C_Rr_j^3,\qquad
 0\leq\mu_j\leq C_Rr_j^2.
\end{equation}
In the real case the charge terms are absent.
By \eqref{first-center-orders} and the rescaled bound
$1-\mu_j\geq C^{-1}$ from \eqref{uniform-margin}, for $r\leq R$,
\[
 \left|\int_0^r\frac{\mu_j-Q_j^2/s^2}{1-\mu_j}(u,s)
                         \frac{\mathrm ds}{s}\right|
 \leq C_R\int_0^r\left(s+\frac{s^3}{j^2}\right)\,\mathrm ds
 \leq C_R.
\]
The value $\partial_vr_j(u,u)$ in \eqref{center-speed}
is bounded above and below by \eqref{outgoing-speed-bounds}.
Thus \eqref{center-speed} gives
\[
 C_R^{-1}\leq-\partial_ur_j\leq C_R.
\]
Integrating \eqref{incoming-scalar-transport} from the center gives
\begin{align}
 \frac{r_jD_{u,j}\phi_j}{-\partial_ur_j}(u,r)
 =\frac1{(-\partial_ur_j)(u,r)}
 \int_0^r\frac{(-\partial_ur_j)(u,s)}s
 \left[\alpha_j-\phi_j
       -\frac{\mathrm i\mathfrak e_jQ_j\phi_j}{1-\mu_j}\right](u,s)
 \,\mathrm ds.                                      \label{center-incoming-average}
\end{align}
Using \eqref{speed-ratio}, $|\alpha_j-\phi_j|(u,s)\leq s/2$,
$|\phi_j|\leq C$, $1-\mu_j\geq C^{-1}$, and
$|\mathfrak e_jQ_j|\leq C_Rs^3/j^2$ from
\eqref{first-center-orders}, we obtain
\[
 \left|\frac{r_jD_{u,j}\phi_j}{-\partial_ur_j}(u,r)\right|
 \leq C_R\int_0^r\left(1+\frac{s^2}{j^2}\right)\,\mathrm ds
 \leq C_Rr.
\]
Finally, integrate \eqref{potential-equation} radially from
$A_{u,j}(u,0)=0$. The bounds $-\partial_ur_j\leq C_R$,
$1-\mu_j\geq C^{-1}$ and $|Q_j(u,s)|\leq C_Rs^3/j$ give
\[
 |A_{u,j}(u,r)|
 \leq 2\int_0^r\frac{|Q_j|(-\partial_ur_j)}{s^2(1-\mu_j)}(u,s)
                     \,\mathrm ds
 \leq\frac{C_R}{j}\int_0^r s\,\mathrm ds
 \leq\frac{C_Rr^2}{j}.
\]
We next bound $\partial_v^2r_j$ up to the center. Fix an incoming
cone $v=\mathrm{const}$ and use $s=r_j$ as its parameter, so that
$\mathrm d/\mathrm ds=(\partial_ur_j)^{-1}\partial_u$ along this cone.
For $s\leq1$, the bracket in \eqref{metric-multiplier} is bounded
by $Cs^2$, using \eqref{first-center-orders} with $R=1$,
$|\alpha_j-\phi_j|\leq s/2$, $|\phi_j|\leq C$ and
$\mathfrak e_j=\mathfrak e/j$.
Together with \eqref{uniform-margin} and
\eqref{outgoing-speed-bounds}, this gives
\[
 \left|\frac{\mathrm d}{\mathrm ds}
          \left(\frac{\partial_v^2r_j}{\partial_vr_j}\right)\right|
 \leq\frac C{s^2}s^2\leq C,\qquad 0<s\leq1.
\]
Fix a point $(u,v)$ with $0<r_j(u,v)\leq R$.
If $r_j(u,v)\leq1$, integrate along its incoming cone from
$s_0=\min\{1,r_j(-jb_j,v)\}$ down to $r_j(u,v)$.
If $s_0=1$, \eqref{critical-derivative-bounds} and
\eqref{outgoing-speed-bounds} bound
$|\partial_v^2r_j/\partial_vr_j|$ there by $C$.
Otherwise the starting point lies on the initial outgoing cone,
where $\partial_vr_j(-jb_j,v)=1/2$ and hence
$\partial_v^2r_j(-jb_j,v)=0$. In either case,
\[
 \left|\frac{\partial_v^2r_j}{\partial_vr_j}(u,v)\right|
 \leq C+C\int_{r_j(u,v)}^{s_0}\mathrm ds\leq C.
\]
Since $\partial_vr_j\geq C^{-1}$, this also bounds
$|\partial_v^2r_j/(\partial_vr_j)^2|$ by $C$.
% Expanded conversion retained for the detailed version:
% At this same point, $\partial_vr_j(u,v)\geq C^{-1}$, so
% \[
%  \left|\frac{\partial_v^2r_j}{(\partial_vr_j)^2}(u,v)\right|
%  =\frac1{\partial_vr_j(u,v)}
%    \left|\frac{\partial_v^2r_j}{\partial_vr_j}(u,v)\right|
%  \leq C.
% \]
If instead $1<r_j(u,v)\leq R$, \eqref{critical-derivative-bounds}
at $(u,v)$ gives
\[
 \left|\frac{\partial_v^2r_j}{(\partial_vr_j)^2}(u,v)\right|
 \leq\frac{C}{r_j(u,v)}\leq C.
\]
Combining the two cases, and extending to the center by continuity,
we obtain a bound independent of both $R$ and $j$:
\begin{equation}\label{first-metric-quotient}
 \left|\frac{\partial_v^2r_j}{(\partial_vr_j)^2}(u,v)\right|\leq C
 \qquad\bigl(0\leq r_j(u,v)\leq R\bigr).
\end{equation}

To complete the Lipschitz bounds, we now estimate $\partial_u^2r_j$
and the first derivatives of $A_{u,j}$. The preceding bounds,
\eqref{first-metric-quotient}, and $|\alpha'_j|\leq1$, applied to
the mass, scalar and mixed-radius equations, already control both
first derivatives of $r_j,\partial_vr_j,\mu_j,\alpha_j,\phi_j$ on $K_T$.
For the remaining derivative $\partial_u^2r_j$, we work at fixed $v$.
The bounds $|r_jD_{u,j}\phi_j/(-\partial_ur_j)|\leq C_Rr_j$
and $C_R^{-1}\leq-\partial_ur_j\leq C_R$ give
\[
 |D_{u,j}\phi_j|\leq C_R.
\]
Substituting this bound and
\eqref{first-center-orders} into \eqref{incoming-mass}, and
using $\mu_j=2m_j/r_j$, yields
\[
 |\partial_um_j|\leq C_Rr_j^2,\qquad
 |\partial_u\mu_j|
 =\left|\frac{2\partial_um_j}{r_j}
              -\frac{\mu_j\partial_ur_j}{r_j}\right|
 \leq C_Rr_j.
\]
Likewise, \eqref{maxwell-equations} with $|\phi_j|\leq C$ and
$\mathfrak e_j=\mathfrak e/j$ gives the first bound below;
\eqref{mixed-radius} with \eqref{first-center-orders},
$1-\mu_j\geq C^{-1}$ and the bounds for
$-\partial_ur_j,\partial_vr_j$ gives the second:
\[
 |\partial_uQ_j|\leq C_Rr_j^2/j,\qquad
 |\partial_u\partial_vr_j|\leq C_Rr_j.
\]
Using these bounds for $\partial_u\mu_j$, $\partial_uQ_j$ and
$\partial_u\partial_vr_j$, together with
\eqref{first-center-orders} and $|\partial_ur_j|\leq C_R$,
we obtain
\[
 \left|\partial_u\left(
       \frac{\partial_vr_j}{r_j}
       \frac{\mu_j-Q_j^2/r_j^2}{1-\mu_j}\right)\right|
 \leq C_R.
\]
To apply this estimate, rewrite \eqref{center-speed} using
$\mathrm ds=\partial_vr_j(u,v')\,\mathrm dv'$:
\begin{equation}\label{first-null-speed-identity}
 \begin{aligned}
 \log(-\partial_ur_j)(u,v)
 &=\log\partial_vr_j(u,u)\\
 &\quad+\int_u^v\left(
       \frac{\partial_vr_j}{r_j}
       \frac{\mu_j-Q_j^2/r_j^2}{1-\mu_j}\right)(u,v')
       \,\mathrm dv'.
 \end{aligned}
\end{equation}
By \eqref{first-center-orders}, the integrand is $O(r_j)$
and vanishes at $v'=u$, so differentiation of the lower endpoint
contributes zero. Also,
$\frac{\mathrm d}{\mathrm du}\partial_vr_j(u,u)
=\partial_v^2r_j(u,u)$, since $\partial_u\partial_vr_j=0$
at the center. Differentiating \eqref{first-null-speed-identity}
at fixed $v$ therefore gives
\[
 \begin{aligned}
 \partial_u\log(-\partial_ur_j)(u,v)
 &=\frac{\partial_v^2r_j}{\partial_vr_j}(u,u)\\
 &\quad+\int_u^v\partial_u\left(
       \frac{\partial_vr_j}{r_j}
       \frac{\mu_j-Q_j^2/r_j^2}{1-\mu_j}\right)(u,v')
       \,\mathrm dv'.
 \end{aligned}
\]
The term $(\partial_v^2r_j/\partial_vr_j)(u,u)$ is bounded by
\eqref{first-metric-quotient} and
\eqref{outgoing-speed-bounds}, and the derivative inside the
integral is bounded as above. Since
$v-u\leq Cr_j(u,v)\leq CR$ and $-\partial_ur_j\leq C_R$,
we conclude that
\[
 |\partial_u\log(-\partial_ur_j)|\leq C_R,\qquad
 |\partial_u^2r_j|\leq C_R\qquad\text{on }K_T.
\]
For the potential, \eqref{potential-equation} and $A_{u,j}(u,u)=0$
give
\[
 A_{u,j}(u,v)=\int_u^v
       \frac{2Q_j\partial_ur_j\,\partial_vr_j}
            {r_j^2(1-\mu_j)}(u,v')\,\mathrm dv'.
\]
The integrand is $O(r_j/j)$ by \eqref{first-center-orders}.
Its $u$ derivative is bounded by $C_R/j$, using the bounds for
$\partial_uQ_j$, $\partial_u\mu_j$, $\partial_u\partial_vr_j$
and $\partial_u^2r_j$ just proved. Differentiation of this
integral thus gives, on $K_T$,
\[
 |\partial_uA_{u,j}|+|\partial_vA_{u,j}|\leq C_R/j.
\]
Thus the fields
\[
 r_j,\quad-\partial_ur_j,\quad\partial_vr_j,\quad
 \mu_j,\quad\alpha_j,\quad\phi_j
\]
are uniformly Lipschitz on $K_T$.

\textit{Step 3. Locally uniform limits.}
Applying Arzel\`a--Ascoli successively on $K_1,K_2,\ldots$ and
taking a diagonal subsequence, we obtain a single subsequence
along which all six fields above converge uniformly on every
$K_T$ to continuous limits. Denote these limits by
$r_\infty,-\partial_ur_\infty,\partial_vr_\infty,
\mu_\infty,\alpha_\infty,\phi_\infty$.
To pass to the limit in \eqref{center-incoming-average}, write
the integral in the same coordinates $(u,v)$:
\[
 \begin{aligned}
 \frac{r_jD_{u,j}\phi_j}{-\partial_ur_j}(u,v)
 &=\frac1{(-\partial_ur_j)(u,v)}\int_u^v
       \frac{(-\partial_ur_j)\partial_vr_j}{r_j}\\
 &\qquad\times\left[\alpha_j-\phi_j
       -\frac{\mathrm i\mathfrak e_jQ_j\phi_j}{1-\mu_j}\right]
       (u,v')\,\mathrm dv'.
 \end{aligned}
\]
For $u<v'\leq v$, we have
\[
 \begin{aligned}
 \frac{(-\partial_ur_j)(u,v')}{(-\partial_ur_j)(u,v)}
       \left(\frac{\partial_vr_j\,|\alpha_j-\phi_j|}{r_j}
       \right)(u,v')
 &\leq C,\\
 \frac{(-\partial_ur_j)(u,v')}{(-\partial_ur_j)(u,v)}
       \left(\frac{\partial_vr_j\,|\mathfrak e_jQ_j\phi_j|}
       {r_j(1-\mu_j)}\right)(u,v')
 &\leq\frac{C_Rr_j(u,v')^2}{j^2}.
 \end{aligned}
\]
Here the ratio is bounded by \eqref{speed-ratio}, and
$\partial_vr_j\leq C$ by \eqref{outgoing-speed-bounds}.
The first line then uses $|\alpha_j-\phi_j|\leq r_j/2$;
the second uses \eqref{first-center-orders}, $|\phi_j|\leq C$,
$1-\mu_j\geq C^{-1}$ and $\mathfrak e_j=\mathfrak e/j$.
Since $r_j(u,v')\leq C(v'-u)$, for $0<d\leq1$ the absolute
contribution from $u<v'<\min\{u+d,v\}$ is at most
\[
 C_R\int_u^{\min\{u+d,v\}}
       \left(1+\frac{(v'-u)^2}{j^2}\right)\,\mathrm dv'
 \leq C_Rd.
\]
On the remaining region $v'-u\geq d$, we have
$r_j(u,v')\geq C^{-1}d$. The uniform field convergence on $K_T$,
together with \eqref{first-center-orders}, therefore gives
uniform convergence of the integrands, with the charged term
tending to zero. Splitting both integrals at $v'=u+d$ and using
the $C_Rd$ bound for each near-center contribution, we obtain
\[
 \left\|
 \frac{r_jD_{u,j}\phi_j}{-\partial_ur_j}
       -\frac{r_\infty\partial_u\phi_\infty}{-\partial_ur_\infty}
 \right\|_{C^0(K_T)}
 \leq\underbrace{2C_Rd}_{0<v'-u<d}
       +\underbrace{o_{j\to\infty}(1)}_{v'-u\geq d}.
\]
For fixed $d$, the last term tends to zero uniformly on $K_T$:
the integration intervals have length at most $T$, and
$1/(-\partial_ur_j)$ converges uniformly and is uniformly bounded.
Letting $j\to\infty$ and then $d\to0$ proves the required uniform
convergence. Here the limiting expression is given by
\eqref{center-incoming-average} with the limiting fields and
zero coupling and charge.
{The $j^{-1}$ bounds for $Q_j,A_{u,j}$ and the scalar and metric
derivative bounds from Step~2, used in \eqref{maxwell-equations}
and the differentiated potential integral, give, on fixed compact sets,
\[
 Q_j,\ A_{u,j},\ \partial_uQ_j,\ \partial_vQ_j,\
 \partial_uA_{u,j},\ \partial_vA_{u,j}\longrightarrow0.
\]}
Both scalar fluxes therefore converge locally uniformly. Passing
to the integrated mass equations identifies the limiting mass
with $m_\infty=\mu_\infty r_\infty/2$ and gives the neutral mass equations.

\textit{Step 4. Positivity of the limiting mass.}
Let $(\widehat u_j,0)$ be a normalized maximum, so that
$|\alpha'_j(\widehat u_j,0)|=1$. We will use this equality to force
a positive lower bound for the integral in \eqref{first-nonzero-annulus}
on a fixed annulus $d\leq r_j\leq R$. For this step and Step~5, set
\begin{equation}\label{neutral-source-integrand}
 \mathcal S_j:=\frac{
  4\pi|\alpha_j-\phi_j|^2(\alpha_j-\phi_j)
          -3\mu_j(\alpha_j-\phi_j)}
      {r_j^2(1-\mu_j)},
\end{equation}
and let $\mathcal E_j$ be the remaining source in
\eqref{scalar-multiplier}, defined by
\[
 D_{u,j}\bigl((\partial_vr_j)^2\alpha'_j\bigr)
 =(-\partial_ur_j)(\partial_vr_j)^2(\mathcal S_j+\mathcal E_j).
\]
By \eqref{outgoing-speed-bounds} and the scaling of $\alpha'$
with bounded initial $C^1$ datum, for $-jb_j\leq w\leq\widehat u_j$,
\[
 \frac{(\partial_vr_j)^2(w,0)}
      {(\partial_vr_j)^2(\widehat u_j,0)}\leq C,
 \qquad
 |\alpha'_j(-jb_j,0)|=\frac{|\alpha'(0,b_j)|}{j}\leq\frac Cj.
\]
Integrating the equation for $(\partial_vr_j)^2\alpha'_j$
from the initial cone gives
\[
 \begin{aligned}
 &[(\partial_vr_j)^2\alpha'_j](\widehat u_j,0)\\
 &=\exp\left(-\mathrm i\mathfrak e_j
      \int_{-jb_j}^{\widehat u_j}A_{u,j}(t,0)\,\mathrm dt\right)
      [(\partial_vr_j)^2\alpha'_j](-jb_j,0)\\
 &\quad+\int_{-jb_j}^{\widehat u_j}
      \exp\left(-\mathrm i\mathfrak e_j
      \int_w^{\widehat u_j}A_{u,j}(t,0)\,\mathrm dt\right)\\
 &\qquad\qquad\times
      [(-\partial_ur_j)(\partial_vr_j)^2
                  (\mathcal S_j+\mathcal E_j)](w,0)\,\mathrm dw.
 \end{aligned}
\]
Since $\mathfrak e_j,A_{u,j}$ are real, both exponentials in this
identity have modulus one. Taking absolute values, dividing by
$(\partial_vr_j)^2(\widehat u_j,0)$ and using the preceding two
bounds therefore yields
\[
 1=|\alpha'_j(\widehat u_j,0)|\leq\frac Cj
 +C\int_{-jb_j}^{\widehat u_j}
       (-\partial_ur_j)(|\mathcal S_j|+|\mathcal E_j|)(w,0)
       \,\mathrm dw.
\]
To estimate the integral on the right, write, in this step only,
\[
 \begin{aligned}
 I_j(a,b)&:=\int_{\substack{-jb_j\leq w\leq\widehat u_j\\
                          a<r_j(w,0)<b}}
           (-\partial_ur_j)|\mathcal S_j|(w,0)\,\mathrm dw,\\
 J_j&:=\int_{-jb_j}^{\widehat u_j}
           (-\partial_ur_j)|\mathcal E_j|(w,0)\,\mathrm dw.
 \end{aligned}
\]
For $0<d\leq1\leq R$, the preceding inequality becomes
\[
 1\leq\frac Cj+C\bigl(I_j(0,d)+I_j(d,R)+I_j(R,\infty)+J_j\bigr).
\]
We will make $I_j(0,d)$, $I_j(R,\infty)$ and $J_j$ small;
the remaining term $I_j(d,R)$ must then stay positive.

For $I_j(0,d)$ and $I_j(R,\infty)$, use
\eqref{first-center-orders} with $R=1$ and
$|\alpha_j-\phi_j|\leq r_j/2$ when $r_j\leq1$, and
\eqref{scalar-bounds} and \eqref{uniform-margin} when $r_j\geq1$.
These give the first estimate for $\mathcal S_j$:
\begin{equation}\label{first-source-radius-bound}
 |\mathcal S_j|\leq
 \begin{cases}
 Cr_j,&0<r_j\leq1,\\
 C/r_j^2,&r_j\geq1.
 \end{cases}
\end{equation}
Substituting this first estimate \eqref{first-source-radius-bound}
in the definition of $I_j$, with $s=r_j(w,0)$ and
$-\mathrm ds=(-\partial_ur_j)\,\mathrm dw$, yields
\begin{align*}
 I_j(0,d)
 &\leq C\int_0^d s\,\mathrm ds\leq Cd^2,\\
 I_j(R,\infty)
 &\leq C\int_R^\infty s^{-2}\,\mathrm ds=\frac CR.
\end{align*}
For $I_j(d,R)$, use \eqref{scalar-bounds} and
\eqref{uniform-margin} in \eqref{neutral-source-integrand}, retaining
$|\alpha_j-\phi_j|$. This gives the second estimate:
\begin{equation}\label{first-source-field-bound}
 |\mathcal S_j|\leq\frac{C|\alpha_j-\phi_j|}{r_j^2}.
\end{equation}
Substituting this second estimate \eqref{first-source-field-bound}
in $I_j(d,R)$ and enlarging the interval of integration along
$v=0$ gives
\[
 I_j(d,R)\leq C\int_{\{v=0,\ d\leq r_j\leq R\}}
                 \frac{|\alpha_j-\phi_j|}{r_j^2}\,\mathrm dr_j.
\]
For $J_j$, use \eqref{scalar-multiplier},
$|\phi_j|+|\alpha_j-\phi_j|\leq C$ from \eqref{scalar-bounds},
the rescaled charge bound $Q_j^2\leq C\mu_jr_j^4/j^2$ from
\eqref{improved-charge}, and $\mathfrak e_j=\mathfrak e/j$,
$1-\mu_j\geq C^{-1}$. These give
\begin{equation}\label{first-source-charge-bound}
 |\mathcal E_j|
 \leq C\left(\frac{Q_j^2}{r_j^4}
       +\frac{|\mathfrak e_jQ_j|}{r_j^2}
       +\mathfrak e_j^2\right)\leq\frac C{j^2}.
\end{equation}
Substitution of \eqref{first-source-charge-bound} in $J_j$ yields
\[
 J_j\leq\frac C{j^2}\int_{-jb_j}^{\widehat u_j}
                  (-\partial_ur_j)(w,0)\,\mathrm dw
 \leq\frac C{j^2}r_j(-jb_j,0)
 \leq\frac Cj r(0,v_*)\leq\frac Cj,
\]
where the original initial radius $r(0,v_*)$ is fixed.
% Expanded substitution retained for the detailed version:
% Putting these four bounds into the inequality for
% $1=|\alpha'_j(\widehat u_j,0)|$, we obtain
% \[
%  \begin{aligned}
%  1&\leq\underbrace{C/j}_{\text{initial value}}
%        +C\Bigl(
%        \underbrace{Cd^2}_{\geq I_j(0,d)}
%        +\underbrace{C\!\int_{\{v=0,\ d\leq r_j\leq R\}}
%              \frac{|\alpha_j-\phi_j|}{r_j^2}\,\mathrm dr_j}_{\geq I_j(d,R)}
%        +\underbrace{C/R}_{\geq I_j(R,\infty)}
%        +\underbrace{C/j}_{\geq J_j}\Bigr).
%  \end{aligned}
% \]
% Thus, after enlarging $C$,
Substituting the bounds for $I_j(0,d)$, $I_j(d,R)$,
$I_j(R,\infty)$ and $J_j$ into the inequality for
$1=|\alpha'_j(\widehat u_j,0)|$ gives, after enlarging $C$,
\begin{equation}\label{first-nonzero-annulus}
 1\leq C(j^{-1}+d^2+R^{-1})
   +C\int_{\{v=0,\ d\leq r_j\leq R\}}
          \frac{|\alpha_j-\phi_j|}{r_j^2}\,\mathrm dr_j .
\end{equation}
Choose $d$ small and then $R$ large enough, also larger than the
radii of the maximizing points. For all large $j$ the annular
integral is bounded below. Hence there are points $(u_j,0)$ with
\[
 d\leq r_j(u_j,0)\leq R,\qquad
 |\alpha_j-\phi_j|(u_j,0)\geq c_{d,R}>0.
\]
The lower bound $\partial_vr_j\geq C^{-1}$ from
\eqref{outgoing-speed-bounds} gives
\[
 C^{-1}(-u_j)
 \leq\int_{u_j}^0\partial_vr_j(u_j,v)\,\mathrm dv
 =r_j(u_j,0)\leq R.
\]
Hence, with $T=CR$ independent of $j$,
\[
 -CR\leq u_j\leq0,\qquad (u_j,0)\in K_T.
\]
After taking a subsequence,
\[
 u_j\longrightarrow u_\infty\in[-T,0].
\]
Uniform convergence on this $K_T$
from Step~3, together with continuity of the limiting fields, yields
\[
 r_\infty(u_\infty,0)\geq d,\qquad
 |\alpha_\infty-\phi_\infty|(u_\infty,0)\geq c_{d,R}>0.
\]
By continuity, there is $v_0\in(u_\infty,0)$ such that
\[
 |\alpha_\infty-\phi_\infty|(u_\infty,v)
 \geq\frac{c_{d,R}}2\qquad(v_0\leq v\leq0).
\]
Since $m_\infty\geq0$, $1-\mu_\infty\geq C^{-1}$ and
$\partial_vr_\infty\geq C^{-1}$, the integrated mass equation
\eqref{outgoing-mass}, with $Q_\infty=0$, yields
\[
 \begin{aligned}
 m_\infty(u_\infty,0)
 &\geq m_\infty(u_\infty,0)-m_\infty(u_\infty,v_0)\\
 &=2\pi\int_{v_0}^0
   \bigl[(1-\mu_\infty)(\partial_vr_\infty)
   |\alpha_\infty-\phi_\infty|^2\bigr](u_\infty,v)\,\mathrm dv\\
 &\geq\frac{\pi c_{d,R}^2}{2C^2}(-v_0)>0.
 \end{aligned}
\]

\textit{Step 5. Convergence of the differentiated fields.}
Fix $T<\infty$. To verify that the limit has the regularity in
Definition~\ref{regularity-definition}, we prove $C^1$ convergence on
$K_T$ of
\[
 \Omega_j,\quad \partial_ur_j,\quad \partial_vr_j,\quad
 \phi_j,\quad r_j\partial_u\phi_j,\quad r_j\partial_v\phi_j.
\]
The first-derivative bounds in Step~2 alone give only Lipschitz
control of the limiting fields. We therefore prove uniform
convergence of the derivatives required above, including the
second derivatives of $r_j$ and the derivatives of
$r_j\partial_u\phi_j$ and $r_j\partial_v\phi_j$; no unweighted
second derivative of $\phi_j$ is required at the center.
Parts~(a)--(b) establish convergence of $\alpha'_j$ and the
quotients needed at the center. Parts~(c)--(d) then recover the
pure second derivatives of $r_j$, and part~(e) completes the
$C^1$ convergence of the displayed fields using their equations.
In this step only, abbreviate
\[
 \lambda_j=\partial_vr_j,\qquad \nu_j=\partial_ur_j,\qquad
 \rho_j=1-\mu_j,\qquad \|f\|_T=\|f\|_{C^0(K_T)}.
\]

\textit{(a) Convergence of $\alpha'_j$.}
Fix $N>T+1$ and $0<d<1$. For $-N\leq w\leq u\leq v$, put
\[
 P_j(u,w;v)=\exp\left(-\mathrm i\mathfrak e_j
                    \int_w^u A_{u,j}(t,v)\,\mathrm dt\right),
 \qquad H_j=(-\nu_j)\lambda_j^2\mathcal S_j.
\]
With $\mathcal S_j,\mathcal E_j$ defined in Step~4, the integrated
equation \eqref{scalar-multiplier} is exactly
\[
 \begin{aligned}
 \alpha'_j(u,v)
 &=B_{j,N}(u,v)+\frac1{\lambda_j(u,v)^2}
       \int_{-N}^u P_j(u,w;v)
       [H_j+(-\nu_j)\lambda_j^2\mathcal E_j](w,v)\,\mathrm dw,\\
 B_{j,N}(u,v)
 &:=P_j(u,-N;v)
       \frac{(\lambda_j^2\alpha'_j)(-N,v)}{\lambda_j(u,v)^2}.
 \end{aligned}
\]
In the uncharged case, $P_j=1$ and $\mathcal E_j=0$.
We first consider the part of the integral with $v-w\geq d$;
the contribution from $v-w<d$ will be estimated below.
Using the uncharged form of the integral term, evaluated with
the $j$th fields, define
\[
 L_{j,N,d}(u,v):=\frac1{\lambda_j(u,v)^2}
       \int_{-N}^{\min\{u,v-d\}}H_j(w,v)\,\mathrm dw.
\]
The integration domain here is the same for every $j$.
On $v-w\geq d$, \eqref{outgoing-speed-bounds} gives
$r_j(w,v)\geq C^{-1}d$. Hence Step~3 and
\eqref{neutral-source-integrand} imply uniform convergence of $H_j$
there and of $\lambda_j^{-2}$ on $K_T$. Since the integration
intervals have length at most $N$, for fixed $N,d$,
\[
 \|L_{j,N,d}-L_{k,N,d}\|_T\longrightarrow0
       \qquad(j,k\to\infty).
\]
We next estimate $\alpha'_j-L_{j,N,d}$.
For the initial value, \eqref{critical-derivative-bounds},
\eqref{outgoing-speed-bounds} and
$r_j(-N,v)\geq C^{-1}(N-T)$ give
\[
 \|B_{j,N}\|_T\leq\frac C{N-T}.
\]
For $P_j$, use $|P_j|=1$ and the potential
bound from Step~2 with $\mathfrak e_j=\mathfrak e/j$:
\[
 |P_j(u,w;v)-1|
 \leq|\mathfrak e_j|\int_w^u|A_{u,j}(t,v)|\,\mathrm dt
 \leq C_N/j^2.
\]
The estimate for $\mathcal S_j$ in \eqref{first-source-radius-bound} and the change
of variable $s=r_j(w,v)$ then give
\[
 \frac1{\lambda_j(u,v)^2}
 \int_{-N}^u |P_j-1|\,|H_j|(w,v)\,\mathrm dw
 \leq\frac{C_N}{j^2}
       \left(\int_0^1 s\,\mathrm ds+\int_1^\infty s^{-2}\,\mathrm ds\right)
 \leq\frac{C_N}{j^2}.
\]
For the integral containing $\mathcal E_j$, apply
\eqref{first-source-charge-bound} and the bounds for $\lambda_j$:
\[
 \frac1{\lambda_j(u,v)^2}\int_{-N}^u
       (-\nu_j)\lambda_j^2|\mathcal E_j|(w,v)\,\mathrm dw
 \leq\frac C{j^2}r_j(-N,v)\leq\frac{C_N}{j^2}.
\]
Substituting these three bounds in the integral formula for $\alpha'_j$ gives
\[
 \sup_{(u,v)\in K_T}
 \left|\alpha'_j(u,v)-\frac1{\lambda_j(u,v)^2}
       \int_{-N}^u H_j(w,v)\,\mathrm dw\right|
 \leq\frac C{N-T}+\frac{C_N}{j^2}.
\]
For the complementary region $v-w<d$, we have $r_j(w,v)\leq Cd$.
The estimate for $\mathcal S_j$ in \eqref{first-source-radius-bound} gives
\[
 \frac1{\lambda_j(u,v)^2}
 \int_{\substack{-N\leq w\leq u\\ v-w<d}}|H_j|(w,v)\,\mathrm dw
 \leq C\int_0^{Cd}s\,\mathrm ds\leq Cd^2.
\]
Combining these two bounds and splitting the integral into
$v-w\geq d$ and $v-w<d$ gives
\[
 \|\alpha'_j-L_{j,N,d}\|_T
 \leq\frac C{N-T}+Cd^2+\frac{C_N}{j^2}.
\]
Applying it to $j$ and $k$ shows explicitly that
\[
 \begin{aligned}
 \|\alpha'_j-\alpha'_k\|_T
 &\leq\frac C{N-T}+Cd^2+C_N(j^{-2}+k^{-2})
       +\|L_{j,N,d}-L_{k,N,d}\|_T,\\
 \limsup_{j,k\to\infty}\|\alpha'_j-\alpha'_k\|_T
 &\leq\frac C{N-T}+Cd^2.
 \end{aligned}
\]
Letting $N\to\infty$ and then $d\to0$ proves uniform convergence
of $\alpha'_j$ on $K_T$, including its center.
% Passing to
% \[
%  \alpha_j(u,v_2)-\alpha_j(u,v_1)
%     =\int_{v_1}^{v_2}(\partial_vr_j)\alpha'_j(u,v)\,\mathrm dv
% \]
% identifies the limit as $\alpha'_\infty$.

\textit{(b) The quotients needed at the center.}
We next need uniform convergence of
$(\alpha_j-\phi_j)/r_j$, $\mu_j/r_j^2$ and
$D_{u,j}\phi_j/(-\nu_j)$, not just convergence at positive radius.
We use these quotients in part~(d) to prove convergence of
$\partial_u^2r_j$, and in part~(e) to prove $C^1$ convergence of
$r_j\partial_u\phi_j$ and $r_j\partial_v\phi_j$.
Write
\[
 \beta_j=\frac{\alpha_j-\phi_j}{r_j},\qquad
 \chi_j=\frac{\mu_j}{r_j^2},\qquad
 \psi_j=\frac{D_{u,j}\phi_j}{-\nu_j},\qquad
 q_j=\frac{Q_j}{r_j^2}.
\]
In the outgoing radial parametrization already used in Step~2,
\eqref{scalar-averages}, \eqref{radial-mass} and
\eqref{center-incoming-average}, respectively, give
\[
 \begin{aligned}
 \beta_j(u,r)&=\int_0^1 t\,\alpha'_j(u,tr)\,\mathrm dt,\\
 \chi_j(u,r)&=\int_0^1 t^2
       [4\pi\rho_j|\beta_j|^2+q_j^2](u,tr)\,\mathrm dt,\\
 \psi_j(u,r)&=\int_0^1\frac{(-\nu_j)(u,tr)}{(-\nu_j)(u,r)}
       \left[\beta_j-
       \frac{\mathrm i\mathfrak e_jQ_j\phi_j}{r_j\rho_j}\right]
                (u,tr)\,\mathrm dt.
 \end{aligned}
\]
For the integrals defining $\chi_j$ and $\psi_j$,
\eqref{first-center-orders} and Step~2 give
\[
 \begin{gathered}
 |q_j|\leq\frac{C_Rr_j}{j},\qquad
 \left|\frac{\mathfrak e_jQ_j\phi_j}{r_j\rho_j}\right|
       \leq\frac{C_Rr_j^2}{j^2},\\
 C_R^{-1}\leq-\nu_j\leq C_R.
 \end{gathered}
\]
On each compact subset of $K_T\cap\{r_\infty>0\}$, the positive
lower bounds for $r_j$ and $-\nu_j$, together with the uniform
convergence in Step~3, give uniform convergence of
$\beta_j,\chi_j,\psi_j$.
Near the center, the formula for $\beta_j$ gives the explicit
estimate
\[
 \left|\beta_j(u,r)-\tfrac12\alpha'_j(u,0)\right|
 \leq\tfrac12\sup_{0\leq s\leq r}
                     |\alpha'_j(u,s)-\alpha'_j(u,0)|.
\]
Uniform convergence in part~(a), continuity of its limit, and
$\lambda_j\geq C^{-1}$ make the right side uniformly small as
$r\to0$ and $j\to\infty$.

By \eqref{first-center-orders} and the uniform Lipschitz bounds
for $\nu_j$ from Step~2, we have, uniformly near the center,
\[
 \rho_j\longrightarrow1,\qquad
 \frac{\nu_j(u,tr)}{\nu_j(u,r)}\longrightarrow1,\qquad
 q_j\longrightarrow0.
\]
Using these limits in the integral formulas for $\chi_j$ and $\psi_j$
gives uniform convergence of $\beta_j,\chi_j,\psi_j$ on $K_T$.
In the radial coordinates $(u,r)$ of the limiting solution, the center
is $r=0$. Substituting the preceding limits into the three integral
formulas gives, as $r\downarrow0$,
\begin{equation}\label{first-center-limits}
 \begin{aligned}
 \frac{\alpha_\infty-\phi_\infty}{r_\infty}(u,r)
 &\longrightarrow\alpha'_\infty(u,0)\int_0^1t\,\mathrm dt
       =\frac{\alpha'_\infty(u,0)}2,\\
 \frac{\mu_\infty}{r_\infty^2}(u,r)
 &\longrightarrow4\pi\left|\frac{\alpha'_\infty(u,0)}2\right|^2
       \int_0^1t^2\,\mathrm dt
       =\frac{\pi|\alpha'_\infty(u,0)|^2}{3},\\
 \frac{\partial_u\phi_\infty}{-\partial_ur_\infty}(u,r)
 &\longrightarrow\frac{\alpha'_\infty(u,0)}2\int_0^1\mathrm dt
       =\frac{\alpha'_\infty(u,0)}2.
 \end{aligned}
\end{equation}

\textit{(c) Convergence of $\partial_v^2r_j$.}
For $G_j:=\partial_v^2r_j/\lambda_j^2$, we follow part~(a):
integrate from $w=-N$ and split the integration interval into
$v-w\geq d$ and $v-w<d$.
Define the two expressions
appearing in \eqref{metric-multiplier} by
\[
 \begin{aligned}
 \mathcal M_j&=\frac{2\mu_j-4\pi|\alpha_j-\phi_j|^2}{r_j^2\rho_j},\\
 \mathcal R_j&=\frac{(-4+4\pi|\alpha_j-\phi_j|^2)Q_j^2/r_j^2
       +8\pi\mathfrak e_jQ_j
             \operatorname{Im}(\phi_j\overline{\alpha_j-\phi_j})}
       {r_j^2\rho_j}.
 \end{aligned}
\]
Equation \eqref{metric-multiplier} is
$\partial_u(\lambda_jG_j)=(-\nu_j)\lambda_j(\mathcal M_j+\mathcal R_j)$,
so
\[
 G_j(u,v)=\frac{(\lambda_jG_j)(-N,v)}{\lambda_j(u,v)}
 +\frac1{\lambda_j(u,v)}\int_{-N}^u
       [(-\nu_j)\lambda_j(\mathcal M_j+\mathcal R_j)](w,v)\,\mathrm dw.
\]
Equations \eqref{critical-derivative-bounds} and
\eqref{outgoing-speed-bounds} bound the initial value; 
\eqref{first-center-orders}, $|\alpha_j-\phi_j|\leq r_j/2$ and
\eqref{scalar-bounds} bound $\mathcal M_j$; the rescaled
\eqref{improved-charge}, together with \eqref{scalar-bounds},
bounds $\mathcal R_j$. Using also $\rho_j\geq C^{-1}$ from
\eqref{uniform-margin}, we obtain
\[
 \begin{gathered}
 \left|\frac{(\lambda_jG_j)(-N,v)}{\lambda_j(u,v)}\right|
 \leq\frac C{N-T},\\
 |\mathcal M_j|\leq
 \begin{cases}C,&r_j\leq1,\\ C/r_j^2,&r_j\geq1,\end{cases}
 \qquad |\mathcal R_j|\leq C/j^2.
 \end{gathered}
\]
We first consider $v-w\geq d$, where $r_j(w,v)\geq C^{-1}d$.
For fixed $N,d$, Step~3 gives uniform convergence of
$(-\nu_j)\lambda_j\mathcal M_j$ there and of $1/\lambda_j(u,v)$ on $K_T$.
Hence the corresponding integrals over $[-N,\min\{u,v-d\}]$
converge uniformly on $K_T$.
For $v-w<d$, changing to $s=r_j(w,v)$ and using
$|\mathcal M_j|\leq C$ gives
\[
 \frac1{\lambda_j(u,v)}
 \int_{\substack{-N\leq w\leq u\\v-w<d}}
       (-\nu_j)\lambda_j|\mathcal M_j|(w,v)\,\mathrm dw
 \leq C\int_0^{Cd}\mathrm ds=Cd.
\]
The bound for $\mathcal R_j$ gives
\[
 \frac1{\lambda_j(u,v)}\int_{-N}^u
       (-\nu_j)\lambda_j|\mathcal R_j|(w,v)\,\mathrm dw
 \leq\frac{Cr_j(-N,v)}{j^2}\leq\frac{C_N}{j^2}.
\]
Together with the initial-value bound, these estimates give
\[
 \sup_{(u,v)\in K_T}\left|
 G_j(u,v)-\frac1{\lambda_j(u,v)}
 \int_{-N}^{\min\{u,v-d\}}
          [(-\nu_j)\lambda_j\mathcal M_j](w,v)\,\mathrm dw
 \right|\leq\frac C{N-T}+Cd+\frac{C_N}{j^2}.
\]
As in part~(a), the uniform convergence of these integrals
and this error bound imply
\[
 \limsup_{j,k\to\infty}\|G_j-G_k\|_T
 \leq\frac C{N-T}+Cd.
\]
Letting $N\to\infty$ and then $d\to0$, and using
$\partial_v\lambda_j=\lambda_j^2G_j$ and the uniform convergence
of $\lambda_j$, we obtain
\[
 \partial_v^2r_j\longrightarrow\partial_v^2r_\infty
       \qquad\text{uniformly on }K_T.
\]

\textit{(d) Convergence of $\partial_u^2r_j$.}
Define
\[
 \mathcal A_j=\frac{\lambda_j}{r_j}
                  \frac{\mu_j-Q_j^2/r_j^2}{\rho_j}
             =\frac{\lambda_j r_j(\chi_j-q_j^2)}{\rho_j}.
\]
The differentiated identity \eqref{first-null-speed-identity},
already obtained in Step~2, reads
\[
 \frac{\partial_u\nu_j}{\nu_j}(u,v)
   =(\lambda_jG_j)(u,u)+\int_u^v\partial_u\mathcal A_j(u,s)\,\mathrm ds.
\]
Part~(c) gives uniform convergence of $(\lambda_jG_j)(u,u)$.
We now prove uniform convergence of the integrand $\partial_u\mathcal A_j$.
The incoming mass equation \eqref{incoming-mass}, the current equation
\eqref{maxwell-equations} and \eqref{mixed-radius}, respectively, give
\begin{align*}
 \frac{\partial_u\mu_j}{r_j}
 &=\frac{4\pi(1-\mu_j)|D_{u,j}\phi_j|^2}{\partial_ur_j}
   +\frac{Q_j^2\partial_ur_j}{r_j^4}
   -\frac{\mu_j\partial_ur_j}{r_j^2},\\
 \frac{\partial_uQ_j}{r_j^2}
 &=-4\pi\mathfrak e_j
       \operatorname{Im}(\phi_j\overline{D_{u,j}\phi_j}),\\
 \frac{\partial_u\partial_vr_j}{r_j}
 &=\frac{\partial_ur_j\,\partial_vr_j}{1-\mu_j}
       \left(\frac{\mu_j}{r_j^2}-\frac{Q_j^2}{r_j^4}\right).
\end{align*}
Part~(b) gives uniform convergence of $\chi_j$ and $\psi_j$.
Using Step~3 and the charge bound in part~(b), we also have,
uniformly on $K_T$,
\[
 \begin{gathered}
 D_{u,j}\phi_j=-\nu_j\psi_j\longrightarrow\partial_u\phi_\infty,\\
 |q_j|\leq\frac{C_Rr_j}{j}\longrightarrow0.
 \end{gathered}
\]
Thus the three displayed derivative quotients converge
uniformly on $K_T$. They are precisely the combinations required
when differentiating $\mathcal A_j$:
\[
 \begin{aligned}
 \partial_u\mathcal A_j
 &=\frac{(r_j\partial_u\lambda_j-\lambda_j\nu_j)
                         (\chi_j-q_j^2)}{\rho_j}\\
 &\quad+\frac{\lambda_j}{\rho_j}
       \left(\frac{\partial_u\mu_j}{r_j}
          -2r_jq_j\frac{\partial_uQ_j}{r_j^2}
          +2q_j^2\nu_j\right)
       +\frac{\lambda_jr_j(\chi_j-q_j^2)\partial_u\mu_j}{\rho_j^2}.
 \end{aligned}
\]
Every term on the right now converges uniformly, including at the
center. Substituting this convergence and part~(c) into the formula
for $\partial_u\nu_j/\nu_j$ proves
\[
 \partial_u^2r_j\longrightarrow\partial_u^2r_\infty
       \qquad\text{uniformly on }K_T.
\]

\textit{(e) $C^1$ convergence of $r_j\partial_u\phi_j$ and $r_j\partial_v\phi_j$.}
The remaining pure incoming derivative is
$\partial_u(r_j\partial_u\phi_j)$. To recover it without dividing
by $r_j$ at the center, set
\[
 W_j=\frac{D_{u,j}(r_j\phi_j)}{\nu_j},\qquad
 Y_j=\partial_uW_j,\qquad
 \mathcal B_j=\frac{\mathrm i\mathfrak e_jQ_j\lambda_j}{r_j\rho_j},
 \qquad h_j(u)=(\lambda_j\alpha'_j)(u,u).
\]
Using $\mathcal A_j$ from part~(d), the differentiated equation
\eqref{pure-incoming-integral} is a linear integral equation for $Y_j$:
\[
 Y_j(u,v)=h_j(u)
       +\int_u^v[-\mathcal A_jY_j+\mathcal F_j](u,s)
             \,\mathrm ds,
\]
where
\[
 \mathcal F_j=(\mathcal A_j+\mathcal B_j)\partial_u\phi_j
       +(\partial_u\mathcal A_j)(\phi_j-W_j)
       +(\partial_u\mathcal B_j)\phi_j.
\]
We verify convergence of its coefficients, forcing term and initial
value before applying Gronwall's inequality. The formulas in part~(d)
and \eqref{first-center-orders} give the bounds for $\mathcal A_j$
and $\partial_u\mathcal A_j$. For $\mathcal B_j$ and
$\partial_u\mathcal B_j$, use \eqref{mixed-radius},
\eqref{incoming-mass}, \eqref{maxwell-equations} and
\eqref{first-center-orders}, with $\mathfrak e_j=\mathfrak e/j$.
Thus, on $K_T$,
\[
 \begin{gathered}
 |\mathcal A_j|\leq C_T r_j,\qquad
 |\partial_u\mathcal A_j|\leq C_T,\\
 |\mathcal B_j|\leq C_T r_j^2/j^2,\qquad
 |\partial_u\mathcal B_j|\leq C_T r_j/j^2.
 \end{gathered}
\]
Moreover, part~(b) and Step~3 give
\[
 \phi_j-W_j=r_j\psi_j,\qquad
 \partial_u\phi_j=-\nu_j\psi_j
                    -\mathrm i\mathfrak e_jA_{u,j}\phi_j.
\]
Both expressions converge uniformly on $K_T$ and satisfy
\[
 |\phi_j-W_j|\leq C_T r_j,\qquad
 |\partial_u\phi_j|\leq C_T.
\]
Substitution into the displayed formula for $\mathcal F_j$, together
with part~(a) for $h_j$, yields
\begin{gather*}
 |\mathcal F_j|\leq C_T r_j,\\
 \mathcal A_j\to\mathcal A_\infty,\quad
 \mathcal F_j\to\mathcal F_\infty
       \quad\text{uniformly on }K_T,\\
 h_j\to h_\infty\quad\text{uniformly on }[-T,0].
\end{gather*}
Applying Gronwall's inequality to the integral equation for $Y_j$, we obtain
\[
 \|Y_j\|_T\leq C_T.
\]
% Expanded Gronwall bound retained for the detailed version:
% \[
%  \|Y_j\|_T
%  \leq\bigl(\|h_j\|_{C^0([-T,0])}+T\|\mathcal F_j\|_T\bigr)
%              e^{T\|\mathcal A_j\|_T}\leq C_T.
% \]
To prove convergence, subtract the two integral equations:
\[
 \begin{aligned}
 (Y_j-Y_k)(u,v)&=(h_j-h_k)(u)\\
 &\quad+\int_u^v\bigl[-\mathcal A_j(Y_j-Y_k)
       +(\mathcal A_k-\mathcal A_j)Y_k
       +\mathcal F_j-\mathcal F_k\bigr](u,s)\,\mathrm ds.
 \end{aligned}
\]
Using $\|Y_k\|_T\leq C_T$ and applying Gronwall's inequality
to this difference equation, we obtain
\[
 \|Y_j-Y_k\|_T\leq C_T\left(
       \|h_j-h_k\|_{C^0([-T,0])}
       +\|\mathcal A_j-\mathcal A_k\|_T
       +\|\mathcal F_j-\mathcal F_k\|_T\right).
\]
The uniform convergence of $h_j$, $\mathcal A_j$ and $\mathcal F_j$
proved above makes the right-hand side tend to zero as $j,k\to\infty$.
Hence $Y_j$ converges uniformly on $K_T$.

We now use this convergence to prove convergence of
$\partial_u(r_j\partial_u\phi_j)$. Differentiating the identity
\[
 r_j\partial_u\phi_j=\nu_j(W_j-\phi_j)
       -\mathrm i\mathfrak e_jr_jA_{u,j}\phi_j
\]
with respect to $u$ gives
\[
 \partial_u(r_j\partial_u\phi_j)
   =(\partial_u\nu_j)(W_j-\phi_j)
      +\nu_j(Y_j-\partial_u\phi_j)
      -\mathrm i\mathfrak e_j\partial_u(r_jA_{u,j}\phi_j).
\]
Parts~(b), (d), the convergence of $Y_j$ and the potential derivative
convergence from Step~3 show that the right side converges uniformly.
For the pure outgoing derivative, use parts~(a)--(c) directly in
\[
 \partial_v(r_j\partial_v\phi_j)
   =(\partial_v\lambda_j)(\alpha_j-\phi_j)
       +\lambda_j^2(\alpha'_j-\beta_j).
\]
The two remaining derivatives are
$\partial_v(r_j\partial_u\phi_j)$ and
$\partial_u(r_j\partial_v\phi_j)$.
Equations \eqref{incoming-scalar-transport} and
\eqref{outgoing-scalar-transport}, respectively, give
\[
 \begin{aligned}
 \partial_v(r_j\partial_u\phi_j)
 &=-\nu_j\partial_v\phi_j
   +\frac{\mathrm i\mathfrak e_jr_jq_j\phi_j\nu_j\lambda_j}{\rho_j}
   -\mathrm i\mathfrak e_j\partial_v(r_jA_{u,j}\phi_j),\\
 \partial_u(r_j\partial_v\phi_j)
 &=-\lambda_jD_{u,j}\phi_j
   -\frac{\mathrm i\mathfrak e_jr_jq_j\phi_j\nu_j\lambda_j}{\rho_j}
   -\mathrm i\mathfrak e_jA_{u,j}r_j\partial_v\phi_j.
 \end{aligned}
\]
By part~(b) and Step~3, the first term in each expression converges
uniformly on $K_T$, and all terms containing $\mathfrak e_j$ tend
uniformly to zero. Consequently, uniformly on $K_T$,
\[
 \begin{aligned}
 \partial_v(r_j\partial_u\phi_j)
 &\longrightarrow-\partial_ur_\infty\,\partial_v\phi_\infty
       =\partial_v(r_\infty\partial_u\phi_\infty),\\
 \partial_u(r_j\partial_v\phi_j)
 &\longrightarrow-\partial_vr_\infty\,\partial_u\phi_\infty
       =\partial_u(r_\infty\partial_v\phi_\infty).
 \end{aligned}
\]
Finally, the mass equations \eqref{incoming-mass}--\eqref{outgoing-mass},
part~(b), and the identity
$\Omega_j^2=-4\nu_j\lambda_j/\rho_j$ give
\[
 \mu_j\longrightarrow\mu_\infty,\qquad
 \Omega_j\longrightarrow\Omega_\infty
       \quad\text{in }C^1(K_T).
\]
Thus the limit has precisely the weighted $C^1$
regularity in Definition~\ref{regularity-definition}.

\textit{Step 6. The integral bound.}
It remains to prove the integral inequality in \eqref{ancient-bounds}.
Fix $v\leq0$ and $N>-v$. The rescaled estimate
\eqref{nonnegative-optical-bound} gives
\[
 \int_{-N}^{v}\frac{(-\partial_ur_j)\mu_j}
                    {r_j(1-\mu_j)}(u,v)\,\mathrm du\leq C.
\]
To apply the convergence proved in Step~5(b), write the integrand as
\[
 \frac{(-\partial_ur_j)\mu_j}{r_j(1-\mu_j)}
 =\frac{(-\partial_ur_j)r_j}{1-\mu_j}\frac{\mu_j}{r_j^2}.
\]
Step~3 and Step~5(b) give uniform convergence of every factor on
$K_N$; the product has continuous value zero at the center.
Thus, for each fixed $N$,
\[
 \int_{-N}^{v}\frac{(-\partial_ur_\infty)\mu_\infty}
                    {r_\infty(1-\mu_\infty)}(u,v)\,\mathrm du
 =\lim_{j\to\infty}\int_{-N}^{v}
       \frac{(-\partial_ur_j)\mu_j}{r_j(1-\mu_j)}(u,v)\,\mathrm du
 \leq C.
\]
The limiting integrand is nonnegative. Monotone convergence as
$N\to\infty$ therefore yields
\[
 \int_{-\infty}^{v}\frac{(-\partial_ur_\infty)\mu_\infty}
                    {r_\infty(1-\mu_\infty)}(u,v)\,\mathrm du
 \leq C,
\]
which is the remaining bound in \eqref{ancient-bounds}.
\end{proof}

For simplicity, from now on we omit the subscript $\infty$ and
write $r,m,\phi,\alpha$ and the other unindexed fields for the
ancient solution just constructed.

\subsection{Past asymptotics of the ancient solution}
\label{subsec:ancient-past}

Throughout this subsection, $(r,m,\phi,\alpha)$ denotes the
ancient limit of Proposition~\ref{ancient-proposition}, in its
double null coordinates $(u,v)$. Neither the original solution
nor a finite member of the rescaled sequence is used.
We next prove the following asymptotic estimates.

\begin{lemma}[Past asymptotics]\label{past-lemma}
There are constants $c\in\mathbb C$ and $\Lambda>0$ such that, for
each fixed $v\leq0$,
\[
 \begin{gathered}
 \mu(u,v)\longrightarrow0,\qquad
 \alpha(u,v),\phi(u,v)\longrightarrow c,\\
 \partial_vr(u,v)\longrightarrow\Lambda
       \quad\hbox{as }u\longrightarrow-\infty.
 \end{gathered}
\]
Moreover, these limits hold uniformly for $\theta u\leq v\leq0$
as $u\to-\infty$, for each fixed $0<\theta<1$.
There also hold
\begin{equation}\label{past-radius-limit}
 \frac{r(u,0)}{-u}\longrightarrow\Lambda,
\end{equation}
and
\begin{equation}\label{past-alpha-derivative}
 \sup_{\theta u\leq v\leq0}|r\alpha'|(u,v)
 \leq C_\theta r(u,0)\int_{-\infty}^u
       \frac{(-\partial_ur)(w,0)\mu(w,0)}{r(w,0)^2}\,\mathrm dw .
\end{equation}
\end{lemma}

\begin{proof}
\textit{Limits on fixed incoming cones.}
Along $v=\mathrm{const}$, radius increases towards the past since
$\partial_ur<0$. The lower bound for $\partial_vr$ in
\eqref{ancient-bounds} gives
\[
 r(u,v)=\int_u^v\partial_vr(u,v')\,\mathrm dv'
       \geq C^{-1}(v-u)\longrightarrow\infty
       \qquad(u\to-\infty).
\]
The neutral incoming mass equation \eqref{incoming-mass}, with
$\mu=2m/r$, gives
\[
 \frac{\mathrm d\mu}{\mathrm d\log r}
 =4\pi(1-\mu)\left|\frac{r\partial_u\phi}{-\partial_ur}\right|^2-\mu.
\]
By \eqref{scalar-bounds} and $0\leq\mu<1$,
\[
 \left|\frac{\mathrm d\mu}{\mathrm d\log r}\right|
 \leq4\pi(1-\mu)
       \left|\frac{r\partial_u\phi}{-\partial_ur}\right|^2
       +\mu\leq C.
\]
On the other hand, \eqref{ancient-bounds} implies
\[
 \int_{r(u,v)}^\infty\frac{\mu}{1-\mu}\,\frac{\mathrm dr}{r}<\infty.
\]
On this fixed cone, set $t=\log r$ and $f(t)=\mu$. If
$f(t_n)\geq\varepsilon>0$ for some $t_n\to\infty$, pass to a
subsequence so that the intervals
$I_n=[t_n-\delta,t_n+\delta]$, $\delta=\varepsilon/(2C)$,
are pairwise disjoint and contained in $[t_0,\infty)$, where
$t_0=\log r(u,v)$. Then
\[
 \begin{gathered}
 f(t)\geq f(t_n)-C|t-t_n|\geq\frac{\varepsilon}{2}
 \quad(t\in I_n),\\
 \int_{t_0}^\infty\frac{f(t)}{1-f(t)}\,\mathrm dt
 \geq\sum_n\int_{I_n}f(t)\,\mathrm dt
 \geq\sum_n\frac{\varepsilon^2}{2C}=\infty,
 \end{gathered}
\]
contradicting the preceding integral bound. Thus $\mu\to0$;
combining this with \eqref{cubic-bound} gives
\[
 \mu(u,v)\to0,\qquad
 |\alpha-\phi|(u,v)\leq C\mu(u,v)^{1/3}\to0.
\]
Using $|\alpha-\phi|\leq C$ from \eqref{scalar-bounds}, the neutral
equations \eqref{alpha-transport} and \eqref{mixed-radius} yield
\[
 |\partial_u\alpha|
 \leq C\frac{(-\partial_ur)\mu}{r(1-\mu)},\qquad
 \partial_u\log\partial_vr=-\frac{(-\partial_ur)\mu}{r(1-\mu)}.
\]
In particular, \eqref{ancient-bounds} gives
\[
 \int_{-\infty}^u
       \bigl(|\partial_u\alpha|
             +|\partial_u\log\partial_vr|\bigr)(w,v)\,\mathrm dw
 \leq C\int_{-\infty}^u
       \frac{(-\partial_ur)\mu}{r(1-\mu)}(w,v)\,\mathrm dw<\infty.
\]
Thus $\alpha$, $\phi$ and $\partial_vr$ have past limits on each
fixed cone. To compare the limits on two fixed cones
$v_1<v_2\leq0$, use \eqref{critical-derivative-bounds},
$\partial_v\alpha=(\partial_vr)\alpha'$, and the bounds for
$\partial_vr$ and $r$ in \eqref{ancient-bounds} and above:
\begin{align*}
 &|\alpha(u,v_2)-\alpha(u,v_1)|
       +|\partial_vr(u,v_2)-\partial_vr(u,v_1)|\\
 &\qquad\leq C\int_{v_1}^{v_2}\frac{\mathrm dv}{r(u,v)}
       \leq \frac{C(v_2-v_1)}{v_1-u}\longrightarrow0 .
\end{align*}
Denote the common limits by $c$ and $\Lambda$.
The lower bound for $\partial_vr$ implies
\[
 \Lambda\geq C^{-1}>0.
\]

\textit{Uniformity on expanding regions.}
We follow the comparison argument of An--Chen--Singh
\cite[Section~4]{AnChenSingh} in self-similar regions, applied here
to the ancient solution as $u\to-\infty$.
For $w\leq u<0$ and $\theta u\leq v\leq0$, the two-sided radius
comparison from \eqref{ancient-bounds}, and outgoing monotonicity
from the neutral equations \eqref{outgoing-mass} and
\eqref{mixed-radius}, give
\begin{align*}
 \frac{r(w,0)}{r(w,v)}
 &\leq C\frac{-w}{v-w}\leq\frac{C}{1-\theta},\\
 m(w,v)&\leq m(w,0),\qquad
 (-\partial_ur)(w,v)\leq(-\partial_ur)(w,0).
\end{align*}
Using $\mu=2m/r$ and $1-\mu\geq C^{-1}$ from
\eqref{ancient-bounds}, these comparisons imply
\begin{align}
 \mu(w,v)&\leq C_\theta\mu(w,0),\notag\\
 \frac{(-\partial_ur)\mu}{r(1-\mu)}(w,v)
 &\leq C_\theta
           \frac{(-\partial_ur)\mu}{r(1-\mu)}(w,0).
                                                        \label{past-cone-comparison}
\end{align}
Integrating \eqref{alpha-transport} and
$\partial_u\log\partial_vr$ from their common past limits, using
\eqref{scalar-bounds} and \eqref{past-cone-comparison}, gives
\begin{align}
 \sup_{\theta u\leq v\leq0}
 \left(|\alpha(u,v)-c|
       +\left|\log\frac{\partial_vr(u,v)}{\Lambda}\right|\right)
 &\leq C_\theta\int_{-\infty}^u
       \frac{(-\partial_ur)\mu}{r(1-\mu)}(w,0)\,\mathrm dw
       \longrightarrow0.                              \label{past-uniform-tail}
\end{align}
The right-hand side tends to zero by \eqref{ancient-bounds}.
Also, \eqref{past-cone-comparison} and \eqref{cubic-bound} give
\[
 \begin{gathered}
 \sup_{\theta u\leq v\leq0}\mu(u,v)\leq C_\theta\mu(u,0)\to0,
 \\
 |\phi-c|\leq|\alpha-c|+C\mu^{1/3}.
 \end{gathered}
\]
This proves all the asserted uniform limits.

Since $r(u,0)=\int_u^0\partial_vr(u,v)\,\mathrm dv$, splitting
at $\theta u$ and using the uniform bound for $\partial_vr$
in \eqref{ancient-bounds} on $[u,\theta u]$ yields
\[
 \left|\frac{r(u,0)}{-u}-\Lambda\right|
 \leq C(1-\theta)
       +\theta\sup_{\theta u\leq v\leq0}
                       |\partial_vr(u,v)-\Lambda|.
\]
First let $u\to-\infty$ and then $\theta\uparrow1$ to obtain
\eqref{past-radius-limit}.

Finally, the past boundary term in \eqref{scalar-multiplier}
satisfies
\[
 |(\partial_vr)^2\alpha'|(w,v)
 \leq\frac{C}{r(w,v)}\longrightarrow0\qquad(w\to-\infty),
\]
by \eqref{critical-derivative-bounds}, \eqref{ancient-bounds}
and the radius comparison above. Hence
\begin{align*}
 (\partial_vr)^2\alpha'(u,v)
 =\int_{-\infty}^u&
 \frac{(-\partial_ur)(\partial_vr)^2}{r^2(1-\mu)}\\
 &\quad\times
 \bigl[4\pi|\alpha-\phi|^2(\alpha-\phi)
                -3\mu(\alpha-\phi)\bigr](w,v)\,\mathrm dw .
\end{align*}
The estimate needed for the numerator is
\[
 \bigl|4\pi|\alpha-\phi|^2(\alpha-\phi)-3\mu(\alpha-\phi)\bigr|
 \leq4\pi|\alpha-\phi|^3+3\mu|\alpha-\phi|\leq C\mu,
\]
using \eqref{cubic-bound} and \eqref{scalar-bounds}.
Using $1-\mu\geq C^{-1}$ and
$C^{-1}\leq\partial_vr\leq C$ from \eqref{ancient-bounds}, followed by
\eqref{past-cone-comparison} and the radius comparison above, gives
\[
 |r\alpha'|(u,v)
 \leq Cr(u,v)\int_{-\infty}^u
                  \frac{(-\partial_ur)\mu}{r^2}(w,v)\,\mathrm dw
 \leq C_\theta r(u,0)\int_{-\infty}^u
                  \frac{(-\partial_ur)\mu}{r^2}(w,0)\,\mathrm dw .
\]
Taking the supremum in $v$ proves \eqref{past-alpha-derivative}.
\end{proof}

\section{Concentration at the mass scale}
\label{sec:mass-concentration}

We use the ancient solution and its past asymptotics from
Section~\ref{sec:first}. The following estimate locates a fixed
positive fraction of the Hawking mass within a fixed multiple of
the mass scale used in the second normalization.

Choose a strictly decreasing sequence $u_n\to-\infty$ such that\footnote{%
For each sufficiently large integer $N$, the continuous function
$\mu(\cdot,0)$ attains a positive maximum on $(-\infty,-N]$:
its limit at $-\infty$ is zero by Lemma~\ref{past-lemma}, while
Proposition~\ref{ancient-proposition} and incoming mass monotonicity
from \eqref{incoming-mass} give positive mass sufficiently far in
the past. Each maximizer also maximizes $\mu$
over its own past. Since these points lie to the left of $-N$,
they tend to $-\infty$; passing to a strictly decreasing subsequence
and relabelling gives $u_n$ satisfying \eqref{past-maxima}.}
\begin{equation}\label{past-maxima}
 \mu(u_n,0)=\sup_{w\leq u_n}\mu(w,0)>0,\qquad
 \mu(u_n,0)\longrightarrow0.
\end{equation}
Set
\[
 R_n=r(u_n,0),\qquad M_n=m(u_n,0).
\]
Incoming mass monotonicity from \eqref{incoming-mass} and
\eqref{past-maxima} give, respectively,
\begin{gather}
 M_n\geq M_1>0,\notag\\
 \frac{R_n}{M_n}=\frac2{\mu(u_n,0)}
                  \longrightarrow\infty.\label{scale-separation}
\end{gather}
There is no assumption that these mass scales have an upper bound.

Let $u_{\mathrm{earlier},n}$ be determined by
\begin{equation}\label{Eqn:defuen}
    r(u_{\mathrm{earlier},n},0)=8R_n.
\end{equation}
Set
\begin{equation}\label{mass-window}
 \mathcal U_n:=\{u<0:R_n\leq r(u,0)\leq3R_n\}.
\end{equation}
We now prove the following concentration estimate.

\begin{lemma}[Concentration at the mass scale]\label{mass-retention-lemma}
There is a fixed $L\geq1$ such that, for all sufficiently large $n$
and every $u\in\mathcal U_n$, the mass at radius $LM_n$ on the
outgoing cone $\{u=\mathrm{constant}\}$ satisfies
\begin{equation}\label{retained-mass}
 \left.m(u,\cdot)\right|_{r=LM_n}
       \geq\tfrac12M_n.
\end{equation}
\end{lemma}

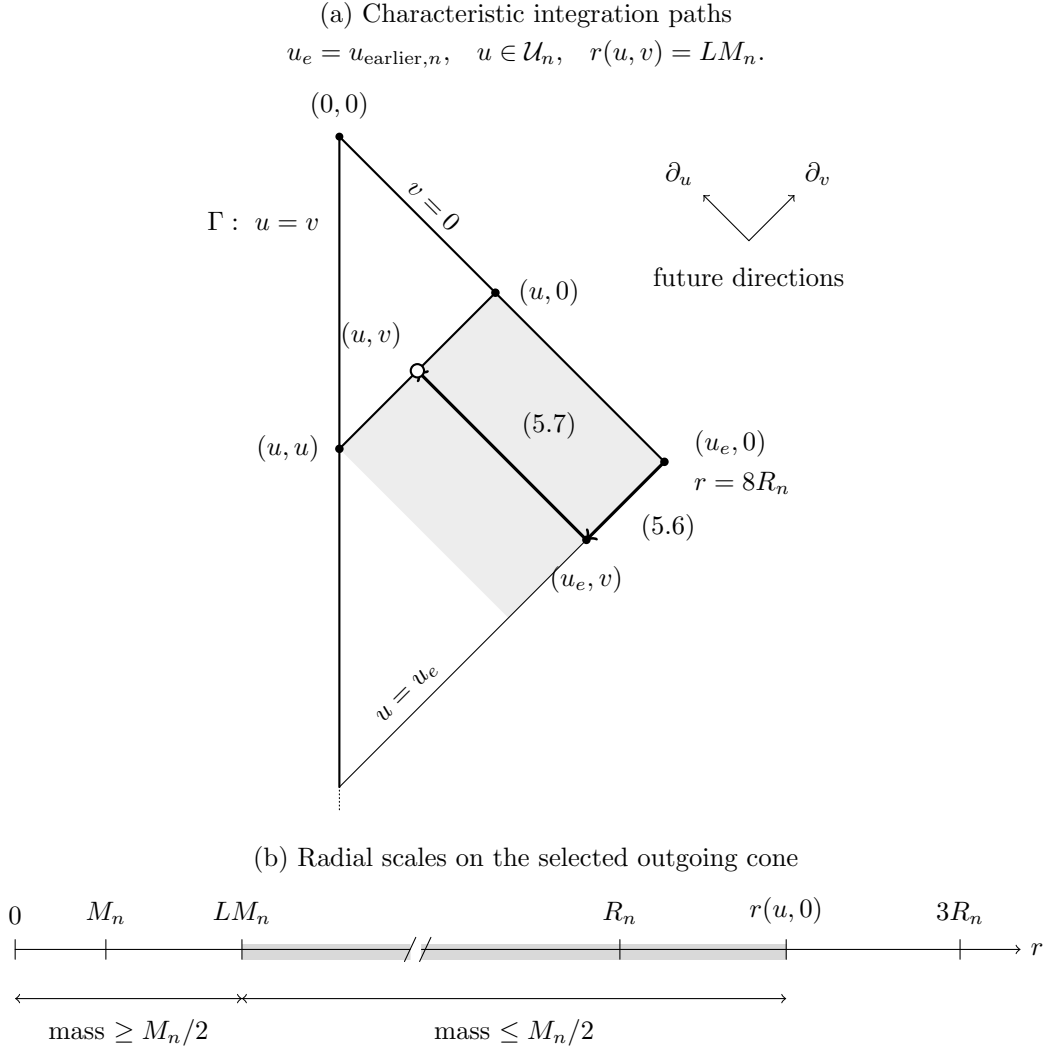
\begin{figure}[htbp]
\centering
{\small (a) Characteristic integration paths\par}
\smallskip
{\small $u_e=u_{\mathrm{earlier},n}$,\quad
 $u\in\mathcal U_n$,\quad $r(u,v)=LM_n$.\par}
\medskip
% Geometry only: x=f(v)-f(u), y=f(v)+f(u), with f strictly increasing.
% Drawing positions f(u_e)=0, f(u)=2.6, f(v)=3.8,
% f(0)=5 preserve the null directions and ordering, not coordinate lengths.
% The center is x=0. All u=constant lines have slope +1;
% all v=constant lines have slope -1. No constant-radius curve is drawn.
% Choose the representative incoming parameter v at radius LM_n on cone u.
\begin{tikzpicture}[x=0.86cm,y=0.86cm,font=\small,
                    line cap=round,line join=round]
 \fill[black!7] (2.6,2.6)--(5,5)--(2.4,7.6)--(0,5.2)--cycle;
 \draw[thick] (0,0)--(0,10);
 \draw[densely dotted] (0,-0.35)--(0,0);
 \fill (0,10) circle (1.5pt);
 \node[above] at (0,10.12) {$(0,0)$};
 \node[left] at (-0.15,8.7) {$\Gamma:\ u=v$};
 \draw[thick] (5,5)--(0,10);
 \node[rotate=-45,above] at (1.25,8.75) {$v=0$};
 \draw (0,0)--(5,5);
 \node[rotate=45,above] at (1.25,1.25) {$u=u_e$};
 \draw[thick] (0,5.2)--(2.4,7.6);
 \fill (0,5.2) circle (1.6pt);
 \node[left] at (-0.15,5.2) {$(u,u)$};
 \fill (2.4,7.6) circle (1.6pt);
 \node[right] at (2.6,7.6) {$(u,0)$};
 \fill (5,5) circle (1.6pt);
 \node[align=left,anchor=west] at (5.3,5.0)
       {$(u_e,0)$\\[3pt]$r=8R_n$};
 \draw[very thick,->] (5,5)--(3.8,3.8);
 \node[below right] at (4.48,4.35) {\eqref{earlier-alpha}};
 \fill (3.8,3.8) circle (1.6pt);
 \node[below] at (3.8,3.55) {$(u_e,v)$};
 \draw[very thick,->] (3.8,3.8)--(1.2,6.4);
 \node[anchor=south west] at (2.65,5.2) {\eqref{mass-scale-alpha}};
 \fill[white] (1.2,6.4) circle (2.5pt);
 \draw[thick] (1.2,6.4) circle (2.5pt);
 \node[above left] at (1.12,6.58) {$(u,v)$};
 \draw[->] (6.3,8.4)--(5.6,9.1) node[above left] {$\partial_u$};
 \draw[->] (6.3,8.4)--(7.0,9.1) node[above right] {$\partial_v$};
 \node[below] at (6.3,8.15) {future directions};
\end{tikzpicture}
\par\bigskip
{\small (b) Radial scales on the selected outgoing cone\par}
\medskip
\begin{tikzpicture}[x=1cm,y=1cm,font=\small,line cap=round]
 \fill[black!15] (3,-0.14) rectangle (10.2,0.07);
 \draw[->] (0,0)--(13.3,0) node[right] {$r$};
 \foreach \x in {0,1.2,3,8,10.2,12.5}
       {\draw (\x,-0.13)--(\x,0.13);}
 \node[above] at (0,0.2) {$0$};
 \node[above] at (1.2,0.2) {$M_n$};
 \node[above] at (3,0.2) {$LM_n$};
 \node[above] at (8,0.2) {$R_n$};
 \node[above] at (10.2,0.2) {$r(u,0)$};
 \node[above] at (12.5,0.2) {$3R_n$};
 \draw[white,line width=4pt] (5.3,-0.2)--(5.3,0.2);
 \draw (5.15,-0.16)--(5.3,0.16);
 \draw (5.35,-0.16)--(5.5,0.16);
 \draw[<->] (0,-0.65)--(3,-0.65);
 \node[below] at (1.5,-0.78) {mass $\geq M_n/2$};
 \draw[<->] (3,-0.65)--(10.2,-0.65);
 \node[below] at (6.6,-0.78) {mass $\leq M_n/2$};
\end{tikzpicture}
\caption{Geometry of Lemma~\ref{mass-retention-lemma}, schematically.
In (a), the shaded characteristic rectangle is
$\{u_e\leq w\leq u,\ u\leq v'\leq0\}$.
The thick arrows indicate the integrations underlying
\eqref{earlier-alpha} and \eqref{mass-scale-alpha}; the open circle is
chosen at radius $LM_n$ on the cone $u$.
Panel (b) shows the retained mass and the outer mass bound after
fixing $L$ and taking $n$ sufficiently large.
Neither panel is to scale; in particular, $LM_n/R_n\to0$.}
\label{mass-concentration-figure}
\end{figure}

Figure~\ref{mass-concentration-figure} shows the characteristic
segments used in the proof and the radial interval on which the
mass difference is estimated.

\begin{proof}
For all large $n$ and $u\in\mathcal U_n$, \eqref{past-radius-limit} gives
\[
 u\geq\tfrac12u_{\mathrm{earlier},n}.
\]
For $u\leq v\leq0$, the earlier intersection therefore satisfies
\[
 cR_n\leq C^{-1}(v-u_{\mathrm{earlier},n})
       \leq r(u_{\mathrm{earlier},n},v)
       \leq r(u_{\mathrm{earlier},n},0)=8R_n.
\]
Here we used \eqref{ancient-bounds} and
$-u_{\mathrm{earlier},n}\asymp R_n$ from
\eqref{past-radius-limit}. Applying
\eqref{past-alpha-derivative} with $\theta=1/2$, followed by
\eqref{past-maxima}, gives
\begin{align*}
 \sup_{u_{\mathrm{earlier},n}/2\leq v\leq0}
       |r\alpha'|(u_{\mathrm{earlier},n},v)
 &\leq C(8R_n)\mu(u_n,0)
       \int_{-\infty}^{u_{\mathrm{earlier},n}}
              \frac{-\partial_ur(w,0)}{r(w,0)^2}\,\mathrm dw\\
 &=C(8R_n)\mu(u_n,0)\frac1{8R_n}
       \leq C\mu(u_n,0).
\end{align*}
Consequently, integration on the earlier outgoing cone yields
\begin{equation}\label{earlier-alpha}
 \begin{aligned}
 |\alpha(u_{\mathrm{earlier},n},v)
        -\alpha(u_{\mathrm{earlier},n},0)|
 &\leq C\mu(u_n,0)
       \int_{r(u_{\mathrm{earlier},n},v)}^{8R_n}
                    \frac{\mathrm ds}{s}\\
 &\leq C\mu(u_n,0)\log(8/c)\leq C\mu(u_n,0).
 \end{aligned}
\end{equation}

The neutral mass equations \eqref{incoming-mass} and
\eqref{outgoing-mass} give, on the segments from
that earlier cone to the outgoing cones with $u\in\mathcal U_n$,
\[
 0\leq m\leq8M_n,\qquad m(u,0)\geq M_n
       \quad\hbox{on the outer cone }v=0.
\]
Indeed, the upper bound follows from
\[
 m(w,v)\leq m(u_{\mathrm{earlier},n},v)
       \leq m(u_{\mathrm{earlier},n},0)
       =4R_n\mu(u_{\mathrm{earlier},n},0)
       \leq4R_n\mu(u_n,0)=8M_n,
\]
for $u_{\mathrm{earlier},n}\leq w\leq u$, while
$u\leq u_n$ gives $m(u,0)\geq M_n$.
The transport equation \eqref{alpha-transport} now gives
\begin{align*}
 |\alpha(u,v)-\alpha(u_{\mathrm{earlier},n},v)|
 &\leq\int_{u_{\mathrm{earlier},n}}^u
       \frac{(-\partial_ur)\,2m\,|\alpha-\phi|}
            {r^2(1-\mu)}(w,v)\,\mathrm dw\\
 &\leq CM_n\int_{r(u,v)}^{r(u_{\mathrm{earlier},n},v)}
                   \frac{\mathrm ds}{s^2}
       \leq\frac{CM_n}{r(u,v)}.
\end{align*}
The second line uses $m\leq8M_n$,
$|\alpha-\phi|\leq C$ from \eqref{scalar-bounds}, and
$1-\mu\geq C^{-1}$ from \eqref{ancient-bounds}.
Combining with \eqref{earlier-alpha}, and writing $r=r(u,v)$,
we obtain
\begin{equation}\label{mass-scale-alpha}
 |\alpha(u,r)-\alpha(u_{\mathrm{earlier},n},0)|
       \leq C\mu(u_n,0)+\frac{CM_n}r.
\end{equation}
For $M_n\leq r\leq r(u,0)$, the scalar average gives
\begin{align*}
 &r\bigl[\phi(u,r)-\alpha(u_{\mathrm{earlier},n},0)\bigr]\\
 &\quad=M_n\bigl[\phi(u,M_n)-\alpha(u_{\mathrm{earlier},n},0)\bigr]
       +\int_{M_n}^r
             [\alpha(u,s)-\alpha(u_{\mathrm{earlier},n},0)]
             \,\mathrm ds,
\end{align*}
and hence
\begin{align*}
 |\phi(u,r)-\alpha(u_{\mathrm{earlier},n},0)|
 &\leq\frac{CM_n}{r}
       +\frac Cr\int_{M_n}^r
              \left(\mu(u_n,0)+\frac{M_n}{s}\right)\,\mathrm ds\\
 &\leq C\mu(u_n,0)+\frac{CM_n}{r}
                 \left(1+\log\frac r{M_n}\right).
\end{align*}
Here the starting value is bounded by \eqref{scalar-bounds},
and the integral by \eqref{mass-scale-alpha}. Using also
$|\alpha-\phi|\leq|\alpha-\alpha(u_{\mathrm{earlier},n},0)|
                  +|\phi-\alpha(u_{\mathrm{earlier},n},0)|$ gives
\begin{align}
 &|\phi(u,r)-\alpha(u_{\mathrm{earlier},n},0)|
       +|\alpha(u,r)-\phi(u,r)|\notag\\
 &\qquad\leq C\mu(u_n,0)+\frac{CM_n}r
                \left(1+\log\frac r{M_n}\right)
       \qquad(r\geq M_n).                       \label{mass-scale-tail}
\end{align}
Fix $L\geq1$. Since $R_n/M_n\to\infty$, for all sufficiently
large $n$ and every $u\in\mathcal U_n$ we have
\[
 M_n\leq LM_n\leq R_n\leq r(u,0)\leq3R_n.
\]
Here $r(u,0)$ and $m(u,0)$ denote the values on the outer incoming
cone $v=0$ in the original double null coordinates; the integrands
below are written in Bondi coordinates. The neutral form of
\eqref{radial-mass} is
\[
 \partial_r m=2\pi(1-\mu)|\alpha-\phi|^2.
\]
Integrating along the fixed outgoing cone, using $0\leq1-\mu\leq1$
and squaring \eqref{mass-scale-tail} with
$(a+b)^2\leq2a^2+2b^2$, gives
\begin{align*}
 m(u,0)-\left.m(u,\cdot)\right|_{r=LM_n}
 &=2\pi\int_{LM_n}^{r(u,0)}
       (1-\mu(u,t))|\alpha(u,t)-\phi(u,t)|^2\,\mathrm dt\\
 &\leq C\mu(u_n,0)^2\bigl(r(u,0)-LM_n\bigr)
       +CM_n^2\int_{LM_n}^{r(u,0)}
          \frac{(1+\log(t/M_n))^2}{t^2}\,\mathrm dt.
\end{align*}
For the first term, $r(u,0)\leq3R_n$ and
$\mu(u_n,0)R_n=2M_n$ imply
\[
 \mu(u_n,0)^2\bigl(r(u,0)-LM_n\bigr)
 \leq3\mu(u_n,0)^2R_n=6M_n\mu(u_n,0).
\]
For the second term, set $t=M_ns$ and integrate by parts to obtain
\[
 M_n^2\int_{LM_n}^{r(u,0)}
       \frac{(1+\log(t/M_n))^2}{t^2}\,\mathrm dt
 \leq M_n\int_L^\infty\frac{(1+\log s)^2}{s^2}\,\mathrm ds
 =M_n\frac{(\log L)^2+4\log L+5}{L}.
\]
% Expanded change of variable and integrations by parts:
% For the second, the change of variable $t=M_ns$ yields
% \begin{align*}
%  M_n^2\int_{LM_n}^{r(u,0)}
%        \frac{(1+\log(t/M_n))^2}{t^2}\,\mathrm dt
%  &=M_n\int_L^{r(u,0)/M_n}\frac{(1+\log s)^2}{s^2}\,\mathrm ds\\
%  &\leq M_n\int_L^\infty\frac{(1+\log s)^2}{s^2}\,\mathrm ds.
% \end{align*}
% Two integrations by parts give
% \begin{align*}
%  \int_L^\infty\frac{(1+\log s)^2}{s^2}\,\mathrm ds
%  &=\frac{(1+\log L)^2}{L}
%        +2\int_L^\infty\frac{1+\log s}{s^2}\,\mathrm ds\\
%  &=\frac{(\log L)^2+4\log L+5}{L}.
% \end{align*}
Combining these estimates, with $C$ independent of $u,n$ and $L$,
we obtain
\begin{equation}\label{retention-tail}
 m(u,0)-\left.m(u,\cdot)\right|_{r=LM_n}
 \leq CM_n\left[\mu(u_n,0)
       +\frac{(\log L)^2+4\log L+5}{L}\right].
\end{equation}
Choose $L$ so large that
\[
 C\frac{(\log L)^2+4\log L+5}{L}\leq\frac14.
\]
Then choose $n$ large enough that
\[
 C\mu(u_n,0)\leq\frac14,\qquad LM_n\leq R_n.
\]
Since $m(u,0)\geq M_n$,
\eqref{retention-tail} gives, uniformly for all $u\in\mathcal U_n$,
\[
 \left.m(u,\cdot)\right|_{r=LM_n}
 \geq m(u,0)-\tfrac12M_n\geq\tfrac12M_n.
\]
\end{proof}

\section{The second normalization and the eternal limit}
\label{sec:second}\label{mass-limit-section}

In this section, we employ the second normalization scheme for the ancient solution obtained in Section~\ref{sec:first}, and construct the associated eternal limit solution in $\mathbb{R}\times (0, \infty)$ for Bondi coordinates $(u, r)$. Notice that each normalized solution retains a positive mass lower bound at a fixed radius scale $r=L$, following Lemma~\ref{mass-retention-lemma}.
%We use the concentration estimate of Section~\ref{sec:mass-concentration} to normalize by the selected Hawking mass and construct an eternal limit with a positive mass lower bound.

\subsection{Normalization and vanishing boundary flux}\label{Subsec:second}

Recall that in Section~\ref{sec:mass-concentration}, we introduce 
\begin{equation*}
    R_n=r(u_n, 0), \qquad M_n=m(u_n, 0)>0
\end{equation*}
with $u_n\to -\infty, M_n/R_n\to 0$ as $n\to\infty$. Also set $ c_n=\alpha(u_{\mathrm{earlier},n},0)$, where $u_{\mathrm{earlier},n}<u_n$ is defined in \eqref{Eqn:defuen}.
%These three reference values belong to the ancient solution; they are constants, not fields in the new coordinates.
Note that by Lemma~\ref{past-lemma}, we have $\lim_{n\to\infty} c_n=c$.
\vspace{2mm}

We first normalize the ancient solutions relative to $M_n$. Introduce the new double null coordinates $(\widetilde u, \widetilde v)$ via
\begin{equation}\label{second-normalization}
 \widetilde u=-\frac{2r(u,0)}{M_n},\qquad
 \widetilde v=\frac{v}{M_n}. 
\end{equation}
Here $\mathrm d\widetilde u/\mathrm d u
=2(-\partial_{u}r)(u,0)/M_n>0$. The null lapse $\widetilde \Omega$ now becomes
\begin{equation*}
    \widetilde\Omega^2
       =\frac{\Omega^2(u,v)}{2(-\partial_{u}r)(u,0)}.
\end{equation*}
Then at each point in new double null coordinates $(\widetilde u, \widetilde v)$, we define the rescaled solutions as below
\begin{equation}\label{Eqn:secondnormalize}
     \widetilde r=\frac{r(u,v)}{M_n},\qquad
 \widetilde m=\frac{m(u,v)}{M_n},\qquad
 \widetilde\phi=\phi(u,v),\qquad
 \widetilde\alpha=\alpha(u,v),
\end{equation}
%All fields on the left are functions of $(\widetilde u,\widetilde v)$, and all fields on the right belong to the ancient solution. 
These rescaled quantities preserve the neutral equations. For sake of simplicity, in the sequel we write $(u,v)$ for
$(\widetilde u,\widetilde v)$ and use the subscript $n$ for the rescaled quantities. %The old coordinates will not occur in the following estimates; their reference values enter only through $R_n,M_n,c_n$.
\vspace{2mm}

Thanks to \eqref{Eqn:secondnormalize}, along the outer incoming null cone $\{v=0\}$, it hold
\begin{equation}\label{Eqn:rnoutbdy}
     \left.r_n\right|_{v=0}(u)=-u/2,\qquad
 \left.(-\partial_ur_n)\right|_{v=0}=\tfrac12.
\end{equation}
And we notice that the scale-invariant quantities keep their bounds as established in \eqref{outgoing-speed-bounds}, \eqref{scalar-bounds} and \eqref{critical-derivative-bounds}, which read as below
\begin{equation}\label{Eqn:scaleinvariantbound1}
    \begin{aligned}
        C^{-1}\le\partial_v r_n\le C, \qquad\quad |\alpha_n|+|\phi_n|+|\alpha_n-\phi_n|+\left|\frac{r_n\partial_u\phi_n}{-\partial_u r_n}\right|+
    |r_n \alpha_n'|+\left|\frac{r_n\partial_v^2 r_n}{(\partial_v r_n)^2}\right|\le C.
    \end{aligned}
\end{equation}

Choose the normalized target interval for $u$ as
\[
 \mathcal I_n=
 \left[-\frac{6R_n}{M_n},-\frac{2R_n}{M_n}\right],
 \qquad\quad |\mathcal I_n|=\frac{4R_n}{M_n}\longrightarrow\infty.
\]
Since $C^{-1}\le \partial_v r_n\le C$, we next parametrize each normalized outgoing null cone by its own area radius $r_n$, thereby obtaining the Bondi coordinates $(u,r)$ for the $n$-th normalized solution.

For convenience, we use $\left.F_n\right|_{v=0}(u)$ to denote the restriction of the quantity $F_n(u,r)$ along the outer incoming null cone $v=0$. We denote the partial derivatives with respect to the Bondi coordinates $(u,r)$ by $\left.\frac{\partial}{\partial u}\right|_r$ and $\left.\frac{\partial}{\partial r}\right|_u$, while retaining the conventions $\partial_u$ and $\partial_v$ for derivatives with respect to the double null coordinates $(u,v)$.

Based on Lemma~\ref{mass-retention-lemma} and its proof, we derive 
\begin{align}
 &0\leq m_n\leq8,\quad 1-\mu_n\geq C^{-1},\quad
 m_n(u,L)\geq\tfrac12,\label{normalized-mass}\\
 &|\alpha_n(u,r)-c_n|
       \leq C M_n/R_n+C/r,\label{normalized-alpha}\\
 &|\phi_n(u,r)-c_n|
        +|\alpha_n(u,r)-\phi_n(u,r)|
       \leq C M_n/R_n+\frac{C(1+\log r)}r
       \quad(r\geq1)\label{normalized-tail}
\end{align}
throughout the spacetime domain $(u, r)\in\{u\in \mathcal I_n, 0< r\le \left.r_n\right|_{v=0}(u) \}$. Here $L\ge 1$ is the fixed constant
from Lemma~\ref{mass-retention-lemma}. Note that the outer radius $\left.r_n\right|_{v=0}(u)$ satisfies
\begin{equation}\label{Eqn:estoutradius}
    \left.r_n\right|_{v=0}(u)\in [R_n/M_n, 3R_n/M_n].
\end{equation}

Integrating \eqref{mixed-radius} in $v$ from the outer boundary $v=0$ now gives
\begin{equation}\label{normalized-speed}
 (-\partial_ur_n)(u,r)=\frac12
 \exp\left[-\int_r^{\,\left.r_n\right|_{v=0}(u)}
             \frac{\mu_n(u,s)}{1-\mu_n(u,s)}\frac{\mathrm ds}{s}\right],
\end{equation}
where we use \eqref{Eqn:rnoutbdy}. In particular, observing that $\mu_n \le Cr^{-1}, 1-\mu_n\ge C^{-1}$ from \eqref{normalized-mass}, we have
\begin{equation}\label{Eqn:prurbound}
   \frac12e^{-C/r} \le(-\partial_ur_n)(u,r)\le \frac12 \qquad\quad \text{for all } r>0. \\[2mm] 
\end{equation}

We proceed to extract an expanding sub-interval of $\mathcal I_n$, within which the boundary flux of $(r_n\partial_v\phi_n/\partial_v r_n)\big|_{v=0}(u)$ tends to $0$ as $n\to\infty$.

The incoming mass equation \eqref{incoming-mass} on $v=0$ reads
\[
 \frac{\mathrm d}{\mathrm du}\left.m_n\right|_{v=0}(u)
 =-\pi\left.(1-\mu_n)
       \left|\frac{r_n\partial_u\phi_n}{-\partial_ur_n}\right|^2
          \right|_{v=0}(u).
\]
This together with \eqref{normalized-mass} yields the bound
\begin{equation*}
    \int_{\mathcal I_n}
       \left.\left|
       \frac{r_n\partial_u\phi_n}{-\partial_ur_n}
       \right|^2\right|_{v=0}(u)\,\mathrm du\leq C.
\end{equation*}
Put
$J_n=|\mathcal I_n|^{1/2}$. For sufficiently large $n$, owing to the fact $|\mathcal I_n|=4R_n/M_n\to\infty$ as $n\to\infty$, it holds
\begin{equation*}
    [s-J_n, s+J_n]\subseteq \mathcal I_n \qquad\quad \text{for all } s\in \left[-\frac{5R_n}{M_n}, -\frac{3R_n}{M_n}\right].
\end{equation*}
It thus follows that
\begin{equation*}
    \int_{-\frac{5R_n}{M_n}}^{-\frac{3R_n}{M_n}}\int_{s-J_n}^{s+J_n}
       \left.\left|\frac{r_n\partial_u\phi_n}{-\partial_ur_n}\right|^2
                   \right|_{v=0}(u)\,\mathrm du \dd s
       =2J_n\int_{-\frac{5R_n}{M_n}-J_n}^{-\frac{3R_n}{M_n}+J_n} \left.\left|\frac{r_n\partial_u\phi_n}{-\partial_ur_n}\right|^2
                   \right|_{v=0}(u) \dd u \le CJ_n.
\end{equation*}
Employing the Integral Mean Value Theorem, we can find $s_n\in  \left[-5R_n/M_n, -3R_n/M_n\right]$ such that
\begin{equation}\label{vanishing-boundary-flux}
 \int_{s_n-J_n}^{s_n+J_n}
       \left.\left|\frac{r_n\partial_u\phi_n}{-\partial_ur_n}\right|^2
                   \right|_{v=0}(u)\,\mathrm du
       \leq\frac{CJ_n}{|\mathcal I_n|}\longrightarrow0. \\[2mm]
\end{equation}

For each $n$, translate the normalized retarded time $u$ by $u\mapsto u+s_n$. %, while leaving $v$ unchanged. For every rescaled quantity, this means
%\[
 %F_n^{\mathrm{new}}(u,v)=F_n^{\mathrm{old}}(u+s_n,v).
%\]
%We henceforth write $u,F_n$ for the translated coordinate and fields, and do not use the superscripts again. In the Bondi chart the same translation reads
%$F_n^{\mathrm{new}}(u,r)=F_n^{\mathrm{old}}(u+s_n,r)$;
%the radius is not translated or rescaled a second time here.
Then the outer radius $\left.r_n\right|_{v=0}(u)$ obeys
\begin{equation*}
     \left.r_n\right|_{v=0}(u)=-\frac{u+s_n}{2},\qquad
 \left.(-\partial_ur_n)\right|_{v=0}=\tfrac12.
\end{equation*}
 and
\eqref{vanishing-boundary-flux} becomes
\begin{equation}\label{vanishing-boundary-flux1}
 \int_{-J_n}^{J_n}
       \left.\left|\frac{r_n\partial_u\phi_n}{-\partial_ur_n}\right|^2
                   \right|_{v=0}(u)\,\mathrm du
       \leq\frac{CJ_n}{|\mathcal I_n|}\longrightarrow0.
\end{equation}
The translated Bondi domains are
\[
 \mathfrak R_n \coloneqq\left\{(u,r): |u|\leq J_n,\ 
                     0<r<-(u+s_n)/2\right\}.
\]
Note that from \eqref{Eqn:estoutradius} we have  $\left.r_n\right|_{v=0}(u)\ge R_n/M_n \to\infty$.
Consequently, for every fixed $T<\infty$ and
$0<\varepsilon<R<\infty$, the rectangle
$[-T,T]\times[\varepsilon,R]$ is contained in $\mathfrak R_n$  whenever $n$ is large enough. %All subsequent limits are taken on such fixed rectangles, not at fixed values of the old null coordinates.

\subsection{Strong convergence at positive radius}
%We now work with the $u$-translated normalized solutions in Bondi coordinates $(u,r)\in \mathfrak R_n$ and aim to establish the strong convergence of an appropriate subsequence of these solutions.

The purpose of this section is to pass the limit (possibly in subsequence) for the normalized solutions defined in \eqref{Subsec:second}, and to verify that this eternal limit is the classical solution to the neutral equations in Bondi coordinates $(u, r)\in \mathbb{R}\times (0,\infty)$. Specifically, we aim to prove

\begin{proposition}[An eternal limit with a positive mass lower bound]\label{eternal-proposition}
A subsequence of the normalized solutions has a classical neutral
complex scalar limit on $\mathbb R\times(0,\infty)$ in Bondi
coordinates $(u,r)$. It admits local double null coordinates and
satisfies
\begin{align}
 &0\leq m\leq8,\qquad 1-\mu\geq C^{-1},\qquad
       m(u,L)\geq\tfrac12,\label{eternal-mass}\\
 &|\alpha|+|\phi|+|\alpha-\phi|
       +\left|\frac{r\partial_u\phi}{-\partial_ur}\right|
       +|r\alpha'|\leq C,\label{eternal-scalar}\\
 &|\alpha(u,r)-c|\leq C/r,\qquad
 |\phi(u,r)-c|+|\alpha(u,r)-\phi(u,r)|
       \leq\frac{C(1+\log r)}r\quad(r\geq1),\label{eternal-tail}\\
 &(-\partial_ur)(u,r)=\tfrac12
       \exp\left[-\int_r^\infty
                \frac{\mu(u,s)}{1-\mu(u,s)}\frac{\mathrm ds}{s}\right],
        \label{eternal-speed}\\
 &\frac{r\partial_u\phi}{-\partial_ur}(u,r)
       =-\frac1{(-\partial_ur)(u,r)}
          \int_r^\infty\frac{(-\partial_ur)(u,s)}s
                          (\alpha-\phi)(u,s)\,\mathrm ds.
                \label{eternal-incoming}
\end{align}
Here $c$ is the constant in Lemma~\ref{past-lemma}.
\end{proposition}
\vspace{2mm}

We still use $\partial_u, \partial_v$ to represent the partial derivatives in the double null coordinates $(u, v)$. By chain rule, the partial derivatives relative to the Bondi coordinates $(u, r)$ can be expressed as
\begin{equation}\label{Eqn:derivative-relation}
     \left.\frac{\partial}{\partial u}\right|_r
 =\partial_u+\frac{-\partial_ur_n}{\partial_vr_n}\partial_v,
 \qquad
 \left.\frac{\partial}{\partial r}\right|_u
 =\frac1{\partial_vr_n}\partial_v.
\end{equation}
Therefore, the corresponding equations for $(m_n, \alpha_n, \phi_n)$ in $(u, r)$ coordinates take the following forms:
\begin{align}
 \left.\frac{\partial m}{\partial r}\right|_u
 &=2\pi(1-\mu)|\alpha-\phi|^2,
       \label{eternal-radial-mass}\\
 \left.\frac{\partial m}{\partial u}\right|_r
 &=2\pi(-\partial_ur)(1-\mu)
       \left[|\alpha-\phi|^2
              -\left|\frac{r\partial_u\phi}{-\partial_ur}\right|^2
       \right],
       \label{eternal-time-mass}\\
 \left.\frac{\partial\alpha}{\partial r}\right|_u
 &=\alpha',\qquad
 \left.\frac{\partial\alpha}{\partial u}\right|_r
       =(-\partial_ur)\alpha'
          +\frac{-\partial_ur}{r}
             \frac{\mu}{1-\mu}(\alpha-\phi),
       \label{eternal-alpha}\\
 \left.\frac{\partial\phi}{\partial r}\right|_u
 &=\frac{\alpha-\phi}{r},\qquad
 \left.\frac{\partial\phi}{\partial u}\right|_r
       =\frac{-\partial_ur}{r}
          \left(\frac{r\partial_u\phi}{-\partial_ur}
                           +\alpha-\phi\right).
       \label{eternal-phi}
\end{align}
And the evolution equations for $\partial_v r_n$ read
\begin{equation}\label{outgoing-speed-chart}
    \begin{aligned}
 \left.\frac{\partial(\partial_vr)}{\partial r}\right|_u
 &=\partial_vr\,
       \frac{\partial_v^2r}{(\partial_vr)^2},
       \\
 \left.\frac{\partial(\partial_vr)}{\partial u}\right|_r
 &=(-\partial_ur)\partial_vr
       \left[
          \frac{\partial_v^2r}{(\partial_vr)^2}
          -\frac{\mu}{r(1-\mu)}
       \right].
\end{aligned}
\end{equation}
Here for simplicity, we omit the subscript ${}_n$ in the equations above.
\vspace{2mm}

We begin by showing that this sequence admits a subsequence
that converges locally uniformly.

\begin{lemma}
\label{eternal-local-compactness}
After passage to a subsequence, the normalized solutions
\[
 m_n,\quad\alpha_n,\quad\phi_n,\quad
 -\partial_ur_n,\quad\partial_vr_n
\]
converge locally uniformly on $\mathbb R\times(0,\infty)$. %Moreover,  all these limits are locally Lipschitz on $\mathbb R\times(0,\infty)$.
\end{lemma}

\begin{proof}
Fix $T<\infty$ and $0<\varepsilon<R<\infty$.
The equations
\eqref{eternal-radial-mass}--\eqref{outgoing-speed-chart}, together with the estimates \eqref{Eqn:scaleinvariantbound1}, \eqref{normalized-mass} and \eqref{Eqn:prurbound} render uniform bounds for the the renormalized quantities $m_n, \alpha_n, \phi_n, \partial_v r_n$ and
their first order derivatives on
$[-T,T]\times[\varepsilon,R]$.
In particular, there holds
\begin{equation}\label{Eqn:uniformbound-pru-mn}
     \left|\left.\frac{\partial m_n}{\partial u}\right|_r\right|
 \leq C.
\end{equation}
Therefore, employing the Arzel\`a--Ascoli theorem, along with a diagonal subsequence argument, we then find a locally uniform convergent subsequence of $(m_n, \alpha_n, \phi_n, \partial_v r_n)$, whose limits exhaust the whole domain $\mathbb{R}\times (0, \infty)$ and are local Lipschitz. 

For simplicity, still use the subscript ${}_n$ to denote this subsequence. We turn to establish the local uniform convergence for $-\partial_ur_n$. Recall \eqref{normalized-speed}:
\begin{equation}\label{normalized-speed1}
 (-\partial_ur_n)(u,r)=\tfrac12
 \exp\left[-\int_r^{\,\left.r_n\right|_{v=0}(u)}
             \frac{\mu_n(u,s)}{1-\mu_n(u,s)}\frac{\mathrm ds}{s}\right],
\end{equation}
By virtue of the bounds \eqref{Eqn:scaleinvariantbound1}, \eqref{normalized-mass}, the integrand in \eqref{normalized-speed1} is at most $Cr^{-2}$. Choose $n$ large enough so that $\left.r_n\right|_{v=0}(u)\ge R_n/M_n>R$ in view of \eqref{Eqn:estoutradius}. This then gives
\begin{equation}\label{Eqn:partialurnint}
    \begin{aligned}
    \int_{r}^{\,\left.r_n\right|_{v=0}(u)} \frac{\mu_n(u,s)}{1-\mu_n(u,s)}\frac{\mathrm ds}{s}=&\int_{r}^{R} \frac{2m_n(u, s)}{s-m_n(u, s)}\frac{\mathrm ds}{s}+O(1)\int_{R}^{\,\left.r_n\right|_{v=0}(u)} \frac{\mathrm ds}{s^2} \\
    =&\int_{r}^{R} \frac{2m_n(u, s)}{s-m_n(u, s)}\frac{\mathrm ds}{s}+ O(R^{-1}).
\end{aligned}
\end{equation}
First let $n\to\infty$ and then let $R\to\infty$, we derive that
\begin{equation}\label{Eqn:limit-partialurn}
    (-\partial_ur_n)(u,r)\to\tfrac12
 \exp\left[-\int_r^{\infty}
             \frac{\mu_n(u,s)}{1-\mu_n(u,s)}\frac{\mathrm ds}{s}\right] \qquad\quad \text{locally uniform in } \mathbb{R}\times (0, \infty).
\end{equation}
%In view of all above estimates, along with $R_n/M_n\xrightarrow{n\to\infty} \infty$, we infer that $m_n, \alpha_n, \phi_n, -\partial_u r_n, \partial_v r_n$ are uniformly bounded and equicontinuous on each fixed rectangle $[-T, T]\times [\varepsilon, R]$. 
\end{proof}

From now on we work with this subsequence and denote the limiting
fields by
\[
 m,\quad\alpha,\quad\phi,\quad-\partial_ur,\quad\partial_vr.
\]
The symbols $\partial_ur$ and $\partial_vr$ here only denote
limiting functions which depends on $(u, r)$. The locally uniform convergence, together with \eqref{Eqn:scaleinvariantbound1}, \eqref{normalized-mass}--\eqref{normalized-tail} as well as $c_n\to c, \, M_n/R_n\to 0$ thus yields \eqref{eternal-mass}, \eqref{eternal-tail} and
\begin{equation*}
    |\alpha|+|\phi|+|\alpha-\phi|
       \leq C.
\end{equation*}
And from \eqref{Eqn:limit-partialurn} we also have that \eqref{eternal-speed} holds true.

%Their identification as derivatives of the radius in a local double null chart is part of the final proof.
%In particular, we will first write the limiting time equations directly in terms of the radial integral appearing in \eqref{eternal-incoming}.

We turn to recover $C^1$-regularity of the limit quantities. Before doing so, we state the following useful convergence lemma, which is a direct consequence of the fundamental theorem of calculus.
\begin{lemma}\label{Lem:calculus}
    Consider a sequence of  functions $f_n\in C^1((a, b))$. Suppose that $f_n\to f$ pointwise in $(a, b)$ and $f'_n\to g$ in $L^1_{\mathrm{loc}}((a, b))$, as $n\to\infty$.  Then if $g\in C^0((a, b))$, we have $f\in C^1((a, b))$ and $f'=g$.
\end{lemma}
\begin{proof}
   It is sufficient to show that $f(x)-f(y)=\int_{x}^y g(t)\dd t$ for any $a<x<y<b$, which directly follows from 
    \begin{align*}
        \left|f(x)-f(y)-\int_{x}^y g(t)\dd t\right|\le &|f(x)-f_n(x)|+|f_n(y)-f(y)|\\&\left|\int_{x}^y (f'_n(t)-g(t))\dd t \right|\to 0 \qquad\quad \text{as } n\to\infty.
    \end{align*}
    This completes the proof.
\end{proof}

Next, we show that $m,\phi,-\partial_ur\in C^1$, based on the vanishing boundary flux condition \eqref{vanishing-boundary-flux1}.
\begin{lemma}
\label{eternal-strong-flux}
For any $T>0$ and $0<\varepsilon<R<\infty$, the normalized incoming scalar derivatives
\[
 \frac{r\partial_u\phi_n}{-\partial_ur_n}
\]
converge strongly in $L^2_u([-T, T])$,
uniformly for $r\in[\varepsilon, R]$, to the continuous
right-hand side of \eqref{eternal-incoming}.
%Their squared moduli converge in time $L^1$, with the same uniformity in radius.

Moreover, $m,\phi,-\partial_ur\in C^1$.
The limiting quantities satisfy
\eqref{eternal-radial-mass}, the radial equation in
\eqref{eternal-phi}, and
\begin{align}
 \left.\frac{\partial\phi}{\partial u}\right|_r
 ={}&\frac{-\partial_ur}r(\alpha-\phi)
      +\frac1r F,
       \label{eternal-bondi-scalar-integral}\\
 \left.\frac{\partial m}{\partial u}\right|_r(u,r)
 ={}&2\pi(-\partial_ur)(1-\mu)
                         |\alpha-\phi|^2
       -\frac{2\pi(1-\mu)}{(-\partial_ur)}
       \left|F\right|^2,
       \label{eternal-bondi-mass-integral}
\end{align}
where 
\begin{equation*}
    F(u, r)\coloneqq -\int_r^\infty
       \frac{(-\partial_ur)(u,s)}s
       (\alpha-\phi)(u,s)\,\mathrm ds.
\end{equation*}
\end{lemma}

\begin{proof}
%\textit{The radial representation and its tail.}
The neutral incoming scalar transport equation \eqref{incoming-scalar-transport} gives, for each $n$,
\[
 \left.\frac{\partial}{\partial r}\right|_u
       (r\partial_u\phi_n)
 =\frac{-\partial_ur_n}{r}(\alpha_n-\phi_n).
\]
Integrating it over $[r, r_n\big|_{v=0}(u)]$ and noting that $\partial_u r_n\big|_{v=0}(u)=-1/2$ from \eqref{Eqn:rnoutbdy}, we obtain
\begin{equation}\label{backwards-scalar}
    \begin{aligned}
 \frac{r\partial_u\phi_n}{-\partial_ur_n}(u,r)
 ={}&
 \frac1{2(-\partial_ur_n)(u,r)}
       \left.
       \frac{r_n\partial_u\phi_n}{-\partial_ur_n}
       \right|_{v=0}(u)
       \\
 &-\frac1{(-\partial_ur_n)(u,r)}
       \int_r^{\,\left.r_n\right|_{v=0}(u)}
       \frac{(-\partial_ur_n)(u,s)}s
       (\alpha_n-\phi_n)(u,s)\,\mathrm ds.
\end{aligned}
\end{equation}
In light of \eqref{vanishing-boundary-flux1}, the first boundary term tends to zero in $L^2_u([-T,T])$, uniformly for
$\varepsilon\leq r\leq R$.

To handle the integral term in \eqref{backwards-scalar}, choose a fixed number
$N>\max\{1,R\}$.
The finite integrals from $r$ to $N$ converge uniformly on
$[-T,T]\times[\varepsilon,R]$ by
Lemma~\ref{eternal-local-compactness}.
For the remaining tail, applying \eqref{normalized-tail} and \eqref{Eqn:prurbound}, we deduce that
\begin{equation}\label{backwards-tail}
    \begin{aligned}
 &\int_N^{\,\left.r_n\right|_{v=0}(u)}
       \frac{(-\partial_ur_n)(u,s)}s
       |\alpha_n-\phi_n|(u,s)\,\mathrm ds
       \\
\leq&
 C\frac{M_n}{R_n}\int_{N}^{\,\left.r_n\right|_{v=0}(u)}  \frac{\dd s}{s}
       +C\int_N^\infty\frac{1+\log s}{s^2}\,\mathrm ds
       \\
 \leq&
 C\frac{M_n}{R_n}\log\frac{3R_n}{M_n}
       +\frac{C(2+\log N)}N  \qquad\quad\text{uniformly in } |u|\le T.
\end{aligned}
\end{equation}
Here we utilize \eqref{Eqn:estoutradius}. Observe that on the RHS of \eqref{backwards-tail}, the first term tends to zero as $n\to\infty$, while the second tends to
zero when $N\to\infty$. 

Meanwhile, according to \eqref{normalized-tail}, the limiting fields satisfy
\[
 |\alpha(u,s)-\phi(u,s)|
 \leq\frac{C(1+\log s)}s\qquad(s\geq1).
\]
This also implies a similar tail bound
\begin{equation}
  \int_N^{\infty}
       \frac{(-\partial_ur)(u,s)}s
       |\alpha-\phi|(u,s)\,\mathrm ds\le C\int_N^{\infty}   \frac{C(1+\log s)}{s^2}\dd s\le \frac{C(2+\log N)}N.
\end{equation}

It follows, first letting $n\to\infty$ with $N$ fixed and then
sending $N\to\infty$, that the integral term in
\eqref{backwards-scalar} converges uniformly in $ [-T, T]\times[\varepsilon, R]$ to the
RHS of \eqref{eternal-incoming}. And this RHS term is continuous as well.

Combining with the vanishing boundary term in \eqref{backwards-scalar}, we obtain 
\begin{equation}\label{Eqn:flux-strongconvergence}
    \frac{r\partial_u\phi_n}{-\partial_ur_n}\to \frac{F}{-\partial_ur} \qquad\quad \text{strongly in } L^2([-T, T]).
\end{equation}
Passing the uniform bound for $ \frac{r\partial_u\phi_n}{-\partial_ur_n}(\cdot, r)$ in \eqref{Eqn:scaleinvariantbound1} for almost sure $u\in[-T, T]$, along with the continuity of $\frac{F}{-\partial_ur}$, we infer
\begin{equation}\label{Eqn:uniform-scalarbound}
    \left| \frac{r\partial_u\phi_n}{-\partial_ur_n}\right|+\left|\frac{F}{-\partial_ur}\right|\le C. \\[2mm]
\end{equation}
Combining \eqref{Eqn:flux-strongconvergence} and \eqref{Eqn:uniform-scalarbound}, by the Cauchy--Schwarz inequality we have
\begin{align*}
\left|\frac{r\partial_u\phi_n}{-\partial_ur_n}\right|^2\to \left|\frac{F}{-\partial_u r}\right|^2, \qquad \frac{r\partial_u\phi_n}{-\partial_ur_n}\to\frac{F}{-\partial_u r}  \qquad\quad \text{in } L^1_u([-T, T]),
\end{align*}
%\begin{align*}
   % &\left\|\left|\frac{r\partial_u\phi_n}{-\partial_ur_n}\right|^2-\left|\frac{F}{-\partial_u r}\right|^2\right\|_{L^1_u([-T, T])} \\
   %\le& \left(\left\|\frac{r\partial_u\phi_n}{-\partial_ur_n}\right\|_{L^2_u([-T, T])}+\left\|\frac{F}{-\partial_u r}\right\|_{L^2_u([-T, T])}\right)\left\|\frac{r\partial_u\phi_n}{-\partial_ur_n}-\frac{F}{-\partial_u r}\right\|_{L^2_u([-T, T])}\to 0, \\
    %&\left\|\frac{r\partial_u\phi_n}{-\partial_ur_n}-\frac{F}{-\partial_u r}\right\|_{L^1_u([-T, T])} \le C\sqrt{T}\left\|\frac{r\partial_u\phi_n}{-\partial_ur_n}-\frac{F}{-\partial_u r}\right\|_{L^2_u([-T, T])}\to 0.
%\end{align*}
uniformly for $\varepsilon\leq r\leq R$.

As a result, applying Lemma~\ref{Lem:calculus} with the evolution equations \eqref{eternal-time-mass} and \eqref{eternal-phi} for the normalized quantities $m_n, \phi_n$ and their limits $m, \phi$, we conclude that $m,\phi\in C^1$, and that 
\eqref{eternal-radial-mass}, the radial equation in
\eqref{eternal-phi}, \eqref{eternal-bondi-scalar-integral}--\eqref{eternal-bondi-mass-integral} hold true for all $(u, r)\in \mathbb{R}\times (0,\infty)$. In particular, by virtue of \eqref{Eqn:uniformbound-pru-mn}, it holds
\begin{equation}\label{Eqn:uniformbound-pru-m}
    \left|\left.\frac{\partial m}{\partial u}\right|_r\right|
 \leq C.
\end{equation}
%This argument passes the full quadratic incoming flux, with no additional term in the mass equation.

It remains to show that $-\partial_u r_n\to-\partial_ur$ in $C^1_{\mathrm{loc}}$. Differentiating \eqref{normalized-speed} in $u, r$ respectively, we get
\begin{align}
    \left.\frac{\partial(-\partial_ur_n)}{\partial r}\right|_u=&(-\partial_ur_n)\cdot  \frac{\mu_n}{r(1-\mu_n)},  \label{Eqn:derivative-prurn1}\\
    \left.\frac{\partial(-\partial_ur_n)}{\partial u}\right|_r%=&(-\partial_ur_n)\cdot  \left(-\frac{\mathrm{d}r_n\big|_{v=0}(u)}{\diff u}\cdot\frac{\mu_n}{r_n(1-\mu_n)}\bigg|_{v=0}(u)-\int_r^{\,\left.r_n\right|_{v=0}(u)} \left.\frac{\partial}{\partial u}\right|_r\left(\frac{\mu_n}{1-\mu_n}\right)\frac{\mathrm ds}{s} \right) \notag \\
    =&(-\partial_ur_n)
    \cdot  \left(\frac{\mu_n}{2r_n(1-\mu_n)}\bigg|_{v=0}(u)-\int_r^{\,\left.r_n\right|_{v=0}(u)}
             \frac{2}{s^2(1-\mu_n)^2}
       \left.\frac{\partial m_n}{\partial u}\right|_r \dd s. \right) \label{Eqn:derivative-prurn2}
\end{align}
Here we use \eqref{Eqn:rnoutbdy}.

Applying Lemma~\ref{Lem:calculus} for \eqref{Eqn:derivative-prurn1} immediately implies that $\left.\partial_r\right|_u(-\partial_ur)\in C^0_{\mathrm{loc}}$ and
\begin{equation*}
    \left.\frac{\partial(-\partial_ur_n)}{\partial r}\right|_u \to  \left.\frac{\partial(-\partial_ur)}{\partial r}\right|_u \qquad\quad\text{in } C^0_{\mathrm{loc}}.
\end{equation*}
Regarding the differentiability of $-\partial_u r$ in $u$, thanks to \eqref{normalized-mass} and \eqref{Eqn:uniformbound-pru-mn}, we have
\begin{equation*}
     \left|\frac{2}{r^2(1-\mu_n)^2}
       \left.\frac{\partial m_n}{\partial u}\right|_r\right|\le \frac{C}{r^2}.
\end{equation*}
Therefore, mimicking the proof for the $C^0_{\mathrm{loc}}$-convergence of $-\partial_u r_n$ as in \eqref{Eqn:partialurnint}, we infer that the RHS of \eqref{Eqn:derivative-prurn2} converges locally uniformly in $\mathbb{R}\times (0,\infty)$. Applying Lemma~\ref{Lem:calculus} again,  we get $\left.\partial_u\right|_r(-\partial_ur)\in C^0_{\mathrm{loc}}$ and
\begin{equation*}
    \left.\frac{\partial(-\partial_ur_n)}{\partial u}\right|_r \to  \left.\frac{\partial(-\partial_ur)}{\partial u}\right|_r \qquad\quad\text{in } C^0_{\mathrm{loc}}.
\end{equation*}
This completes the proof.
\end{proof}

We move on to prove that $\alpha,\partial_vr\in C^1$.

\begin{lemma}
\label{eternal-differentiated-convergence}
The normalized quantities
\[
 \alpha'_n,\qquad
 \frac{\partial_v^2r_n}{(\partial_vr_n)^2}
\]
converge locally uniformly on $\mathbb R\times(0,\infty)$. These limits are continuous and given by
\begin{equation}\label{Eqn:limit-alphaprime}
    \alpha'=\left.\frac{\partial \alpha}{\partial r}\right|_u, \qquad\quad  \frac1{\partial_vr}
       \left.\frac{\partial(\partial_vr)}{\partial r}\right|_u,
\end{equation}
respectively. Moreover, $\alpha,\partial_vr\in C^1$, and
\eqref{eternal-alpha} and 
\eqref{outgoing-speed-chart} hold true.
\end{lemma}

\begin{proof}
Fix $T<\infty$ and $0<\varepsilon<R<\infty$ and choose any point  $(u, r)\in [-T, T]\times [\varepsilon, R]$. Parametrize the incoming characteristic emanating from $(u, r)$ by $\gamma_n(s)=(u_n(s), s)$ in Bondi coordinates, with
\begin{equation*}
   u_n(r)=u, \qquad\quad \frac{\diff u_n(s)}{\diff s}=\frac{1}{\partial_u r_n(u_n(s), s)}.
\end{equation*}
This gives for any function $F=F(u, r)$,
\begin{align}\label{Eqn:dbyds}
    \frac{\diff}{\diff s} F(u_n(s), s)=\frac{1}{\partial_u r_n}\left.\frac{\partial F}{\partial u}\right|_{r}(u_n(s), s)+\left.\frac{\partial F}{\partial r}\right|_{u}(u_n(s), s)=\frac{\partial_u F}{\partial_u r_n}(u_n(s), s),
\end{align}
where we make use of \eqref{Eqn:derivative-relation} as well.

Recall \eqref{Eqn:prurbound}:
\begin{equation*}
    0<\frac12 e^{-C/r} \le -\partial_u r_n
 \leq \frac12.
\end{equation*}
Thus we infer that 
\begin{equation*}%\label{Eqn:boundrn1}
  2|s-r|\le |u_n(s)-u|\le 2e^{C/\varepsilon}|s-r| \qquad\quad \text{for any } s\ge \varepsilon. 
\end{equation*}
As a result, for any fixed $N>R$, we have 
\begin{equation}\label{Eqn:boundrn1}
 |u_n(s)|\le 2e^{C/\varepsilon}N+T\qquad\quad \text{for all } s\in [\varepsilon, N].
\end{equation}
Now select $n$ large enough so that 
\begin{equation*}
    J_n\ge 2e^{C/\varepsilon}N+T, \qquad\quad s_n\le -2(e^{C/\varepsilon}+1)N-T.
\end{equation*}
This ensures that 
\begin{equation*}
 \{ \gamma_n(s):\ s\in [\varepsilon, N]\}\subset  \mathfrak{R}_n=\left\{(w,s): |w|\leq J_n,\ 
                     0<s<-(w+s_n)/2\right\}.
\end{equation*}
Moreover, in light of \eqref{Eqn:boundrn1}, along with the locally uniform convergence of $-\partial_u r_n\to -\partial_u r$, we infer that $u_n(s)$  converges uniformly in $ [\varepsilon, N].$
%These endpoint radii tend to infinity uniformly over the fixed target rectangle.
\vspace{2mm}

Applying \eqref{scalar-multiplier} from Lemma~\ref{multiplier-lemma} and using \eqref{Eqn:dbyds}, we obtain
\begin{align*}
    \frac{\diff}{\diff s} \left[(\partial_v r_n)^2\alpha'_n(u_n(s), s)\right]=F_{\alpha'_n}(u(s), s),
\end{align*}
where $F_{\alpha'_n}$ is given by 
\begin{align*}
   F_{\alpha'_n}(u, r)=  -\frac{(\partial_vr_n)^2}
      {r^2(1-\mu_n)}\cdot
 \Bigl[
 4\pi|\alpha_n-\phi_n|^2(\alpha_n-\phi_n)
       -3\mu_n(\alpha_n-\phi_n)
 \Bigr](u, r).
\end{align*}
Integrating the above equation over $s\in[r, N]$ gives
\begin{equation}\label{Eqn:alphanprime}
    \begin{aligned}
 \alpha'_n(u, r)
 ={}&
 \frac{(\partial_vr_n)^2(u_n(N), N)}
      {(\partial_vr_n)^2(u, r)}\,\alpha'_n(u_n(N),v)+\frac1{(\partial_vr_n)^2(u,r)}
 \int_{r}^{N} F_{\alpha_n'}(u_n(s), s)\,\mathrm ds.
\end{aligned}
\end{equation}
Employing \eqref{Eqn:scaleinvariantbound1}, the first term in \eqref{Eqn:alphanprime} can be controlled by $CN^{-1}$. Combining with  \eqref{Eqn:boundrn1}, the uniform convergence of $u_n(s)$ in $[\varepsilon, N]$, as well as the locally uniform convergence of the normalized quantities and the locally uniform boundedness of their derivatives, we derive
\[
 \limsup_{n,k\to\infty}
 \sup_{\substack{|u|\leq T\\ \varepsilon\leq r\leq R}}
 |\alpha'_n(u,r)-\alpha'_k(u,r)|
 \leq\frac CN.
\]
Letting $N\to\infty$ proves local uniform convergence of
$\alpha'_n$.

Proceeding as in the same fashion for  \eqref{metric-multiplier} in Lemma~\ref{multiplier-lemma}, we get the local uniform convergence of $\partial_v^2r_n/(\partial_vr_n)^2$. 

Finally, applying Lemma~\ref{Lem:calculus} for equations of $\alpha_n', \partial_v r_n$ as in \eqref{eternal-alpha}, \eqref{outgoing-speed-chart}, we conclude that \eqref{Eqn:limit-alphaprime} hold true, and that $\alpha, \partial_vr\in C^1$, which obey the desired differential equations \eqref{eternal-alpha}, \eqref{outgoing-speed-chart}.
\end{proof}

We are now prepared to prove Proposition~\ref{eternal-proposition}.

\begin{proof}[Proof of Proposition~\ref{eternal-proposition}]
Lemmas~\ref{eternal-local-compactness}, \ref{eternal-strong-flux},
and \ref{eternal-differentiated-convergence} give
\[
 m,\phi,\alpha,-\partial_ur,\partial_vr
 \in C^1\bigl(\mathbb R\times(0,\infty)\bigr)
\]
in $(u,r)$ coordinates. %We first construct local double null coordinates $(u ,v)$ and identify the limiting null derivatives.   We then establish the regularity needed to verify the neutral equations.

By \eqref{eternal-speed},
\[
 \left.\frac{\partial(-\partial_ur)}{\partial r}\right|_u
 =\frac{(-\partial_ur)\mu}{r(1-\mu)}.
\]
Combining this identity with the limiting form of
\eqref{outgoing-speed-chart}, we obtain
\[
 \left.\frac{\partial}{\partial u}\right|_r
       \frac1{\partial_vr}
 =
 \left.\frac{\partial}{\partial r}\right|_u
       \frac{-\partial_ur}{\partial_vr}.
\]
Thus the $C^1$ one-form
\begin{equation}\label{null-one-form}
 \frac{\mathrm dr+(-\partial_ur)\,\mathrm du}{\partial_vr}
\end{equation}
is closed and admits a local $C^2$ primitive $v$. Since $\partial_r\big|_uv=(\partial_v r)^{-1}>0$, the map $(u,r)\mapsto(u,v)$ has a local $C^2$ inverse.
In the inverse chart,
\[
 \mathrm dr=(\partial_ur)\,\mathrm du
                 +(\partial_vr)\,\mathrm dv,
\]
so the limiting functions $\partial_ur$ and $\partial_vr$ are
the null derivatives of $r=r(u,v)$. In particular, the quotient
identified in \eqref{Eqn:limit-alphaprime} is
$\partial_v^2r/(\partial_vr)^2$.

Now the change of coordinates gives
\begin{equation}\label{Eqn:change-of-coor}
    \partial_u
 =\left.\frac{\partial}{\partial u}\right|_r
   +(\partial_ur)\left.\frac{\partial}{\partial r}\right|_u,
 \qquad
\partial_v
 =(\partial_vr)\left.\frac{\partial}{\partial r}\right|_u.
\end{equation}
Applying these identities to the radial scalar equation in \eqref{eternal-phi} and
\eqref{eternal-bondi-scalar-integral}, we find
\begin{equation}\label{eternal-proof-incoming-integral}
 (r\partial_u\phi)(u,r)
 =F(u, r)=-\int_r^\infty
       \frac{(-\partial_ur)(u,s)}s
       (\alpha-\phi)(u,s)\,\mathrm ds.
\end{equation}
This is \eqref{eternal-incoming}, with $\partial_u\phi$ now
interpreted as the null derivative. The same change of coordinates
identifies \eqref{eternal-bondi-scalar-integral} and
\eqref{eternal-bondi-mass-integral} with \eqref{eternal-phi} and
\eqref{eternal-time-mass}, respectively.
\vspace{2mm}

We next prove that $r\partial_u\phi$ and $r\partial_v\phi$ are
$C^1$ at positive radius. Fix a compact $u$-interval. For $r\geq1$,
\eqref{eternal-time-mass} and \eqref{eternal-speed} give
\[
 \left|
 \left.\frac{\partial(-\partial_ur)}{\partial u}\right|_r
 \right|\leq\frac Cr,
\]
while the scalar equations, the boundedness of the scalar quantities,
and $|r\alpha'|\leq C$ give
\[
 \left|
 \left.\frac{\partial(\alpha-\phi)}{\partial u}\right|_r
 \right|\leq\frac Cr.
\]
The constants may be chosen uniformly on the fixed $u$-interval.
It follows that, with $s$ held fixed,
\[
 \left|
 \frac{\partial}{\partial u}
 \left[
   \frac{(-\partial_ur)(u,s)}s(\alpha-\phi)(u,s)
 \right]
 \right|\leq\frac C{s^2},\qquad s\geq1.
\]
The differentiated integrand is continuous, and the bound is
integrable at infinity. We may therefore differentiate
\eqref{eternal-proof-incoming-integral} under the integral sign.
This gives a continuous $u$-derivative of $r\partial_u\phi$ at
fixed $r$, and hence of $r\partial_u\phi/(-\partial_ur)$.
Differentiation in $r$, using \eqref{eternal-speed}, yields
\[
 \left.\frac{\partial}{\partial r}\right|_u
 \left(\frac{r\partial_u\phi}{-\partial_ur}\right)
 =\frac1r\left[
       \alpha-\phi
       -\frac{\mu}{1-\mu}
        \frac{r\partial_u\phi}{-\partial_ur}
 \right].
\]
Thus $r\partial_u\phi/(-\partial_ur)$, and consequently
$r\partial_u\phi$, belong to $C^1$ at positive radius.
The identity
\begin{equation}\label{eternal-proof-outgoing-scalar}
 r\partial_v\phi=(\partial_vr)(\alpha-\phi)
\end{equation}
gives the same regularity for $r\partial_v\phi$.

We now verify the neutral equations in the $(u,v)$ coordinates.
Define the null lapse
\[
 \Omega^2=\frac{4(-\partial_ur)\partial_vr}{1-\mu}.
\]
Since $-\partial_ur$, $\partial_vr$, and $1-\mu$ are positive,
we have $\Omega>0$ and $\Omega\in C^1$.
Transforming the Bondi equations by the chain rule gives the two
neutral mass equations and
\begin{equation}\label{eternal-proof-null-transport}
 \partial_u\partial_vr
 =\frac{\partial_ur\,\partial_vr}{r}\frac{\mu}{1-\mu},
 \qquad
 \partial_u\alpha
 =\frac{-\partial_ur}{r}\frac{\mu}{1-\mu}(\alpha-\phi).
\end{equation}
Differentiating \eqref{eternal-proof-outgoing-scalar} in $u$ and
substituting \eqref{eternal-proof-null-transport}, we obtain
\[
 \partial_u(r\partial_v\phi)
 =-(\partial_vr)\partial_u\phi,
\]
which is the neutral scalar wave equation.
Similarly, differentiating
\[
 \frac{\partial_ur}{\Omega^2}
 =-\frac{1-\mu}{4\partial_vr},
 \qquad
 \frac{\partial_vr}{\Omega^2}
 =\frac{1-\mu}{4(-\partial_ur)}
\]
in $u$ and $v$, respectively, and using the neutral mass equations
and \eqref{eternal-proof-null-transport}, we find
\[
 \partial_u\left(\frac{\partial_ur}{\Omega^2}\right)
 =-\frac{4\pi r|\partial_u\phi|^2}{\Omega^2},
 \qquad
 \partial_v\left(\frac{\partial_vr}{\Omega^2}\right)
 =-\frac{4\pi r|\partial_v\phi|^2}{\Omega^2}.
\]

For the mixed lapse equation, we first pass to the limit in the
integrated neutral metric multiplier identity on a compact incoming
characteristic segment. Lemmas~\ref{eternal-local-compactness},
\ref{eternal-strong-flux}, and \ref{eternal-differentiated-convergence}
give local uniform convergence of the multiplier fields, sources,
and characteristic curves. The limiting integral identity has a
continuous integrand, so it yields the continuous derivative
\begin{equation}\label{eternal-proof-metric-multiplier}
 \partial_u\left(\frac{\partial_v^2r}{\partial_vr}\right)
 =\frac{(-\partial_ur)\partial_vr}{r^2(1-\mu)}
       \left[2\mu-4\pi|\alpha-\phi|^2\right].
\end{equation}
The mass relation and the radial equations also give
\begin{equation}\label{eternal-proof-lapse-derivative}
 \partial_v\log\Omega
 =\frac12\frac{\partial_v^2r}{\partial_vr}
  +\frac{2\pi\,\partial_vr}{r}|\alpha-\phi|^2.
\end{equation}
By \eqref{eternal-proof-null-transport} and
\eqref{eternal-proof-outgoing-scalar},
\begin{equation}\label{eternal-proof-scalar-product}
 \partial_u\left(
       \frac{\partial_vr}{r}|\alpha-\phi|^2\right)
 =\frac{(-\partial_ur)\partial_vr}{r^2(1-\mu)}
       |\alpha-\phi|^2
  -2\operatorname{Re}
       (\partial_u\phi\,\overline{\partial_v\phi}).
\end{equation}
We may therefore differentiate \eqref{eternal-proof-lapse-derivative}
in $u$. Substituting \eqref{eternal-proof-metric-multiplier} and
\eqref{eternal-proof-scalar-product} cancels the terms containing
$|\alpha-\phi|^2$ and gives
\[
 \partial_u\partial_v\log\Omega
 =\frac{\Omega^2\mu}{4r^2}
  -4\pi\operatorname{Re}
       (\partial_u\phi\,\overline{\partial_v\phi}).
\]
The limit is therefore a classical neutral solution at every
positive radius.

Finally, the bound for $r\alpha'$ passes to the limit by \eqref{Eqn:scaleinvariantbound1} and  Lemma~\ref{eternal-differentiated-convergence}, while
the incoming scalar bound for $\partial_u\phi/(-\partial_u r)$ follows from \eqref{Eqn:uniform-scalarbound}  and \eqref{eternal-proof-incoming-integral}.
This proves all the asserted estimates, completing the proof of Proposition~\ref{eternal-proposition}.
\end{proof}

\subsection{Constant asymptotic scalar coefficient and mass}
\label{asymptotic-data-section}

In this section, we continue to study the asymptotic behaviors of the eternal limit solution obtained in Proposition~\ref{eternal-proposition}, along future null infinity $\{u=\text{const}, \, r=\infty\}$  in Bondi coordinates
$(u,r)$. This will be crucially used for the stationary comparison in Section~\ref{sec:rigidity}.

\begin{lemma}\label{asymptotic-lemma}
For the eternal limit solution of Proposition~\ref{eternal-proposition},
the limits
\begin{equation}\label{asymptotic-constants}
 c_1=\lim_{r\to\infty}r(\phi(u,r)-c)\in\mathbb C,\qquad
 M=\lim_{r\to\infty}m(u,r),\qquad \tfrac12\leq M\leq8
\end{equation}
exist and are independent of $u$. Furthermore, the following estimates hold
\begin{equation}\label{Eqn:imporved-bounds-phi-c}
    \left|r(\phi(u, r)-c)
       -c_1\right|+|m(u, r)-M|+r|\alpha(u, r)-c|\leq C/r \qquad\quad \text{for all } u\in \mathbb{R}, r\ge 2.
\end{equation}
\end{lemma}

\begin{proof}
First, in light of \eqref{eternal-mass} and \eqref{eternal-radial-mass}, we infer that
\begin{equation}\label{Eqn:radialmass}
    \left.\frac{\partial m}{\partial r}\right|_u =2\pi(1-\mu)|\alpha-\phi|^2\geq0, \qquad\quad 0\le m\le 8.
\end{equation}
This implies that $M(u)=\lim_{r\to\infty}m(u,r)$ exists and $0\le M(u)\le 8$.
The lower mass bound in Proposition~\ref{eternal-proposition}
gives $M(u)\geq\tfrac12$.

We proceed to derive improved bounds for $\alpha-c$ and $r(\phi-c)$. Fix $u\in\mathbb{R}, r\ge 2$, and
parametrize the incoming characteristic emanating from $(u, r)$ by $\gamma_(s)=(u(s), s)$ in Bondi coordinates, with
\begin{equation*}
   u(r)=u, \qquad\quad \frac{\diff u(s)}{\diff s}=\frac{1}{\partial_u r(u(s), s)}.
\end{equation*}
Hence, for any function $F=F(u, r)$ we get
\begin{align}\label{Eqn:dbyds1}
    \frac{\diff}{\diff s} F(u(s), s)=\frac{1}{\partial_u r}\left.\frac{\partial F}{\partial u}\right|_{r}(u(s), s)+\left.\frac{\partial F}{\partial r}\right|_{u}(u(s), s)=\frac{\partial_u F}{\partial_u r}(u(s), s),
\end{align}
where we employ \eqref{Eqn:change-of-coor}. 

Passing the limit for \eqref{Eqn:prurbound} gives
\begin{equation}\label{Eqn:prurbound1}
     0<\frac12 e^{-C/s} \le -\partial_u r(w, s)
 \leq \frac12 \qquad\quad \text{for all } (w, s)\in\mathbb{R}\times (0, \infty).
\end{equation}
from which we deduce
\begin{equation*}%\label{Eqn:boundrn1}
  -2C(u-s)\le u(s)-u\le -2(u-s) \qquad\quad \text{for any } s\ge r. 
\end{equation*}
This indicates that $\gamma(s)=(u(s), s)$ approaches the past null infinity as $s\to\infty$.

In view of \eqref{alpha-transport} and \eqref{Eqn:dbyds1}, the transport equation of $\alpha$ along $\gamma(s)$ reads
\begin{equation}\label{Eqn:transport-a-prime}
    \frac{\diff }{\diff s} \alpha(u(s), s)= -\frac{1}{s}
             \frac{\mu}{1-\mu}(\alpha-\phi)(u(s), s).
\end{equation}
 Integrating it from $s=r$ to $s=\infty$, together with the fact $\alpha(u(s), s)\to c$ as $s\to\infty$ and \eqref{eternal-mass}, \eqref{eternal-tail}, we find 
\[
 |\alpha(u,r)-c|\leq\int_{r}^{\infty} \frac{C(1+\log s)}{s^3} \dd s \le \frac{C(1+\log r)}{r^2} \qquad\quad \text{for all } r\ge2.
\]
Since this bound is integrable in $r$, the radial scalar equation
\begin{equation}\label{Eqn:radial-scalar}
     \left.\frac{\partial }{\partial r}\right|_u\bigl[r(\phi-c)\bigr]=\alpha-c
\end{equation}
shows that $r(\phi(u,r)-c)$ has a limit $c_1(u)$ as $r\to\infty$.
Integrating from $r=2$ and using the uniform bound for $|\phi|$ in
\eqref{eternal-scalar}, we also obtain
\begin{equation}\label{Eqn:improved-est-phi-c}
     |r(\phi(u,r)-c)|+|c_1(u)|\leq C
\end{equation}

We move to improve the decay estimate for $\alpha-c$. 
Rewrite \eqref{Eqn:transport-a-prime} as
\begin{equation*}
    \frac{\diff }{\diff s} \left(\alpha(u(s), s)-c\right)= -\frac{1}{s}
             \frac{\mu}{1-\mu}(\alpha-c)(u(s), s)+\frac{1}{s}
             \frac{\mu}{1-\mu}(\phi-c)(u(s), s).
\end{equation*}
Thus, we derive the below exact integral formula
\begin{align}
 \alpha(u,r)-c=-\int_r^\infty&
 \exp\left[\int_r^s
       \frac{\mu(u(t),t)}{1-\mu(u(t),t)}\frac{\mathrm dt}{t}\right]
 \frac{\mu(u(s),s)}{1-\mu(u(s),s)}
 \frac{\phi(u(s),s)-c}{s}\,\mathrm ds .
        \label{exterior-incoming}
\end{align}
The bound $\mu/(1-\mu)\leq C/r$ from \eqref{eternal-mass} implies that the exponential factor is uniformly bounded. Along with \eqref{Eqn:improved-est-phi-c},  this thus implies
\[
  |\alpha(u,r)-c|\leq C\int_r^\infty s^{-3}\,\mathrm ds
  \leq \frac{C}{r^2}.
\]
The radial scalar equation \eqref{Eqn:radial-scalar},
\eqref{eternal-incoming}, and the mass equation \eqref{Eqn:radialmass} then give
\begin{equation} \label{asymptotic-estimates}
    \begin{aligned}
 \left|r(\phi(u,r)-c)
       -c_1(u)\right|&\leq C/r,\\
 \left|\alpha(u,r)-\phi(u,r)
       +\frac{c_1(u)}{r}\right|&\leq C/r^2,\\
 \left|\frac{r\partial_u\phi}{-\partial_ur}(u,r)
       -\frac{c_1(u)}{r}\right|&\leq C/r^2,\\
 0\leq M(u)-m(u,r)&\leq C/r.
\end{aligned}
\end{equation}
Here we used
$(-\partial_ur)(u,r)=1/2+O(r^{-1})$, which follows from
\eqref{Eqn:prurbound1}.

Consequently, incorporating with \eqref{eternal-time-mass} and \eqref{eternal-phi}, this implies
\begin{align*}
   \left.\frac{\partial m}{\partial u}\right|_r
 =&2\pi(-\partial_ur)(1-\mu)
       \left[|\alpha-\phi|^2
              -\left|\frac{r\partial_u\phi}{-\partial_ur}\right|^2
       \right]=O(r^{-3})\\
  \left.\frac{\partial}{\partial u}\right|_r[r(\phi-c)]
 =&(-\partial_ur)
       \left(\frac{r\partial_u\phi}{-\partial_ur}+\alpha-\phi\right)
       =O(r^{-2}),
\end{align*}
For any $u_1,u_2\in\mathbb R$, integration at fixed radius therefore
gives
\[
  \bigl|r(\phi(u_2,r)-\phi(u_1,r))\bigr|
    \leq \frac{C|u_2-u_1|}{r^2},
  \qquad
  |m(u_2,r)-m(u_1,r)|\leq\frac{C|u_2-u_1|}{r^3}.
\]
Letting $r\to\infty$, we conclude that $c_1(u_2)=c_1(u_1)$ and
$M(u_2)=M(u_1)$. Denoting these constants by $c_1$ and $M$, the
claimed estimate follows from \eqref{asymptotic-estimates} and
$|\alpha-c|\leq C/r^2$.
\end{proof}

\section{Stationary comparison and rigidity}
\label{sec:rigidity}\label{rigidity-section}

We now use the asymptotic data obtained in
Section~\ref{asymptotic-data-section} to rule out the eternal
solution constructed in Proposition~\ref{eternal-proposition}.
The argument is based on comparison with a stationary solution. More precisely, we aim to prove

\begin{proposition}\label{rigidity-proposition}
There is no classical neutral complex scalar solution on
$\mathbb R\times(0,\infty)$ satisfying
\eqref{eternal-mass}--\eqref{eternal-incoming}.
\end{proposition}

\subsection{A stationary exterior with the same asymptotic data}
We first construct the stationary solution by prescribing its data at
infinity. Throughout this section, stationarity means independence of
$u$ at fixed $r$ in Bondi coordinates $(u,r)$.
\begin{lemma}\label{stationary-exterior-lemma}
For sufficiently large $R\ge \max(4M, 2)$, there exists a stationary neutral solution $\left(m_{\mathrm{s}}(r), \phi_{\mathrm{s}}(r), \alpha_{\mathrm{s}}(r), (-\partial_u r)_{\mathrm{s}}(r)\right)$ on an exterior region $\{r\ge R\}$, satisfying
\begin{equation}
\begin{aligned}
 \frac{\mathrm dm_{\mathrm{s}}}{\mathrm dr}
 &=2\pi\left(1-\frac{2m_{\mathrm{s}}}{r}\right)
             |\alpha_{\mathrm{s}}-\phi_{\mathrm{s}}|^2,    \\
 \frac{\mathrm d\alpha_{\mathrm{s}}}{\mathrm dr}
 &=-\frac1r\frac{2m_{\mathrm{s}}}
                    {r-2m_{\mathrm{s}}}
             (\alpha_{\mathrm{s}}-\phi_{\mathrm{s}}),\\
 \frac{\mathrm d}{\mathrm dr}
       (\alpha_{\mathrm{s}}-\phi_{\mathrm{s}})
 &=-\frac{\alpha_{\mathrm{s}}-\phi_{\mathrm{s}}}
             {r(1-2m_{\mathrm{s}}/r)}, 
    \label{stationary-equations}
\end{aligned}
\end{equation}
\begin{align}
     (-\partial_ur)_{\mathrm{s}}(r)
 &=\frac12\exp\left[-\int_r^\infty
       \frac{2m_{\mathrm{s}}(s)}
            {s-2m_{\mathrm{s}}(s)}
        \frac{\mathrm ds}{s}\right], \label{Eqn:integral-ru}\\
     \alpha_{\mathrm{s}}-c=&-\int_r^\infty
 \exp\left[\int_r^s
       \frac{2m_{\mathrm{s}}(t)}{t-2m_{\mathrm{s}}(t)}\frac{\mathrm dt}{t}\right]
 \frac{2m_{\mathrm{s}}(s)}{s-2m_{\mathrm{s}}(s)}
 \frac{\phi_{\mathrm{s}}(s)-c}{s}\,\mathrm ds   \label{Eqn:integralalpha}
\end{align}
and 
\begin{equation}\label{Eqn:conditioninfty}
  \phi_{\mathrm{s}}\to c, \qquad\quad  r(\phi_{\mathrm{s}}-c)\to c_1, \qquad\quad m_{\mathrm{s}}\to M, \qquad\quad (-\partial_u r)_{\mathrm{s}}\to \frac12 \qquad\quad \text{as } r\to \infty.
\end{equation}
Here $c$ is the constant in Lemma~\ref{past-lemma}, while $c_1, M$ are defined from Lemma~\ref{asymptotic-lemma}.
\end{lemma}
\begin{proof}
We conduct the standard Banach fixed point argument to solve \eqref{stationary-equations} backward from $r=\infty$ with the imposed boundary conditions \eqref{Eqn:conditioninfty}.  For a sufficiently large fixed radius $R$, consider a function space
\begin{equation}\label{stationary-class}
    X_R\coloneqq \left\{m=m(r)\in C^0([R, \infty)): \ \frac{M}{2}\le m(r)\le M, \quad M-m(r)
       \leq\frac{2\pi e^2|c_1|^2}{r} \right\}
\end{equation}
equipped with the distance
\begin{equation*}
    d_{X_R}(m_1, m_2)\coloneqq \sup\limits_{r\geq R} r|m_1(r)-m_2(r)|.
\end{equation*}
It is easy to check that $(X_R, d_{X_R})$ forms a complete metric space and
\begin{equation}\label{Eqn:XRbpundedness}
    d_{X_R}(m_1, m_2)\le 2\pi e^2|c_1|^2 \qquad\quad\text{for all } m_1, m_2\in X_R.
\end{equation}

We aim to solve the integral equation
\begin{align}
 m_{\mathrm{s}}(r)=M-2\pi|c_1|^2\int_r^\infty&
 \left(1-\frac{2m_{\mathrm{s}}(s)}s\right)\frac1{s^2}
 \cdot\exp\left[
       2\int_s^\infty
       \frac{2m_{\mathrm{s}}(t)}{t-2m_{\mathrm{s}}(t)}\frac{\mathrm dt}{t}\right]\mathrm ds\eqqcolon \mathscr{L}(m_{\mathrm{s}})(r).
        \label{stationary-mass}
\end{align}

Let first $R\geq 4M$. For any $m_1, m_2\in X_R$, direct computations imply that for any $s\ge R$,
\begin{align*}
     2\int_s^\infty
       \frac{2m_{1}(t)}{t-2m_{1}(t)}\frac{\mathrm dt}{t}\le& \int_s^{\infty} \frac{4M}{(t-2M)t} \dd t\le \int_s^{\infty} \frac{8M}{t^2}\dd t\le 2,  \\
    \left|\frac{2m_1(s)}{s-2m_1(s)} 
       -\frac{2m_2(s)}{s-2m_2(s)}\right|
       &=\frac{2s|m_1(s)-m_2(s)|}{(s-2m_1(s))(s-2m_2(s))}\le\frac8{s^2}\sup_{t\geq R}t|m_1(t)-m_2(t)|,
\end{align*}
As a consequence, utilizing \eqref{Eqn:XRbpundedness}, for all $r\ge R$ we derive
\begin{align*}
    0\le M-\mathscr L(m_1)(r)
       \le& 2\pi |c_1|^2e^2\int_r^\infty \frac1{s^2}=\frac{2\pi e^2|c_1|^2}{r}\le \frac{M}2, \\
     \left|\mathscr L(m_1)(r)-\mathscr L(m_2)(r) \right|\le& 2\pi|c_1|^2e^2 \int_r^{\infty} \frac{2|m_1(s)-m_2(s)|}{s}\frac{\dd s}{s^2}\\
     &+2\pi|c_1|^2e^2\int_r^\infty \frac{1}{s^2}\left(\exp\left(\int_s^\infty \frac{16 d_{X_R}(m_1, m_2)}{t^2}\dd t \right) -1\right) \dd s \\
     \le& \left(\frac{C}{R^3}+\left(e^{\frac{Cd_{X_R}(m_1, m_2)}{R}}-1\right)^2\right) d_{X_R}(m_1, m_2)\le CR^{-2}d_{X_R}(m_1, m_2).
\end{align*}
provided that $R\ge \max(4M, 4\pi e^2|c_1|^2/M)$.

Choosing $R$ large enough, we obtain that $\mathscr{L}: X_R\to X_R$ and $\mathscr{L}$ is a contraction on $X_R$. The contraction mapping theorem then gives a fixed point $m_s\in X_R$, which is the unique solution to \eqref{stationary-mass}.

Now we define
\begin{align}
 \phi_{\mathrm{s}}(r)
 &=c+c_1\int_r^\infty\frac1{s^2}
       \exp\left[\int_s^\infty
       \frac{2m_{\mathrm{s}}(t)}{t-2m_{\mathrm{s}}(t)}\frac{\mathrm dt}{t}\right]\mathrm ds,\notag\\
    \alpha_{\mathrm{s}}(r)
 &=\phi_{\mathrm{s}}(r)+r\frac{\mathrm d\phi_{\mathrm{s}}}{\mathrm dr}, \label{stationary-fields}\\
 (-\partial_ur)_{\mathrm{s}}(r)
 &=\frac12\exp\left[-\int_r^\infty
       \frac{2m_{\mathrm{s}}(s)}
            {s-2m_{\mathrm{s}}(s)}
        \frac{\mathrm ds}{s}\right]. \notag
\end{align}
A double null chart for these fields is obtained by setting
\[
 v=u+\int_R^r
       \frac{\mathrm ds}{(-\partial_ur)_{\mathrm{s}}(s)}.
\]
It has $(\partial_vr)_{\mathrm{s}}
=(-\partial_ur)_{\mathrm{s}}$.
Differentiating the integral identities \eqref{stationary-mass} and \eqref{stationary-fields} in $r$ then gives  \eqref{stationary-equations}. Hence these are the stationary 
neutral equations in Bondi coordinates $(u,r)$. Integrating the equation for $\diff \alpha_r/\diff r$ also yields \eqref{Eqn:integralalpha}.

Furthermore, owing to the fact that $m_{\mathrm{s}}\in X_R$, applying L'H\^opital's rule, we can verify that when $r$ approaches $\infty$,
\begin{align*}
     &\phi_{\mathrm{s}}\to c, \qquad\quad m_{\mathrm{s}}\to M, \qquad\quad (-\partial_u r)_{\mathrm{s}}\to \frac12,\\
     &\lim\limits_{r\to\infty} r(\phi_{\mathrm{s}}-c)=\lim\limits_{r\to\infty} \frac{-c_1r^{-2}\exp\left[\int_r^\infty
       \frac{2m_{\mathrm{s}}(s)}{s-2m_{\mathrm{s}}(s)}\frac{\mathrm ds}{s}\right]}{-r^{-2}}=c_1,
\end{align*}
which are exactly the asymptotics in \eqref{Eqn:conditioninfty} as $r=\infty$.
\end{proof}

Fix this stationary solution on the exterior region $\{r\ge R\}$.
Combining with \eqref{Eqn:imporved-bounds-phi-c}, we obtain that 
\begin{equation}\label{primitive-bound}
 |r(\phi(u,r)-c)|
       +|r(\phi_{\mathrm{s}}(r)-c)|\leq C \qquad\quad\text{for all } u\in\mathbb{R}, r\ge R.
\end{equation}

\subsection{Exterior equality}
\label{exterior-equality-section}
We now use the common asymptotic data to show that the eternal
solution agrees with the stationary solution at all sufficiently
large radius.
\begin{lemma}\label{exterior-equality-lemma}
There exists $R_1>R$ sufficiently large such that,
\begin{equation*}
    (m, \alpha, \phi, -\partial_u r)(u, r)=(m_{\mathrm{s}}, \alpha_{\mathrm{s}}, \phi_{\mathrm{s}}, (-\partial_u r)_{\mathrm{s}})(r) \qquad\quad \text{for any } u\in\mathbb{R}, \, r\ge R_1.
\end{equation*}
\end{lemma}

\begin{proof}
Throughout this proof, all suprema are over the domain
$\{u\in\mathbb R$ and $r\geq R_1\}$. By \eqref{asymptotic-estimates},
\eqref{stationary-class}, and \eqref{primitive-bound}, we have
\[
 \sup|r(\phi-\phi_{\mathrm{s}})|+
 \sup r|m-m_{\mathrm{s}}|\le C.
\]
Note also that the difference of the two mass-ratio coefficients is
bounded by
\begin{equation*}
     \left|\frac{\mu}{1-\mu}(u,s)
       -\frac{2m_{\mathrm{s}}(s)}
                    {s-2m_{\mathrm{s}}(s)}\right|
       \leq\frac C{s^2}\sup r|m-m_{\mathrm{s}}|.
\end{equation*}
As a result, in light of \eqref{exterior-incoming} and \eqref{Eqn:integralalpha}, using \eqref{eternal-mass}, \eqref{primitive-bound}, and the fact that $m_{\mathrm{s}}\in X_R$, we are able to bound $\alpha(u, r)-\alpha_{\mathrm{s}}(r)$ by
\begin{equation}\label{exterior-alpha}
    \begin{aligned}
 |\alpha(u,r)-\alpha_{\mathrm{s}}(r)|
 \leq{}&\frac{CM}{r^2}\sup|r(\phi-\phi_{\mathrm{s}})|
       +\frac{C}{r^3}\sup r|m-m_{\mathrm{s}}|+\left(e^{\frac{C\sup r|m-m_{\mathrm{s}}|}{r^2}}-1\right)\cdot\frac{CM}{r^2} \\
      \leq{}&\frac{CM}{r^2}\sup|r(\phi-\phi_{\mathrm{s}})|
       +\frac{C}{r^3}\sup r|m-m_{\mathrm{s}}|.
\end{aligned}
\end{equation}
Here we use the fact that $r\ge R_1\ge R\ge 4M$.
\vspace{2mm}

Regarding $\phi-\phi_{\mathrm{s}}$, employing \eqref{eternal-phi} and \eqref{stationary-fields},  we have
\begin{equation*}
    \left.\frac{\partial}{\partial r}\right|_u \big(r(\phi-\phi_{\mathrm{s}})\big)=\alpha-\alpha_{\mathrm{s}}.
\end{equation*}
Noting that $r(\phi-c), r(\phi_{\mathrm{s}})$ share the same limit at $r=\infty$ from Lemma~\ref{asymptotic-lemma} and Lemma~\ref{stationary-exterior-lemma}, the integration from $r=\infty$ then gives
\begin{equation}\label{Eqn:rphi-phis}
     r(\phi(u,r)-\phi_{\mathrm{s}}(r))
       =-\int_r^\infty
             (\alpha(u,s)-\alpha_{\mathrm{s}}(s))\,\mathrm ds.
\end{equation}
Substituting \eqref{exterior-alpha} into this, we get
\begin{equation}\label{exterior-scalar-absorption}
 \sup|r(\phi-\phi_{\mathrm{s}})|
 \leq\frac{CM}{R_1}\sup|r(\phi-\phi_{\mathrm{s}})|
       +\frac{C}{R^2_1}\sup r|m-m_{\mathrm{s}}|.
\end{equation}

Proceeding similarly as above, based on the associated equations for $m,m_{\mathrm{s}}$ in \eqref{eternal-radial-mass}, \eqref{stationary-equations} and the fact that $\lim_{r\to\infty}m(u, r)=\lim_{r\to\infty}m_{\mathrm{s}}( r)=M$, we derive the estimate for $m-m_{\mathrm{s}}$ as below:
\begin{equation}\label{Eqn:estm-ms}
    |m(u, r)-m_{\mathrm{s}}(r)|\le C\int_r^{\infty} \left|(1-\mu)|\alpha-\phi|^2(u, s)
       -\left(1-\frac{2m_{\mathrm{s}}}s\right)
          |\alpha_{\mathrm{s}}-\phi_{\mathrm{s}}|^2(s)
 \right| ds
\end{equation}
Observe that from \eqref{Eqn:imporved-bounds-phi-c},  \eqref{stationary-fields} and \eqref{primitive-bound}, it holds
\begin{equation*}
    |\alpha-\phi|+|\alpha_{\mathrm{s}}-\phi_{\mathrm{s}}|\le\frac{C}r.
\end{equation*}
Combining with \eqref{exterior-alpha}, we obtain\footnote{Here we use the inequality $$\left||a|^2-|b|^2\right|=|\text{Re}((a-b)(\overline{a+b}))|\le |a-b|\cdot|a+b| \qquad\quad\text{for any } a,b\in\mathbb{C}.$$}
\begin{equation}\label{Eqn:est-integrand-m-ms}
    \begin{aligned}
 &\left|(1-\mu)|\alpha-\phi|^2
       -\left(1-\frac{2m_{\mathrm{s}}}r\right)
          |\alpha_{\mathrm{s}}-\phi_{\mathrm{s}}|^2
 \right|\\
 \le&\left(|\alpha-\alpha_{\mathrm{s}}|+|\phi-\phi_{\mathrm{s}}|\right)\cdot \frac{C}{r}+\frac{|m-m_{\mathrm{s}}|}{r}\cdot \frac{C}{r^2} \\
 \leq&\frac{C}{r^2}\sup|r(\phi-\phi_{\mathrm{s}})|
       +\frac{C}{r^4}\sup r|m-m_{\mathrm{s}}|.
\end{aligned}
\end{equation}
Plugging it into \eqref{Eqn:estm-ms}, we hence derive
\begin{equation}\label{exterior-mass-absorption}
 \sup r|m-m_{\mathrm{s}}|
 \leq C\sup|r(\phi-\phi_{\mathrm{s}})|
       +\frac{C}{R^2_1}\sup r|m-m_{\mathrm{s}}|.
\end{equation}
Choosing $R_1$ large enough to absorb the last term on the RHS of \eqref{exterior-mass-absorption}, we get
\begin{equation}\label{Eqn:rm-ms1}
    \sup r|m-m_{\mathrm{s}}|\le C\sup|r(\phi-\phi_{\mathrm{s}})|.
\end{equation}
Insert this bound in \eqref{exterior-scalar-absorption}
and enlarge $R_1$ if necessary, we deduce 
\begin{equation*}
    \sup|r(\phi-\phi_{\mathrm{s}})|=0.
\end{equation*}
Back to \eqref{Eqn:rm-ms1} and \eqref{exterior-alpha}, we also get
\begin{equation*}
   \sup r|m-m_{\mathrm{s}}|= \sup|\alpha-\alpha_{\mathrm{s}}|=0.
\end{equation*}
This together with \eqref{eternal-speed} proves $-\partial_u r=(-\partial_u r)_{\mathrm{s}}$ when $r\ge R_1$, which finishes the proof.
%In Bondi coordinates $(u,r)$ the metric is
%\[
% g=-\frac{4(-\partial_ur)^2}{1-\mu}\,\mathrm du^2 -\frac{4(-\partial_ur)}{1-\mu}\,\mathrm du\,\mathrm dr+r^2\,\mathrm d\sigma_{\mathbb S^2}^2.\]
%Thus these equalities identify the scalar field and the spacetime metric on the exterior. No choice of parametrization of the other null coordinate is involved in this comparison.
\end{proof}

\subsection{Interior equality and the rigidity contradiction}
\label{interior-equality-section}
The exterior equality provides matching data from which we propagate
the comparison inward. This also shows that the stationary solution
extends to every positive radius.
\begin{lemma}\label{stationary-equality-lemma}
The stationary solution constructed in Lemma~\ref{stationary-exterior-lemma} extends to $r\in (0, \infty)$, which coincides with the eternal solution from Proposition~\ref{eternal-proposition} for all $(u, r)\in \mathbb{R}\times (0,\infty)$.
\end{lemma}
\begin{proof}
Continue the stationary equations inward on their maximal interval where $r>0$ and $1-2m_{\mathrm{s}}/r>0$, and denote its left endpoint by $R_m\ge0$. The continuation criterion gives either $R_m=0$ or
\begin{equation}\label{Eqn:conditionRm}
    R_m>0, \qquad\quad 1-2m_{\mathrm{s}}(R_m)/R_m=0.
\end{equation}
Define
\begin{equation*}
    r_m=\inf \{r\in(R_m, \infty): (m, \alpha, \phi)(u, s)=(m_{\mathrm{s}}, \alpha_{\mathrm{s}}, \phi_{\mathrm{s}})(s) \quad\text{for all } u\in\mathbb{R}, \, r\le s<\infty  \}.
\end{equation*}
In view of Lemma~\ref{exterior-equality-lemma}, we have $r_m\in [R_m, R_1]$. If $r_m>R_m$, from the continuity we have
\begin{equation*}
     (m, \alpha, \phi)(u, r)=(m_{\mathrm{s}}, \alpha_{\mathrm{s}}, \phi_{\mathrm{s}})(r) \quad\text{for all } u\in\mathbb{R}, \, r\ge r_m.
\end{equation*}
Choose $\delta\in (0, r_m-R_m)$. Hence, applying \eqref{Eqn:rphi-phis}, \eqref{Eqn:estm-ms} as derived in the proof of Lemma~\ref{exterior-equality-lemma}, along with \eqref{Eqn:est-integrand-m-ms}, we deduce for any $r\in[r_m-\delta, r_m]$,
\begin{equation}
    \begin{aligned}
         |\phi(u,r)-\phi_{\mathrm{s}}(r)|
       \le& C_{r_m-\delta}\int_r^{r_m}
             \sup_{u\in\mathbb R}|\alpha(u,s)-\alpha_{\mathrm{s}}(s)|\,\mathrm ds,\\
         |m(u, r)-m_{\mathrm{s}}(r)|\le& C\int_r^{r_m} \left|(1-\mu)|\alpha-\phi|^2(u, s)
       -\left(1-\frac{2m_{\mathrm{s}}}s\right)
          |\alpha_{\mathrm{s}}-\phi_{\mathrm{s}}|^2(s) \right| ds \\
        \le& C_{r_m-\delta} \int_{r}^{r_m}   \sup_{u\in\mathbb R}
       \bigl(|\phi(u,s)-\phi_{\mathrm{s}}(s)|
            +|m(u,s)-m_{\mathrm{s}}(s)|
            +|\alpha(u,s)-\alpha_{\mathrm{s}}(s)|\bigr) \,\mathrm ds.
    \end{aligned}
\end{equation}
Comparing \eqref{exterior-incoming} and \eqref{Eqn:integralalpha}, for any $r\in[r_m-\delta, r_m]$ we obtain
\begin{equation*}
    |\alpha(u, r)-\alpha_{\mathrm{s}}(r)|\le C_{r_m-\delta} \int_{r}^{r_m}  \sup_{u\in\mathbb R}
       \bigl(|\phi(u,s)-\phi_{\mathrm{s}}(s)|
            +|m(u,s)-m_{\mathrm{s}}(s)|
            \bigr) \,\mathrm ds.
\end{equation*}
Taking the supremum over real $u$ for all above estimates, we conclude that for any $r\in[r_m-\delta, r_m]$
\begin{align}
 &\sup_{u\in\mathbb R}
       \bigl(|\phi(u,r)-\phi_{\mathrm{s}}(r)|
            +|m(u,r)-m_{\mathrm{s}}(r)|
            +|\alpha(u,r)-\alpha_{\mathrm{s}}(r)|\bigr)
                                                        \notag\\
 &\quad\leq C_{r_m-\delta}\int_r^{r_m}
       \sup_{u\in\mathbb R}
       \bigl(|\phi(u,s)-\phi_{\mathrm{s}}(s)|
            +|m(u,s)-m_{\mathrm{s}}(s)|
            +|\alpha(u,s)-\alpha_{\mathrm{s}}(s)|\bigr)
       \,\mathrm ds           \label{inward-uniqueness}
\end{align}
%These bounded suprema are measurable, since continuity permits taking the supremum over rational times. The integral form of Gronwall's inequality, with reversed radial orientation, proves equality throughout the strip.
It then follows from Gronwall's inequality that 
\begin{equation*}
     (m, \alpha, \phi)(u, r)=(m_{\mathrm{s}}, \alpha_{\mathrm{s}}, \phi_{\mathrm{s}})(r) \quad\text{for all } u\in\mathbb{R}, \, r\in [r_m-\delta, r_m].
\end{equation*}
This contradicts to the definition of $r_m$.

Therefore, we must have $r_m=R_m$. If $r_m=R_m>0$, from \eqref{eternal-mass} and \eqref{Eqn:conditionRm} we have 
\begin{equation*}
  C^{-1}\le 1-\mu(u, R_m)=1-2m_{\mathrm{s}}(R_m)/R_m=0,
\end{equation*}
again a contradiction. This gives that $r_m=R_m=0$ and thus completes the proof.
\end{proof}

\begin{proof}[Proof of Proposition~\ref{rigidity-proposition}]
By Lemma~\ref{stationary-equality-lemma}, the eternal solution in Proposition~\ref{eternal-proposition} is stationary
for all $(0, \infty)$. If $\alpha-\phi$ is nonzero at $r_0\in (0,\infty)$, from \eqref{Eqn:integral-ru} we have 
\[
 |\alpha(r)-\phi(r)|
       \geq|\alpha(r_0)-\phi(r_0)|\,\frac{r_0}{r}
             \qquad\quad \text{for all } 0<r<r_0,
\]
because $0<1-2m/r\leq1$. This contradicts
\eqref{eternal-scalar} when taking $r\to 0$. Hence $\alpha-\phi=0$ everywhere.
The mass equation \eqref{eternal-radial-mass} then makes $m$ constant. Incorporating with $\mu=2m/r\le 1$ yields $m=0$,
which contradicts $m(u,L)\geq1/2$.
\end{proof}

\section[Endpoint regularity and continuation]{Endpoint regularity and continuation}
\label{endpoint-section}\label{sec:continuation}
This section devotes to completing the proofs of \hyperref[thm:charged]{Theorem~B} and \hyperref[thm:real]{Theorem~C}. Return to the original solution under the hypotheses of \hyperref[thm:charged]{Theorem~B} or \hyperref[thm:real]{Theorem~C}, with the subscript ${}_{\mathrm{orig}}$.  If $|\alpha'_{\mathrm{orig}}|$
were unbounded, Propositions~\ref{ancient-proposition}
and~\ref{eternal-proposition} would produce an eternal neutral
solution with positive mass, contrary to
Proposition~\ref{rigidity-proposition}. Hence
\begin{equation}\label{original-alpha-bound}
 K=\sup_{\mathcal D(0,v_*)}|\alpha'_{\mathrm{orig}}|<\infty.
\end{equation}
Now we drop the subscript ${}_{\mathrm{orig}}$ and retain the charged terms. This uniform bound \eqref{original-alpha-bound}
%Constants may also depend on the bound in \eqref{original-alpha-bound}.
allows us to verify the hypothesis of
\cite[Theorem~6.2]{AnTan} and obtain endpoint regularity from
the resulting extension. 

\begin{proposition}\label{endpoint-proposition}
The original solution in \hyperref[thm:charged]{Theorem~B}
or \hyperref[thm:real]{Theorem~C} extends, with the prescribed
characteristic data, to a regular $C^1$ solution on
$\mathcal D(0,v_*+\xi)$ for some $\xi>0$.
In particular, it has all the traces in
Definition~\ref{regularity-definition} on the closed original
triangle. Moreover, for $u_0<v<v_*$,
\begin{equation}\label{vanishing-optical-tail}
 \int_{u_0}^{v}\frac{-\partial_ur}{r}
       \frac{\mu+Q^2/r^2}{1-\mu}(u,v)\,\mathrm du
 \leq CK^4r(u_0,v)^2\leq CK^4(v_*-u_0)^2.
\end{equation}
\end{proposition}

\begin{proof}
On the regular center before $v_*$,
\[
 \frac{\mathrm d}{\mathrm du}\phi(u,u)
   =\partial_vr(u,u)\alpha'(u,u).
\]
Thus $|\alpha'|\leq K$ and \eqref{outgoing-speed-bounds}
give $|\phi(u,u)|\leq CK$.
The scalar averages \eqref{scalar-averages} then yield
\[
 |\alpha|+|\phi|\leq CK,\qquad
 |\alpha-\phi|\leq \frac K2r.
\]
Integrating \eqref{radial-charge} and then \eqref{radial-mass}
from the center, using $0\leq\mu<1$ and the uniform
bound for $r$, and enlarging $K$ if necessary, we obtain
\begin{align}
 |Q|&\leq CK^2\int_0^r s^2\,\mathrm ds
       \leq CK^2r^3,\notag\\
 0\leq m&\leq C\int_0^r(K^2s^2+K^4s^4)\,\mathrm ds
       \leq CK^4r^3,\notag\\
 \mu&\leq CK^4r^2.
 \label{endpoint-orders}
\end{align}
Since $r(u,v)\leq C(v-u)$ by
Proposition~\ref{optical-proposition}, we arrive at
\[
 \sup_{u_0\leq u\leq v<v_*}\mu(u,v)
 \leq CK^4(v_*-u_0)^2\longrightarrow0
 \qquad\text{as }u_0\uparrow v_*.
\]
An--Tan's extension theorem \cite[Theorem~6.2]{AnTan} applies
to $C^1$ characteristic data and a solution regular on every
earlier triangle whenever this limiting supremum is sufficiently
small. The prescribed data extend through $v_*$, so the theorem
gives a $C^1$ extension agreeing with those data; uniqueness
identifies it with the original solution on $\mathcal D(0,v_*)$.
The real case is included by taking $\phi$ real and $Q=A_u=0$.

The regularity in \cite[Definition~2.2]{AnTan} includes
$\partial_ur,\partial_vr,\phi,r\partial_u\phi,
r\partial_v\phi,Q/r\in C^1$.
The mass constraints
\eqref{incoming-mass}--\eqref{outgoing-mass} and
\eqref{endpoint-orders} give $\mu\in C^1$, hence
$\Omega\in C^1$ by \eqref{basic-definitions}.
Radial averaging of \eqref{radial-charge} gives a continuous
$Q/r^3$. Integrating \eqref{potential-equation} from the center
first gives a continuous $A_u$; \eqref{maxwell-equations} then
gives continuity of $(\partial_uQ)/r^2$.
These quotient limits make the potential integrand and its
incoming derivative continuous up to the center, so differentiation
under the integral gives $A_u\in C^1$.
Restricting the extension to the closed original triangle gives
all the traces in Definition~\ref{regularity-definition}.

Finally, \eqref{endpoint-orders} and \eqref{uniform-margin} yield
\begin{align*}
 \int_{u_0}^{v}\frac{-\partial_ur}{r}
       \frac{\mu+Q^2/r^2}{1-\mu}(u,v)\,\mathrm du
 &\leq CK^4\int_{u_0}^{v}(-\partial_ur)r(u,v)\,\mathrm du=\frac{CK^4}{2}r(u_0,v)^2.
\end{align*}
This along with $r(u_0,v)\leq C(v-u_0)$ gives
\eqref{vanishing-optical-tail}, and hence finishes the proof. 
\end{proof}

%Together with \eqref{original-alpha-bound},Proposition~\ref{endpoint-proposition} completes the proofs of \hyperref[thm:charged]{Theorem~B} and \hyperref[thm:real]{Theorem~C}.

\appendix
\section[Charged characteristic existence]{Charged characteristic existence}
\label{characteristic-appendix}

We give the two local constructions used in the continuation
argument. The first includes the regular center and keeps exactly
one outgoing derivative of the characteristic datum.

\subsection[A uniform construction at the center]{A uniform construction at the center}

\begin{proposition}\label{center-existence}
For a fixed bound on the $C^1$ norm of the initial scalar datum and
on $|\mathfrak e|$, there is $\delta>0$ such that every admissible
datum on $[0,b]$, $0<b\leq\delta$, with
$\partial_vr(0,v)=1/2$ gives a regular charged solution on
$\{0\leq u\leq v\leq b\}$. The solution is unique in the regular
class, and its $C^1$ bounds are uniform for those data.
\end{proposition}

\begin{proof}
Consider pairs of the original quantities $\alpha,\partial_vr$,
continuous on the triangle with jointly continuous outgoing
derivatives, with their prescribed initial values.
Use the norm
\begin{equation}\label{characteristic-norm}
 \|\alpha\|_\infty+\|\partial_v\alpha\|_\infty
       +\|\partial_vr\|_\infty+\|\partial_v^2r\|_\infty.
\end{equation}
This space is complete by the fundamental theorem of calculus on
each outgoing interval. Work in a closed ball with
$1/4\leq\partial_vr\leq3/4$ and fixed bounds for the other three
terms. The ball is chosen from the initial norm bound.

For each input pair reconstruct
\begin{align}
r(u,v)&=\int_u^v\partial_vr(u,s)\,\mathrm ds,\notag\\
 \phi(u,v)&=\frac1{r(u,v)}
\int_u^v\alpha(u,s)\partial_vr(u,s)\,\mathrm ds,\notag\\
 Q(u,v)&=4\pi\mathfrak e\int_u^v
       r\partial_vr\,\operatorname{Im}
     (\phi\overline{\alpha-\phi})(u,s)\,\mathrm ds.
         \label{center-reconstruction}
\end{align}
Set their center values by continuity. Obtain $m$ from the linear
equation
\begin{equation}\label{linear-mass-reconstruction}
 \partial_vm+\frac{4\pi\partial_vr}{r}|\alpha-\phi|^2m
       =\left(2\pi|\alpha-\phi|^2+\frac{Q^2}{2r^2}\right)\partial_vr,
 \qquad m(u,u)=0.
\end{equation}
Set $\mu=2m/r$. Reconstruct the incoming radius component and the
potential by
\begin{align}
 \partial_ur(u,v)
 &=-\partial_vr(u,u)
 \exp\left[\int_u^v\frac{\partial_vr}{r}
               \frac{\mu-Q^2/r^2}{1-\mu}(u,s)\,\mathrm ds\right],
               \notag\\
 A_u(u,v)
&=\int_u^v\frac{2Q\partial_ur\,\partial_vr}{r^2(1-\mu)}(u,s)
                  \,\mathrm ds .       \label{center-metric-reconstruction}
\end{align}
At this stage the first line labels a reconstructed component.
Its equality with the actual incoming derivative of the reconstructed
radius will follow at the fixed point.

The averages in \eqref{center-reconstruction} give
\begin{equation}\label{iterate-orders}
 \begin{gathered}
 C^{-1}(v-u)\leq r\leq C(v-u),\qquad
 |\alpha-\phi|\leq C(v-u),\\
 |Q|+m\leq C(v-u)^3,\qquad
 \mu+|A_u|\leq C(v-u)^2,\qquad
 |\partial_ur|\leq C .
 \end{gathered}
\end{equation}
The mass is nonnegative by the positive forcing in
\eqref{linear-mass-reconstruction}. Decreasing $\delta$ ensures
$\mu\leq1/4$, so all reconstructions are defined.

The quotient expressions also have jointly continuous center
limits at every iterate. On the right sides of the next display,
the fields are evaluated at $(u,u)$:
\begin{align}
 \frac{\alpha-\phi}{r}
 &\longrightarrow\frac{\partial_v\alpha}{2\partial_vr},\notag\\
 \frac Q{r^3}
 &\longrightarrow\frac{2\pi\mathfrak e}{3\partial_vr}
          \operatorname{Im}(\alpha\overline{\partial_v\alpha}),\notag\\
 \frac\mu{r^2}
&\longrightarrow\frac{\pi|\partial_v\alpha|^2}
        {3(\partial_vr)^2},\qquad
 \frac{\partial_v\mu}{r}
\longrightarrow\frac{2\pi|\partial_v\alpha|^2}{3\partial_vr}.
         \label{iterate-limits}
\end{align}
Rescale each average to the unit interval to obtain these limits.
They require continuity of the first outgoing derivatives only.

Define the updated $\alpha$ and $\partial_vr$ by integrating their
incoming equations \eqref{alpha-transport} and \eqref{mixed-radius}
from their fixed initial values. The ordinary scalar source is
\[
 -\frac{\partial_ur}{r}\frac{\mu-Q^2/r^2}{1-\mu}(\alpha-\phi)
       -\mathrm i\mathfrak e A_u\alpha
       -\frac{\mathrm i\mathfrak e Q\partial_ur}{r(1-\mu)}\phi.
\]
It is $O((v-u)^2)$, and its outgoing derivative is $O(v-u)$.
The outgoing-speed source is $O(v-u)$, and its outgoing derivative
is bounded. The latter has center limit
$-\pi(\partial_vr)|\partial_v\alpha|^2/3$; the scalar source
derivative has center limit zero. Thus the map preserves the
continuity required by \eqref{characteristic-norm}.

For completeness, its difference estimates have the same orders.
If two input pairs differ by $d$ in \eqref{characteristic-norm},
their reconstructed radii differ by $C(v-u)d$, their scalar
differences $\alpha-\phi$ by $C(v-u)d$, and their charges and masses
by $C(v-u)^3d$. Their mass ratios and potentials differ by
$C(v-u)^2d$, and their incoming speeds by $Cd$.
The scalar sources and their outgoing derivatives consequently
differ by $C(v-u)^2d$ and $C(v-u)d$.
The speed sources and their outgoing derivatives differ by
$C(v-u)d$ and $Cd$. The quotient limits
\eqref{iterate-limits} justify the same estimates at the center.
Integration in $u$ makes the map a contraction with constant
$C\delta$ in \eqref{characteristic-norm}. A further decrease of
$\delta$ also preserves the chosen ball. The contraction mapping
theorem gives a fixed point.

We next verify the remaining equations and the regularity at
the center. At the fixed point, $\partial_u\alpha$ and the
incoming derivative of the outgoing speed are continuous sources.
Differentiating the first integral in
\eqref{center-reconstruction} gives the actual $\partial_ur$.
Its outgoing derivative and its center value agree with the
reconstructed component in \eqref{center-metric-reconstruction},
so they are equal.

Differentiating the scalar average in the incoming parameter gives
\begin{align}
 r\partial_u\phi(u,v)
 ={}&\partial_vr(u,u)\,[\phi(u,v)-\phi(u,u)]\notag\\
 &+\int_u^v\left[
       \partial_vr\,\partial_u\alpha
       +(\alpha(u,s)-\phi(u,v))\partial_u\partial_vr
             \right](u,s)\,\mathrm ds .
                                                    \label{local-incoming-average}
\end{align}
It has the continuous center limit
$\partial_u\phi=\partial_v\alpha/2
=(\partial_vr)\alpha'/2$.
The $\alpha$ equation then gives the covariant wave equation.
Parameter differentiation of the charge integral and the linear
mass equation gives continuous $\partial_uQ,\partial_um$, both
of order $(v-u)^2$.

The incoming Maxwell defect is
\[
 \partial_uQ+4\pi\mathfrak e r^2
                        \operatorname{Im}(\phi\overline{D_u\phi}).
\]
Its outgoing derivative is zero by the two ordered scalar
transports, and its center value is zero.
It therefore vanishes. With this current equation, direct
differentiation gives
\begin{align}
 &\partial_v\left[
       \partial_um-\frac{2\pi r^2(1-\mu)|D_u\phi|^2}{\partial_ur}
                    -\frac{Q^2\partial_ur}{2r^2}\right]\notag\\
 &\quad=-\frac{4\pi r|\partial_v\phi|^2}{\partial_vr}
       \left[
       \partial_um-\frac{2\pi r^2(1-\mu)|D_u\phi|^2}{\partial_ur}
                    -\frac{Q^2\partial_ur}{2r^2}\right].
                                                    \label{incoming-constraint-defect}
\end{align}
The bracket has zero center value, so the incoming mass constraint
holds as well.

Now differentiate the incoming-speed reconstruction in $u$.
The constraints just proved and \eqref{iterate-limits} make the
derivative of its exponent integrand continuous at the center.
The center speed has derivative $\partial_v^2r(u,u)$.
Thus $\partial_ur$ is $C^1$. Differentiating the potential integral
gives $A_u\in C^1$, with both derivatives of order $v-u$.
The current equations give $Q/r\in C^1$.

\begin{samepage}
To obtain the pure incoming weighted scalar derivative, use
\eqref{incoming-ratio-equation}. The constraints and
\eqref{iterate-limits} give continuous incoming derivatives of its
coefficients. Its center value
$\phi(u,u)$ has continuous derivative $\partial_vr\,\alpha'$.
To state the differentiated integral equation explicitly, all the
integrands below are evaluated at $(u,s)$:
\begin{equation}
\begin{aligned}
 &\partial_u\left(\frac{D_u(r\phi)}{\partial_ur}\right)(u,v)
       =(\partial_vr\,\alpha')(u,u)\\
 &\quad+\int_u^v\Biggl\{
 -\frac{\partial_vr}{r}\frac{\mu-Q^2/r^2}{1-\mu}
       \partial_u\left(\frac{D_u(r\phi)}{\partial_ur}\right)\\
 &\qquad\quad+
 \left[\frac{\partial_vr}{r}\frac{\mu-Q^2/r^2}{1-\mu}
       +\frac{\mathrm i\mathfrak e Q\partial_vr}{r(1-\mu)}\right]
       \partial_u\phi\\
 &\qquad\quad+
 \partial_u\left[\frac{\partial_vr}{r}
                        \frac{\mu-Q^2/r^2}{1-\mu}\right]
       \left(\phi-\frac{D_u(r\phi)}{\partial_ur}\right)
 +\partial_u\left[\frac{\mathrm i\mathfrak e Q\partial_vr}
                              {r(1-\mu)}\right]\phi
       \Biggr\}\,\mathrm ds .
\end{aligned}
                                      \label{pure-incoming-integral}
\end{equation}
\end{samepage}
There is no lower-endpoint term because the undifferentiated
integrand vanishes at the center. Its real coefficient is $O(r)$
with incoming derivative $O(1)$; its imaginary coefficient is
$O(r^2)$ with incoming derivative $O(r)$. Moreover,
$\phi-D_u(r\phi)/\partial_ur=O(r)$ and $\partial_u\phi$ is continuous.
Thus the inhomogeneous integrand in
\eqref{pure-incoming-integral} is $O(r)$ and continuous at the center.
The elementary theory of linear integral equations gives a unique
continuous solution. Difference quotients of
\eqref{incoming-ratio-equation} satisfy the same integral estimates
and converge to it, proving that it is the actual incoming
derivative. No derivative of $\alpha'$ is required.
The identity
\[
 r\partial_u\phi
       =\partial_ur\left(\frac{D_u(r\phi)}{\partial_ur}-\phi\right)
                    -\mathrm i\mathfrak e rA_u\phi
\]
proves $r\partial_u\phi\in C^1$.
The outgoing weighted derivative is $C^1$ directly from
$r\partial_v\phi=(\partial_vr)(\alpha-\phi)$.
Finally, the mass relation recovers a positive $C^1$ lapse.
This establishes Definition~\ref{regularity-definition}.

The same contraction proves local uniqueness in bounded norm balls.
Successive characteristic overlaps give uniqueness in the regular
class. Every bound and the choice of $\delta$ depend only on the
specified $C^1$ norm and $|\mathfrak e|$. This norm includes the
value of the scalar datum as well as its first outgoing derivative.
\end{proof}

\subsection[Characteristic construction at positive radius]{Characteristic construction at positive radius}

\begin{proposition}\label{positive-radius-existence}
Compatible $C^1$ characteristic side data in a region where
$r$, $-\partial_ur$, $\partial_vr$, and $1-\mu$ are bounded below
by positive constants determine a unique regular charged solution
on a sufficiently small characteristic rectangle.
A compact terminal segment with such bounds and $C^1$ traces
admits a common positive extension width when the original outgoing
data have a compatible $C^1$ continuation.
\end{proposition}

\begin{proof}
Integrate the five original quantities
\[
 r,\quad\partial_ur,\quad\phi,\quad rD_u\phi,\quad A_u
\]
in the outgoing direction, using the defining first derivative
identities, the mixed radius equation, the ordered incoming scalar
transport, and the potential equation.
Integrate
\[
 \partial_vr,\quad r\partial_v\phi,\quad Q,\quad m
\]
in the incoming direction, using their respective equations.
Their right sides are smooth functions of these nine quantities
on the stated field ranges. The uniform-norm integral map is a
contraction on a sufficiently small rectangle.

The fields have continuous derivatives in the directions of
integration. Differentiate the outgoing integral equations in the
incoming parameter. The incoming derivatives of the other group
are already its known sources, so the remaining derivatives solve
a linear outgoing integral system. Differentiating the incoming
equations in the outgoing parameter gives the corresponding linear
incoming system. This proves $C^1$ regularity from $C^1$ side data.

The difference between the actual incoming derivative of the
reconstructed radius and its incoming component has zero outgoing
derivative. Once it vanishes, the same is true for the difference
between the actual $D_u\phi$ and its reconstructed component divided
by $r$: the curvature terms cancel by
\eqref{potential-equation}. Both defects have zero side data.
The outgoing Maxwell defect
\[
 \partial_vQ-4\pi\mathfrak e r^2
                   \operatorname{Im}(\phi\overline{\partial_v\phi})
\]
has zero incoming derivative. The outgoing mass defect
\[
 \partial_vm-\frac{2\pi r^2(1-\mu)|\partial_v\phi|^2}{\partial_vr}
                    -\frac{Q^2\partial_vr}{2r^2}
\]
satisfies a homogeneous incoming linear equation with coefficient
$-4\pi r|D_u\phi|^2/\partial_ur$. Their side values vanish.
Thus all reduced equations hold.

For a compact terminal segment, choose one bounded neighborhood
of its field values, with uniform positive lower bounds for the quantities in the hypotheses. Over the fixed
finite incoming interval, differences from the terminal side obey
a Gronwall estimate bounded by $C\eta\exp(Cu_1)$, where $\eta$
is the new outgoing width. The transverse derivative systems are
linear with bounded coefficients; their changes are bounded by
$C[\omega(\eta)+\eta]\exp(Cu_1)$, where $\omega$ is the modulus
of continuity of the prescribed first derivatives.
Choosing $\eta$ small keeps the construction in this fixed
neighborhood. A finite covering gives a strip of common positive
width, with an overlap in the original solution. Uniqueness follows
from the same integral estimates.
\end{proof}

To check that the construction gives the full spherical system,
recover the lapse from
\[
 \Omega^2=-\frac{4\partial_ur\,\partial_vr}{1-\mu}.
\]
The two mass equations and the mixed radius equation imply
\begin{align}
 \partial_u\left(\frac{\partial_ur}{\Omega^2}\right)
 &=-\frac{4\pi r|D_u\phi|^2}{\Omega^2},&
 \partial_v\left(\frac{\partial_vr}{\Omega^2}\right)
 &=-\frac{4\pi r|\partial_v\phi|^2}{\Omega^2},\notag\\
 \partial_u\partial_v\log\Omega^2
 &=-8\pi\operatorname{Re}(D_u\phi\overline{\partial_v\phi})
   +\frac{2\partial_ur\,\partial_vr}{r^2(1-\mu)}
           \left(\frac{2Q^2}{r^2}-\mu\right).
                                                    \label{recovered-einstein}
\end{align}
The mixed lapse identity uses the differentiated metric equation
and the mixed scalar equation, whose derivatives were established
above. These are precisely the remaining Einstein equations.

\clearpage

%\begingroup

%\section*{Uses of generative artificial intelligence}
%OpenAI tools were used in exploratory mathematical discussions and in the preparation and editing of this paper. The authors are responsible for the mathematical arguments, verification of the claims, and final text.
%\par
%\endgroup

% Author confirmation of the declaration below is required before submission.
%\section*{Declaration of interests}
%The authors declare no competing interests.
% References embedded to keep this preprint self-contained.
% Bibliographic data are also preserved in announcement_references.bib.
\providecommand{\bysame}{\leavevmode\hbox to3em{\hrulefill}\thinspace}
\providecommand{\MR}{\relax\ifhmode\unskip\space\fi MR }
\providecommand{\MRhref}[2]{%
  \href{http://www.ams.org/mathscinet-getitem?mr=#1}{#2}
}
\providecommand{\href}[2]{#2}

\end{document}